\documentclass[preprint,aps,prd,showpacs,onecolumn]{revtex4}
\usepackage{amsmath}
\usepackage{amsfonts}
\usepackage{amssymb}
\usepackage{bm}
\usepackage[dvips]{epsfig}
\usepackage[dvips]{graphics}
\usepackage{graphics}
\usepackage{float}
\usepackage{dcolumn}
\usepackage{ifpdf}
\usepackage{hyperref}
\usepackage{dcolumn}
\usepackage{multirow}

\newcommand{\sech}{\mathrm{sech}}
\newcommand{\be}{\begin{equation}}
\newcommand{\ee}{\end{equation}}
\newcommand{\bes}{\begin{subequations}}
\newcommand{\ees}{\end{subequations}}
\newcommand{\ben}{\begin{eqnarray}}
\newcommand{\een}{\end{eqnarray}}

\begin{document}
\title{Trapping Spin-$0$ particles on $p-$balls in $(D,1)$ dimensions}
\author{R. Casana$^{1}$, A. R. Gomes$^{2}$, F. C. Simas$^{1}$}
\affiliation{
$^1$ Departamento de F\'\i sica, Universidade Federal do Maranh\~ao (UFMA)
Campus Universit\'ario do Bacanga, 65085-580, S\~ao Lu\'\i s, Maranh\~ao, Brasil\\
$^2$ Departamento de F\' isica, Instituto Federal de Educa\c c\~ao, Ci\^encia e
Tecnologia do Maranh\~ao (IFMA), 65025-001, S\~ao Lu\'\i s, Maranh\~ao, Brasil
}


\begin{abstract}

p-Balls are topological defects in $(D,1)$ dimensions constructed with
$\mathcal{M}\ge 1$ scalar fields which depend radially on only $2 \le p\le D-2$ spatial
dimensions. Such defects are characterized by an action that breaks
translational invariance and are inspired on the physics of a brane with
$D-p$ extra dimensions. Here we consider the issue of localization of bosonic
states described by a scalar field $\Phi$ sufficiently weak to not disturb
sensibly the defect configuration. After describing the general formalism, we
consider some specify examples with $\mathcal{M}=1,2$ and $3$, looking for some region
of parameters where bound and resonant bosonic states can be found. We investigate the
way the influence of the defect structure, number of radial dimensions and
coupling between the fields are related to the occurrence of bound and
resonant states.

\end{abstract}



\maketitle


\section{Introduction}

The study of topological defects \cite{vilenkin,sutcliffe} is important due
to their mathematical properties and connection to several areas of physics
such as quark confinement \cite{phenom}, cosmology \cite{cosm} and condensed
matter \cite{condmat}. From the point of view of multidimensional spacetimes,
one can cite  for instance the vortex \cite{vortex} for superconductor
physics in $(2,1)$ dimensions, the magnetic monopole \cite{monopole}
connected to cosmology in $(3,1)$ dimensions and brane models
\cite{brane-origins} in general $(p,1)$ dimensions. The simplest example of a
topological defect is the kink \cite{kink}, where the solution interpolates
between two different vacua. The kink is extended by the concept of branes
with one extra dimension
\cite{thick1,thick2,thick3,thick4,thick5,thick6,thick7,thick8,thick9,thick10,thick_rev},
where the brane structure is a result of an action with a dynamical scalar
field. The tentative of solving the hierarchy and the constant cosmological
problems with one extra dimension always faced a fine-tuning problem
\cite{brane-origins}. Some tentative to evade this problem included to
consider more than one extra dimension. The literature has several
interesting examples of topological defects with larger  {codimension
number}. In 6 dimensions with codimension 2 one can cite gravity localization
on strings \cite{6dim} and baby-Skyrmion branes \cite{skyr}. Higher
codimension topological defects are studied in \cite{highercodim}.

In general codimension-1 brane models can be easily treated using the
first-order formalism \cite{thick1}. In such cases the scalar field potential
and metric can be described by first-order differential equations in terms of
a function $W$ called fake superpotential. This name refers to the flat
spacetime analog where a true superpotential is introduced to find BPS states
\cite{bps}. The  BPS formalism for some defects with radial symmetry
constructed with a scalar field  was introduced in Refs. \cite{bmm1,bmm2}.
Recently, inspired in brane models with codimension-2, {such study was
extended  for the} case of two coupled scalar fields in $(3,1)$-dimensions
with the aim to {investigate in absence of gravity resonances and
localization of } particles with spin-$0$ \cite{tube_scalar} and spin-$1/2$
\cite{tube_fermion} in axial symmetric topological defects. {Such
analysis has shown similar results with the previous works dealing} with
gravity and particle localization and resonances \cite{brane_ress} on branes
with codimension 1.

{In presence of gravity, the lack of a  first-order formalism for
brane-models with codimension higher than 1 make the analysis of field
localization and resonance much more involved, needing in general numerical
analysis for finding the way the scalar fields depend on the extra dimensions.
However, by neglecting gravity effects, in this work we show that is
possible to implement a first-order formalism to describe topological defects
generated by scalar fields with pure radial dependence. Consequently, we
analyze the effects in the trapping of spin-$0$ fields due to the higher
codimension of the topological defect.}

This paper is presented in the following way: In Section II we consider the
general fist-order formalism for a $(D,1)-$dimensional flat spacetime where
$\mathcal{M}\ge1$ scalar fields depend radially on $p$ spatial dimensions, with $2 \le
p\le D-2$. Next we apply the formalism for a certain number of specific
cases. Then, in Sects \ref{sec_1field}, \ref{sec_2field} and \ref{sec_3field}
we consider defects formed respectively by one, two and three field models,
looking for some aspects of localization of a spin-0 field in each system,
and numerically investigating the occurrence of bound states and resonance
effects. Our main conclusions concerning to a comparative analysis of the
influence of $p$, $D$ and $\mathcal{M}$ on the number and intensity of resonances are
presented in Sect. \ref{sec_concl}.


\section{General formalism}


We start with the action
\begin{eqnarray}
S=\int dt d^D x  \bigg(\sum_{i=1}^\mathcal{M}\frac12 \partial_\alpha \phi^i
\partial^\alpha \phi^i - V(\phi^1,...,\phi^\mathcal{M}) \bigg)
\label{action}
\end{eqnarray}
with $\alpha=0,...,D$. The $(D,1)$-dimensional cartesian coordinates will be
separated in $D-p$-dimensions $(x^1,x^2,...,x^{D-p})$ {where the fields can be located and the remaining} $p$ transverse dimensions
$(x^{D-p+1},...,x^D)$, with $2 \le p\le D-2$, where the defect will be formed.

{The potential is chosen to be}
\begin{equation}
V(\phi ^{1},...,\phi ^{\mathcal{M}})=\frac{1}{2r^{N}}\sum_{i=1}^{\mathcal{M}}\left( W_{\phi
^{i}}\right) ^{2},  \label{potential}
\end{equation}%
{where we have used a simplified notation:}
\begin{equation}
W_{\phi _{i}}=\frac{\partial W}{\partial {\phi _{i}}},~W_{\phi _{i}\phi
_{j}}=\frac{\partial ^{2}W}{\partial \phi _{i}\partial \phi _{j}}, \label{xx}
\end{equation}
{and so on. The explicit dependence of the potential on} $r$,
\begin{equation}
r=\sqrt{\sum_{i=D-p+1}^{D}(x^{i})^{2}},
\end{equation}%
{follows closely and generalizes for $\mathcal{M}$ scalar fields the construction
of Ref. \cite{bmm1,bmm2}, initially motivated for avoiding the
Derrick-Hobart's theorem \cite{Hob}-\cite{Raj}. We also suppose that the scalar fields $\phi^{i}$ depend only on $r$.}

{The equations of motion for the scalar fields read}
\begin{equation}
\square\phi^{i}+\frac{1}{r^{N}}\sum_{j=1}^{\mathcal{M}}W_{\phi ^{j}}
W_{\phi ^{j}\phi ^{i}}=0,\;\;i=1,...,\mathcal{M},  \label{eom_gen0}
\end{equation}
{and the ones describing static solutions are}
\begin{equation}
\nabla ^{2}\phi ^{i}=\frac{1}{r^{N}}\sum_{j=1}^{\mathcal{M}}W_{\phi ^{j}}
W_{\phi ^{j}\phi ^{i}}
,\;\;i=1,...,\mathcal{M},  \label{eom_gen}
\end{equation}%
where $\nabla ^{2}$ is the $p-$dimensional Laplacian, defined by
\begin{equation}
\nabla ^{2}\phi ^{i}=\frac{1}{r^{p-1}}\frac{d}{dr}\left( r^{p-1}\frac{d\phi
^{i}}{dr}\right) .  \label{nabla2}
\end{equation}%
The energy density is given by
\begin{equation}
\rho (r)=\frac{1}{2}\sum_{i=1}^{\mathcal{M}}\left[\left(\nabla \phi
^{i}\right)^{2} +\frac{1}{r^{N}} \left(W_{\phi^{i}}\right)^{2}\right],  \label{en1}
\end{equation}
{and the total energy of the defect in the transverse volume is}
\begin{equation}
E=\int dx^{D-p+1}...dx^{D}\,\rho (r). \label{en2}
\end{equation}%

{In order to describe the system via first-order differential equations, we implement the BPS formalism such that the total
energy can be written as}
\begin{equation}
E=\frac{2\pi ^{p/2}}{\Gamma (p/2)}\frac{1}{2}\sum_{i=1}^{\mathcal{M}}\int dr~r^{p-1}%
\left[ \left( \frac{d\phi ^{i}}{dr}\mp \frac{1}{r^{N/2}}W_{\phi ^{i}}\right)
^{2}\pm \frac{2}{r^{N/2}}\frac{d\phi ^{i}}{dr}W_{\phi ^{i}}\right] .
\end{equation}%
{By setting as null the squared term we get the set of first-order differential equations}%
\begin{equation}
\frac{d\phi ^{i}}{dr}=\pm \frac{1}{r^{N/2}}W_{\phi ^{i}},i=1,...,\mathcal{M},
\label{phi_D0}
\end{equation}%
{whose solutions we label as }$\phi _{\pm }^{i}(r)${. Whenever
these first-order equations are satisfied and using (\ref{xx}), the total energy
reads}%
\begin{equation}
E=\pm \frac{2\pi ^{p/2}}{\Gamma (p/2)}\sum_{i=1}^{\mathcal{M}}\int dr~r^{p-1}\frac{1}{%
r^{N/2}}\frac{dW}{dr}.
\end{equation}%
{At this point we observe the integrand can be transformed in a total
derivative whether $N=2p-2$. In this way the BPS energy reads}
\begin{equation}
E_{_{BPS}}=\frac{2\pi ^{p/2}}{\Gamma (p/2)}|W(r\rightarrow \infty )-W(r=0)|,
\end{equation}
and from Eq. (\ref{phi_D0}) the corresponding BPS equations read
\begin{equation}
\frac{d\phi ^{i}}{dr}=\pm \frac{1}{r^{p-1}}W_{\phi ^{i}},i=1,...,\mathcal{M}.
\label{phi_D}
\end{equation}

{A similar result about parameter $N$ can be obtained by considering scaling properties of the scalar fields in the energy density (\ref{en2}). Firstly, we define the vectors $\vec\phi=(\phi^1,...,\phi^\mathcal{M})$ and $\vec
r=(x^{D-p+1},...,x^D)$. We make the scaling transformation $\vec r \to\lambda\vec r$ and $\vec\phi(\vec
r)\to\vec\phi(\lambda\vec r)$,} corresponding to a change in the energy given
by $E\to E^\lambda$, and impose $(\partial
E^\lambda)/(\partial\lambda)|_{\lambda=1}=0$. This leads to the following
restrictions on $N$ and $p$: i) for $p=1$, $N=0$; ii) for $p=2$, $N=2$.
{Additionally by imposing the equality} of the gradient and potential
parts of $E$, we get iii) for $p\ge3$, $N=2p-2$. {A similar result was}
previously found in \cite{bmm1}, for the case where $D=p$.

{Hence, we have shown the system (\ref{eom_gen}) admits topological
solutions obtained from the set of self-dual equations (\ref{phi_D}) which
minimize the system energy (\ref{en2}).}

The topological character of the solutions $\phi^i_\pm$ can be demonstrated
following closely Ref. \cite{bmm1}, with the difference that there one has
$p=D$. For $(1,1)$-dimensions we have $\mathcal{M}$ conserved currents $j^{i\mu}=\epsilon^{\mu\nu}\partial_\nu\phi^i$, with $\mu=0,1$ and $i=1,...\mathcal{M}$. This results in $\mathcal{M}$ conserved quantities
$\sigma^i=d\phi^i/dx$, such that $\rho=\sum_i(\sigma^i)^2$ is the energy of the field
configuration \cite{bbrito} and the topological charge
$Q_T=\int_{-\infty}^{\infty}dx \rho$ is also the total energy of the
solution. However, for the class of defects described here one must be in
$(D,1)$-dimensions with $D\ge4$, with the scalar fields depending on $p\ge2$
spatial dimensions. For the minimum case, with $D=4$ and $p=2$ we have $\mathcal{M}$
current tensors $j^{i\mu_1\mu_2}=\epsilon^{\mu_1\mu_2\mu_3}
\partial_{\mu_3}\phi^i$, where each $\mu_1,\mu_2,\mu_3$ can assume the values $0,1,2$.
We have $\partial_{\mu_1}j^{i\mu_1\mu_2}=0$. For each scalar field this gives the set of two
conserved densities $\sigma^{i k_1}=j^{i 0k_1}$, where $k_1=1,2$. The scalar
quantity $\rho=\sigma_{ik_1}\sigma^{ik_1}=j_{i0k_1}j^{i0k_1}=\epsilon_{i0k_1k_2}
\epsilon^{i0k_1k_3}\partial^{k_2}\phi^i\partial_{k_3}\phi^i=-\sum_i(d\phi^i/dr)^2$ can be
used to define the topological charge as $Q_T=\int d\vec{r} \rho
=-\Omega\Delta W$, which coincides with the energy of the defect in the
transverse volume. Finally for general $(D,1)$-dimensions with $p\ge 2$
transverse dimensions, we have $\mathcal{M}$ current tensors $j^{i\mu_1\mu_2...\mu_p}
=\epsilon^{\mu_1\mu_2...\mu_p\mu_{p+1}}\partial_{\mu_{p+1}}\phi^i$
with $\partial_{\mu_1}j^{i\mu_1\mu_2...\mu_p}=0$. This gives, for each scalar field, the set of $p$
conserved densities $\sigma^{ik_1k_2...k_{p-1}}=j^{i0k_1k_2...k_{p-1}}$. The
scalar quantity $\rho=\sigma_{ik_1k_2...k_{p-1}}\sigma^{ik_1k_2...k_{p-1}}
=(-1)^p(p-1)!\sum_i(d\phi^i/dr)$ leads to the topological charge $Q_T=\int d\vec{r}
\rho=(-1)^p(p-1)!\Omega_p\Delta W$, which coincides with the energy
density of the defect in the transverse volume.

{In the following we show that the solutions $\phi^i_\pm(r)$ are
stable under radial and time-dependent fluctuations. For such a purpose, we
follow the procedure realized in Ref. \cite{bbb} for domain walls with two
scalar fields. Thus, we construct the function}
\be
\phi^k(r,t) = \phi^k_\pm(r) + \sum_n \eta^k_n (r) e^{i\omega_n t}.
\ee
{By substituting it in Eq. (\ref{eom_gen0}) and keeping only linear terms in the fluctuations} $\eta^k_n (r)$, we get
\be
\label{eom-M}
\left( - \nabla^2 + \frac1{r^{2p-2}} \mathbb{M}\right)
{\bf \eta}_n = \omega_n^2 {\bf\eta}_n,
\ee
{where we have defined the matrix $\mathbb{M}$ and the eigenvector ${\bf\eta}_n$ by }
\be
\!\!\mathbb{M}=
\begin{pmatrix}
V_{\phi^1\phi^1} & V_{\phi^1\phi^2} & \hdots & V_{\phi^1\phi^\mathcal{M}}\\[0.1cm]
V_{\phi^2\phi^1} & V_{\phi^2\phi^2} & \hdots & V_{\phi^2\phi^\mathcal{M}}\\[0.1cm]
\vdots & \vdots & \hdots & \vdots\\[0.1cm]
V_{\phi^\mathcal{M}\phi^1} & V_{\phi^\mathcal{M}\phi^2} & \hdots & V_{\phi^\mathcal{M}\phi^\mathcal{M}}
\end{pmatrix}, \; {\bf\eta}_n=
\begin{pmatrix}
 \eta^1_n \\[0.1cm]
\eta^2_n \\[0.1cm]
\vdots\\[0.1cm]
\eta^\mathcal{M}_n
\end{pmatrix}.
\ee

{We have verified that}
\be
\frac1{r^{2p-2}} \mathbb{M} = \pm\frac{d\mathbb{G}}{dr} \pm \frac{p-1}r \mathbb{G} + \mathbb{G}^2,
\ee
{in this section, upper and lower signals are for, respectively,
$BPS$ solutions $\phi_\pm^i$ of Eq. (\ref{phi_D}) and we have defined the matrix $\mathbb{G}$ }
\be
\mathbb{G}=
\frac1{r^{p-1}}
\begin{pmatrix}
W_{\phi^1\phi^1} & W_{\phi^1\phi^2} & \hdots & W_{\phi^1\phi^\mathcal{M}}\\[0.1cm]
W_{\phi^2\phi^1} & W_{\phi^2\phi^2} & \hdots & W_{\phi^2\phi^\mathcal{M}}\\[0.1cm]
\vdots & \vdots & \hdots & \vdots\\[0.1cm]
W_{\phi^\mathcal{M}\phi^1} & W_{\phi^\mathcal{M}\phi^2} & \hdots & W_{\phi^\mathcal{M}\phi^\mathcal{M}}
\end{pmatrix}
\ee

Now this in Eq. (\ref{eom-M}) leads to the useful factorization
\be
\label{eomS+S-}
\widehat A_\pm \widehat B_\pm {\bf\eta}_n= \omega_n^2 {\bf\eta}_n,
\ee
where the operators are defined as
\be
\label{Apm}
\widehat A_\pm = - {\bf 1} \frac{d}{dr} \mp \mathbb{G} - \frac{(p-1)}r {\bf 1}
\ee
and
\be
\label{Bpm}
\widehat B_\pm =  {\bf 1} \frac{d}{dr} \mp \mathbb{G}.
\ee
Note that, for  one scalar field (i.e., $\mathcal{M}=1$), this factorization differs from the presented in Ref. \cite{bmm1}. The advantage is that now one can verify explicitly that these operators are such that $\widehat A_\pm^\dagger = \widehat B_\pm$, that is,
in $p$ spatial dimensions,
\be
\int dr r^{p-1} (\widehat A_\pm \psi)^\dagger \psi = \int dr r^{p-1} \psi^\dagger \widehat B_\pm  \psi,
\ee
provided we impose the boundary condition
\be
r^{p-1} \psi^\dagger \psi|_{r=0}^\infty=0,
\ee
a condition valid if $\psi$ are square-integrable bound states (not scattering states). From this analysis we can rewrite Eq. (\ref{eomS+S-}) as
\be
\widehat B_\pm^\dagger \widehat B_\pm {\bf\eta}_n
= \omega_n^2
{\bf\eta}_n,
\ee
which means that the $\omega_n^2$ are eigenvalues of a non-negative operator $\widehat B_\pm^\dagger \widehat B_\pm$.
This proves that negative eigenvalues are absent and that the $p-$balls which satisfy the set of first-order equations given by Eq. (\ref{phi_D}) are stable. The lowest bound state is given by the zero-mode, identified as $\widehat B_\pm {\bf\eta}_n = 0$, which gives $\eta^i_n=c W_{\phi^i}$ for the $\mathcal{M}$ components of ${\bf\eta}_n$, where $c$ is the normalization constant, such that
\be
\int dr r^{p-1} {\bf\eta}_n^\dagger {\bf\eta}_n = 1.
\ee
Further, note that the presence of an explicit dependence with $r$ in $\widehat A_\pm$ and its absence in $\widehat B_\pm$ (compare Eqs. (\ref{Apm}) and (\ref{Bpm})) introduces an asymmetry necessary for the condition $\widehat A_\pm^\dagger = \widehat B_\pm$ to be valid. In this way there is an extension of the usual symmetric form of factorization for problems with $p=1$, the $(1,1)$-dimensional kink being an archetype (see Eq. (3.5) from Ref. \cite{bbb}). Specific factorizations of the Hamiltonian where also attained in other contexts, for instance for quantum systems with position-dependent masses \cite{hott_ho}.

Now let us turn to the search of explicit solutions for the dependence in the
radial dimension of the scalar fields. A convenient way to solve Eq.
(\ref{phi_D}) is to make a change of variables $d\xi=1/r^{p-1} dr$, or
equivalently
\be
\label{xir_p2}
\xi(r)=\ln (r/r_{0}), \; p=2
\ee
or
\be
\label{xir_p}
\xi(r)=\frac1{p-2} \biggl( -\frac1{r^{p-2}} + \frac1{r_0^{p-2}}\biggr), \; p=3,4,....
\ee
This coordinate transformation turns Eq. (\ref{phi_D})  in
\begin{eqnarray}
\label{phixi}
\frac{d\phi}{d\xi}&=&W_\phi.
\end{eqnarray}
After solving this equation for $\phi(\xi)$, and back to $r$ variable, explicit expressions for the scalar field and energy density can be easily attained. Now to form a topological defect one must chose a function $W(\phi^1,\phi^2,...\phi^p)$ with $W(r\to \infty)\neq W(r=0)$. A convenient choice is a field $\phi^i$ with a kink-like pattern in $r$ around a finite value $r_0$ and the remaining other fields $\phi^j$, $j\neq i$ with kink or bell-shape pattern around the same value of $r$. In a terminology from the literature we could say the field $\phi^i$ forms the defect whereas the other fields are responsible for its internal structure.

As we saw, from the $D+1$ spacetime dimensions, the topological defect lives in $p$ of them. Now we want to consider how a spin-0 particle living in the full $D+1$-dimensional spacetime can be effectively trapped by the topological defect in a form of bound or resonant states.  Then we consider a scalar field $\Phi$ in a region where it is formed  a radial defect constructed with the $\mathcal{M}$ scalar fields $\phi^i$. In the present analysis we neglect the backreaction on the topological defect by considering that the interaction between the scalar fields is sufficiently weak in comparison to the self-interaction that generates the defect. In the following, we designate $\Phi$ as the weak field and $\phi^i, i=1,2,...,\mathcal{M}$ the strong ones. We write the following action describing the system as
\begin{eqnarray}
S_1 = \int{ dt d^Dx\bigg(\frac{1}{2} \partial_\beta\Phi \partial^\beta\Phi - \frac{\eta}{2} F(\phi^1,...,\phi^p) \Phi^2 \bigg)}.
\label{action_Phi}
\end{eqnarray}
with $\beta=0,1,...,D$. Here $F(\phi^1,...,\phi^p)$ is the coupling between the weak field $\Phi$ and the topological defect. The equation of motion of the scalar field $\Phi$ is
\begin{eqnarray}
\partial_\mu \partial^\mu \Phi - \nabla^2_T \Phi + \eta F(\phi^1,...,\phi^p) \Phi = 0,
\label{eq_motion_Phi}
\end{eqnarray}
where in the former expression we decomposed the $D+1$-dimensional d'Alembertian between the $D-p$ transverse dimensions and the $p$ transverse dimensions. That is $\partial_\mu \partial^\mu = \Box$, with $\mu=0,...,D-p$ and $\nabla^2_T$ is a $p$-dimensional Laplacian. Now considering that the strong fields $\phi^1,...,\phi^p$ depend only in the radial direction, we have a coupling $F(\phi^1,...,\phi^p)=F(r)$. We restrict our discussion to functions $F(r)$ finite for all values of $r$, with $\eta \lim_{r\to 0}F(r)=0$.

We decompose the scalar field $\Phi$ as
\begin{eqnarray}
\Phi(t,x^1,x^2,...,x^D) = \sum_{n\jmath}  \xi_{n\jmath}(t,x^1,...,x^{D-p}) \varsigma_{n,\jmath}(r) Y_\jmath(\varphi,\theta^1,...,\theta^{p-2}) = \sum_{n\jmath} \Phi_{n\jmath},
\end{eqnarray}
where $\jmath$ is related to the angular momentum eigenvalue. Here we have changed the transverse coordinates
$(x^{D-p+1},...,x^D)$ from the cartesian to the generalized spherical coordinates $(r, \varphi,\theta^1,...,\theta^{p-2})$, with $r$ defined previously and
\begin{eqnarray}
\varphi&=&\tan^{-1}(x^{D-p+2}/x^{D-p+1})\\
\theta^1&=&\tan^{-1}(\sqrt{(x^{D-p+1})^2+(x^{D-p+2})^2}/x^{D-p+3})\\
&&...\\
\theta^{p-2}&=&\tan^{-1}(\sqrt{(x^{D-p+1})^2+...+(x^{D-1})^2}/x^{D}).
\end{eqnarray}
The $p$-dimensional Laplacian is given by
\be
\nabla^2_T=\frac1{r^{p-1}}\partial_r (r^{p-1} \partial_r) - \frac1{r^2}\widehat L_p^2,
\ee
where $\widehat L_p$ is the $p$-dimensional angular momentum operator, given by (we set $\theta^1=\theta, \theta^2=w$ to ease notation)
\begin{eqnarray}
\widehat L_2^2&=&-\partial_\varphi^2,\\
\widehat L_3^2&=&-\biggl[\frac1{\sin^2\theta} \partial^2\varphi +  \frac1{\sin\theta} \partial_\theta (\sin\theta \partial_\theta) \biggr],\\
\widehat L_4^2&=&-\frac1{\sin^2 w}\biggl[\partial_w(\sin^2w\partial_w)+\frac1{\sin^2\theta} \partial^2_\varphi +  \frac1{\sin\theta} \partial_\theta (\sin\theta \partial_\theta) \biggr].
\end{eqnarray}
In general, for $p\ge3$ we have
\be
\widehat L_p^2=-\biggl[\sum_{i=2}^p \biggl( \prod_{j=i+1}^p \frac1{\sin^2\theta_j}  \biggr) \frac1{\sin^{i-2}\theta_i} \partial_{\theta_i} (\sin^{i-2} \theta_i \partial_{\theta_i})  \biggr].
\ee
Now the field $\xi_{n\jmath}(t,x^1,...,x^{D-p})$ satisfies the $(D-p)+1$-dimensional Klein Gordon equation
\begin{eqnarray}
\big(\Box + M^2_{n\jmath} \big)\xi_{n\jmath}(t,x^1,...,x^{D-p}) = 0,
\label{KG_eq}
\end{eqnarray}
and the amplitude $\varsigma_{n\jmath}(r)$ satisfies the radial Schr\"odinger-like equation
\begin{eqnarray}
-\varsigma''_{n\jmath}(r)-\frac{p-1}{r} \varsigma'_{n\jmath}(r) + V_{sch}(r) \varsigma_{n\jmath}(r) = M^2_{n\jmath} \varsigma_{n\jmath}(r),
\label{eq_rho}
\end{eqnarray}
with the Schr\"odinger potential given by
\begin{equation}
V_{sch}(r) = \frac{\jmath(\jmath + p-2)}{r^2} + \eta F(r).
\label{pot_schr}
\end{equation}

By requiring that Eq. (\ref{eq_rho}) defines a self-adjoint differential operator in  $r\in [0, +\infty)$, the Sturm-Liouville theory establishes the orthonormality condition for the components $\varsigma_{n,\jmath}(r)$
\be
\int  dr\,r^{p-1} \varsigma_{n'\jmath}(r) \varsigma_{n\jmath}(r) =  \delta_{n
n'}.
\label{orthonormality}
\ee
The spherical harmonics of degree $\jmath$ satisfy (see Ref. \cite{fe} for a
general treatment of spherical harmonics with  general number of dimensions)
\be
\label{sp_harm}
\widehat L_p^2 Y_\jmath(\varphi,\theta^1,...,\theta^{p-2})={\jmath(\jmath +
p-2)} Y_\jmath(\varphi,\theta^1,...,\theta^{p-2})
\ee
and are polynomials of degree $\jmath$ with variables restricted to the unit
$(p-1)$-sphere, which satisfy the orthonormality condition
\be
\int_{S^{p-1}}   Y_{\jmath'}(\varphi,\theta^1,...,\theta^{p-2})
Y_\jmath(\varphi,\theta^1,...,\theta^{p-2}) = \delta_{\jmath \jmath'},
\ee
which means that spherical harmonics of different orders are orthogonal.
Given a particular value of $p$, Eq. (\ref{sp_harm}) is solved by separation
of variables. Some examples are
\begin{itemize}

\item For $p=2$, $\jmath\equiv m=0,1,2,...$ and $Y_m=e^{im\varphi}$. This means
    that $Y_m$ is associated with the eigenvalue $m^2$ and carries angular
    momentum $m$. The index $m$ labels the irreducible representations of
    $SO(2)$.

\item  For $p=3$, $\jmath\equiv \ell=0,1,2,...$ and $Y_\ell
    (\varphi,\theta)=\sum_{m=-\ell}^\ell Y_{\ell m}(\varphi,\theta)$, with
    $Y_{\ell m}(\theta,\varphi) = \Theta_{\ell m}(\theta)
    \Psi_m(\varphi)$. Here $\Psi_m(\varphi)=e^{im\varphi}$ and $\Theta_{\ell m}(\theta)$ satisfies the following differential equation
\begin{eqnarray}
\cot \theta \, \frac{d \Theta(\theta)}{d \theta} & + & \frac{d^2
\Theta(\theta)}{d \theta^2} + \bigg( \ell(\ell+1)-\frac{m^2}{\sin^2
\theta}\bigg) \Theta(\theta) = 0.
\label{eq_aux_p3}
\end{eqnarray}
This means that $Y_{\ell m}$ is associated with the eigenvalue $\ell(\ell
+1)$ and carries angular momentum $\sqrt{\ell(\ell +1)}$. The index $\ell$
labels the irreducible representations of $SO(3)$ whereas $m$ labels the
corresponding representations of the subgroup $SO(2)$. For each $\ell$
there are $2\ell+1$ linearly independent spherical harmonics corresponding
to the various values of $m$. Therefore the irreducible representations of
$SO(3)$ based on $Y_{\ell m}$ are $(2\ell +1)$ dimensional \cite{avery1}.
\end{itemize}

For the general case, the irreducible representations of $SO(p)$ based on
hyperspherical harmonics have dimension given by \cite{avery2}
\be
\dim=\frac{(p+2\jmath -2) (p+\jmath -3)!}{\jmath! (p-2)!}
\ee
and we have an orthonormal set of hyperspherical harmonics which have extra indices
that are labels of the irreducible representations of the following chain of
subgroups of $SO(p)$:
\be
SO(p)\supset SO(p-1) ...\supset SO(2).
\ee
Let us illustrate how this works with one more example. The generalization
for even larger values of $p$ demands additional work but is straightforward.
\begin{itemize}
\item For $p=4$ an orthonormal set of hyperspherical harmonics have extra
    indices that are labels of the irreducible representations of the
    following chain of subgroups of $SO(4)$:
\be
SO(4)\supset SO(3) \supset SO(2).
\ee

We have $Y_\jmath (\varphi,\theta, w)=\sum_{\ell=0}^{\jmath}
\sum_{m=-\ell}^\ell Y_{\jmath \ell m}(\varphi,\theta,w)$, with the
following specific constructions:
\begin{enumerate}
\item[i)] $\jmath=0\implies \dim=1$. Then $\ell=0,m=0$ which gives $Y_{\jmath\ell m}=Y_{0,0,0}$.

\item[ii)] $\jmath=1\implies \dim=4$. Then if $\ell=0$ then $m=0$. If $\ell=1$ then $m=0,\pm1$. This gives the four possibilities for $Y_{\jmath\ell m}$.

\item[iii)] $\jmath=2\implies \dim=9$. Then if $\ell=0$ then $m=0$. If $\ell=1$ then $m=0,\pm1$. If $\ell=2$ then $m=0,\pm1,\pm2$. This gives the nine possibilities for $Y_{\jmath\ell m}$.

\item[iv)] In general, given $\jmath$, we have $\ell=0,1,...,\jmath$ ($\jmath+1$ possibilities) and $m=-\ell,...,\ell$ ($2\ell+1$ possibilities), resulting in $\dim=(\jmath+1)^2$ possible constructions for $Y_{\jmath\ell m}$.

\end{enumerate}

One can make the decomposition $Y_{\jmath\ell m}=\mathcal{W}_{\jmath\ell}(w)Y_{\ell m}(\theta,\varphi)$, where $Y_{\ell m}(\theta,\varphi)$ are the usual spherical harmonic described in the $p=3$ case, and  $\mathcal{W}_{\jmath\ell}(w)$ satisfies the following differential equation
\be
2\cot \mathcal{W}'_{\jmath\ell}(w)+\mathcal{W}''_{\jmath\ell}(w)-\ell(\ell+1)
\mathcal{W}_{\jmath\ell}(w)=\jmath(\jmath+1)\mathcal{W}_{\jmath\ell}(w),
\ee
\end{itemize}
where prime means derivative with respect to the argument.

Now the action given by the Eq. $(\ref{action_Phi})$ can be integrated in the $(x^{D-p+1},...,x^{D})$ dimensions
to give
\begin{eqnarray}
S_1 = \int{ dt dx^1...dx^{D-p} \bigg(\frac{1}{2} \partial_\mu\Phi_{n\jmath} \partial^\mu\Phi_{n\jmath} - M_{n\jmath}^2 \Phi_{n\jmath}^2 \bigg)},
\label{action_Phi_int}
\end{eqnarray}
which shows that $\Phi_{n\jmath}$ is a massive $(D-p+1)$-dimensional Klein-Gordon field with mass $M_{n\jmath}$.

In order to investigate numerically the massive states, firstly we consider the region near the origin $(r \ll r_0)$. Since we are considering only functions $\eta F(r)$ finite, the Schr\"odinger-like potential for $\jmath=0$  reads\begin{equation}
-\varsigma _{n0}^{\prime \prime }(r)-\frac{p-1}{r}\varsigma _{n0}^{\prime
}(r)=\left( M_{n0}^{2}-V^{(0)}\right) \varsigma _{n0}(r),
\end{equation}%
{{where $V^{(0)}=\lim_{r\rightarrow 0}\eta F(r)$, whose nonsingular solution at $r=0$ is }}
\begin{equation}
\varsigma _{n0}(r)=r^{1-\frac{p}{2}}J_{\frac{p}{2}-1}\left( r\sqrt{%
M_{n0}^{2}-V^{(0)} }\right). \label{Dp_rho_rsmall_0}
\end{equation}

On the other hand, for $\jmath\ge1$ the Schr\"odinger-like potential is dominated by the contribution  of  the angular momentum proportional to $1/r^{2}$,
\begin{eqnarray}
V_{sch}(r) = \frac{\jmath (\jmath + p-2 )}{r^2}, r\ll r_0
\label{pot_V(r)}
\end{eqnarray}
and the nonsingular solutions in $r=0$, given by
\begin{equation}
\varsigma _{n\jmath }^{\left( 0\right) }(r)=r^{1-\frac12 p} J_{[\frac12 (2\jmath+p-2) ]}(M_{n\jmath}r).\ \ \jmath \geq 1.
\label{Dp_rho_rsmall}
\end{equation}

{Both functions (\ref{Dp_rho_rsmall_0}) and (\ref{Dp_rho_rsmall}) are used as an input for the numerical method.}  From this approximation we can calculate
$\varsigma(r_{min})$ and $d\varsigma/dr(r_{min})$, to be used for the
Runge-Kutta method to determine $\varsigma(r)$ from the Schr\"odinger-like
equation.  We define the probability for finding scalar modes with mass
$M_{n\jmath}$ and angular momentum inside the $p-$ball of radius $r_0$ as
is \cite{liu}
\begin{eqnarray}
P = \frac{\int_{r_{min}}^{r_0} dr \,r^{p-1} \, |\varsigma_{n\jmath}(r)|^2 }{\int_{r_{min}}^{r_{max}} dr \,r^{p-1} \, |\varsigma_{n\jmath}(r)|^2 },
\label{probability}
\end{eqnarray}
here $r_{min} \ll r_0$ is used as the initial condition and $r_{max}$ is the characteristic box length used for the normalization procedure, being a value where the Schr\"odinger potentials are close to zero and where the massive modes $\varsigma(r)$ oscillate as planes waves. Resonances are characterized by peaks in the plots of $P$ as a function of $M_{n\jmath}$. The thinner is a peak, the longer is the lifetime of the resonance. This finishes the part of the general formalism. In the remaining of this work we will solve some specific examples with one, two and three scalar fields.

The number of parameters involved in the models considered here
led us to make some restrictions in order to better identify the effect
of the number of transverse dimensions for the occurrence of
bound and/or resonant states.  For instance, we have identified
that an increasing in $\jmath$ reduces the possibility of the occurrence
of bound states. {{Case $\jmath=0$ is special, since in this case we have $\lim_{r\to 0} V_{sch}(r)$ is finite or even zero,}} in opposition to $\lim_{r\to 0} V_{sch}(r)=\infty$ for $\jmath\ge1$ . Then without loosing generality we have chosen to study
states with $\jmath=0$ and $\jmath=2$.

\section{A one-field model}
\label{sec_1field}
In this section we will consider the model \cite{bmm1}
\be
W_\phi(\phi)=\lambda \bigg(\phi^{(q-1)/q} -\phi^{(q+1)/q}\bigg),
\ee
with $q=1,3,5...$. The first-order equation, described by Eq. (\ref{phixi}), has solution given by
\begin{eqnarray}
\phi(\xi)&=&\tanh^q(\lambda \xi/q).
\end{eqnarray}
The case $q=1$ corresponds to the usual kink solution of the $\phi^4$ model in the variable $\xi$.
Back to $r$ variable, explicit expressions for the scalar field and energy density can be easily attained:
\begin{eqnarray}
\phi (r) &=&\,\mathrm{tanh}^q(\eta_p), \\
\rho(r) &=& \frac{\lambda^2}{r^{2p-2}}\tanh^{2q-2}(\eta_p)\sech^4(\eta_p),  \notag
\end{eqnarray}%
with
\be
\label{eta_xi}
\eta_p=\frac\lambda q \xi(r)
\ee
with $\xi(r)$ given by Eq. (\ref{xir_p2}), for $p=2$ or (\ref{xir_p}), for $p=3, 4,...$.
\begin{figure}
\scalebox{0.9}{\includegraphics[{angle=0,width=6.0cm}]{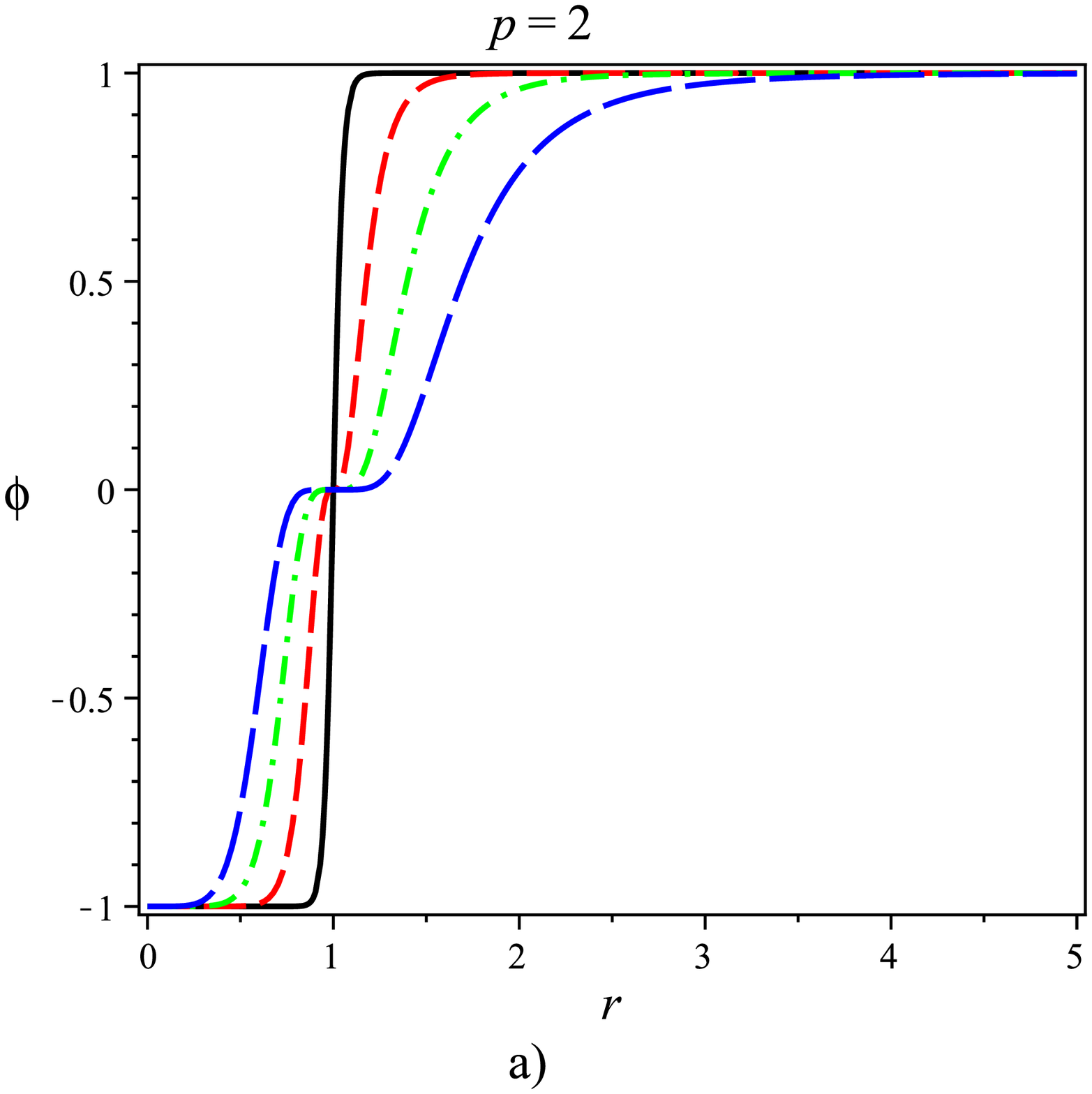}
\includegraphics[{angle=0,width=6.0cm}]{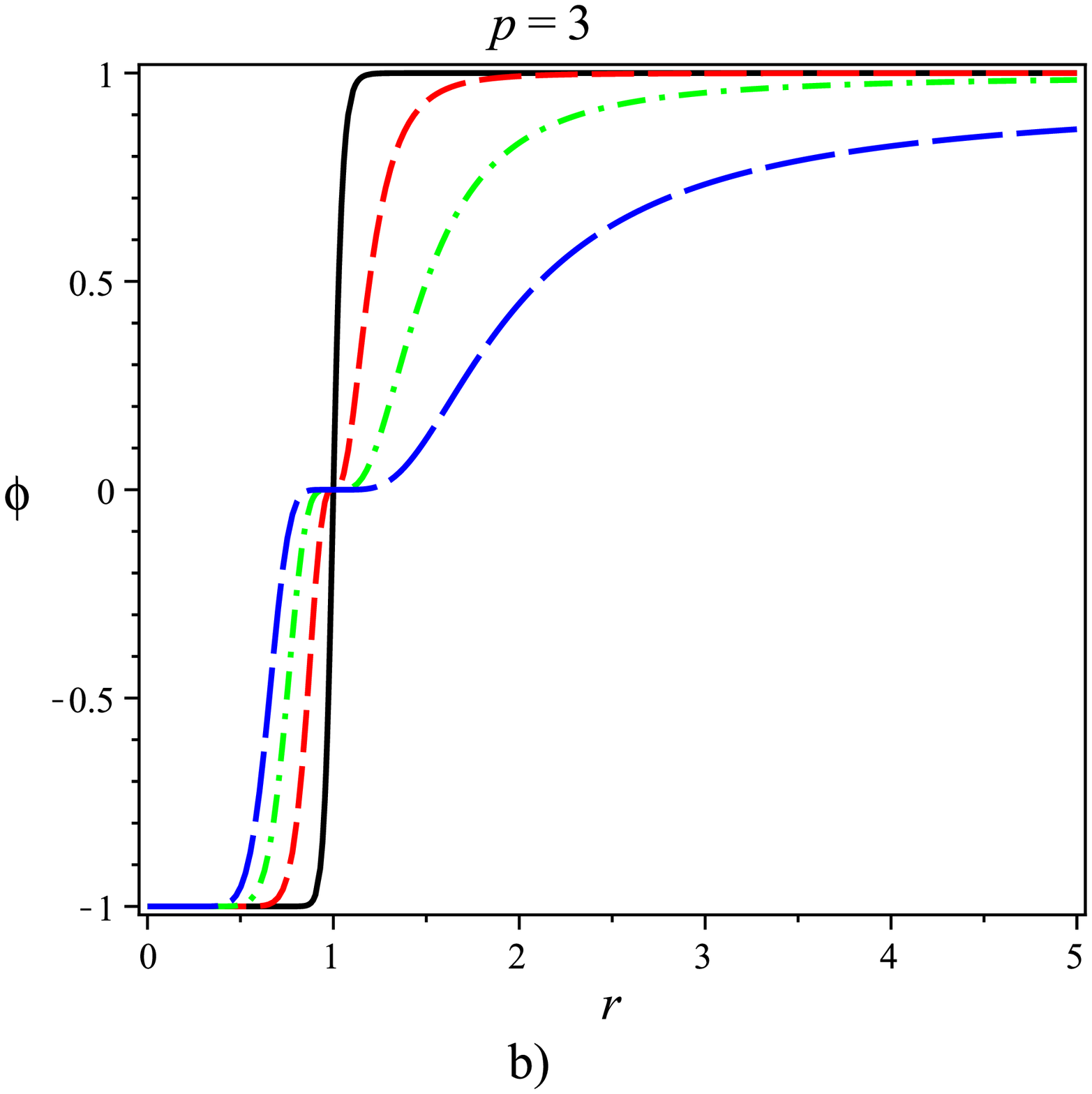}
\includegraphics[{angle=0,width=6.0cm}]{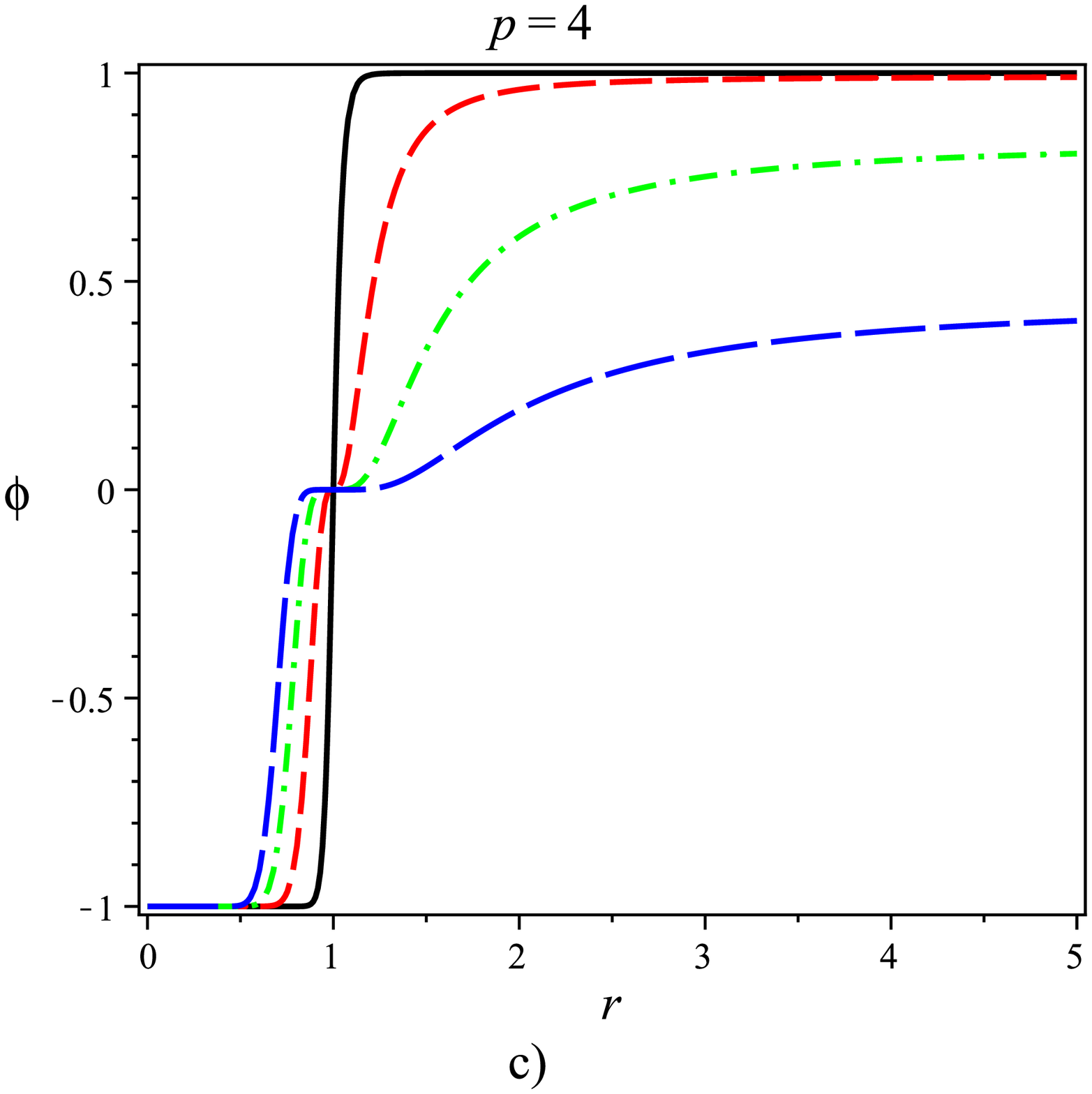}
}
\caption{A one-field model for $p-$balls in $(D,1)$-dimensions: function $\protect\phi(r)$. We fix $r_0=1$, $\lambda=30$. We have a) $p=2$, b) $p=3$ and c) $p=4$. Curves are for $q=1$ (black), 3 (red), 5 (green) and 7 (blue). }
\label{Dp_q_field}
\end{figure}
Fig. \ref{Dp_q_field} shows plots of $\phi(r)$ for fixed $\lambda, r_0$ and several values of $q$ and $p$. The scalar field $\phi(r)$ interpolates between zero and $\phi_c$, with i) $\phi_c=1$ for either $q=1$ or $p=2$ and ii) $\phi_c=\tanh^q[\lambda /(q(p-2)r_0^{p-2})]$ for $q>1$ and $p\neq 2$. For $q\neq1$ the larger is $p$, the lower is $\phi_c$. We note from the figure that the $\phi(r)$ configurations is now of two kinks connected  at $r=r_0$ with a flat region around $r_0$ that grows with $q$. The internal kink runs from $r=0$ to $r=r_0$ and has a compacton character whereas the external kink goes from $r=r_0$ to $r\to\infty$ and is a semi-compacton. As $p$ increases we see that the internal kink (for $r<r_0$) has its thickness reduced whereas the external kink (for $r>r_0$) has its thickness increased.
\begin{figure}
\scalebox{0.9}{\includegraphics[{angle=0,width=6.0cm}]{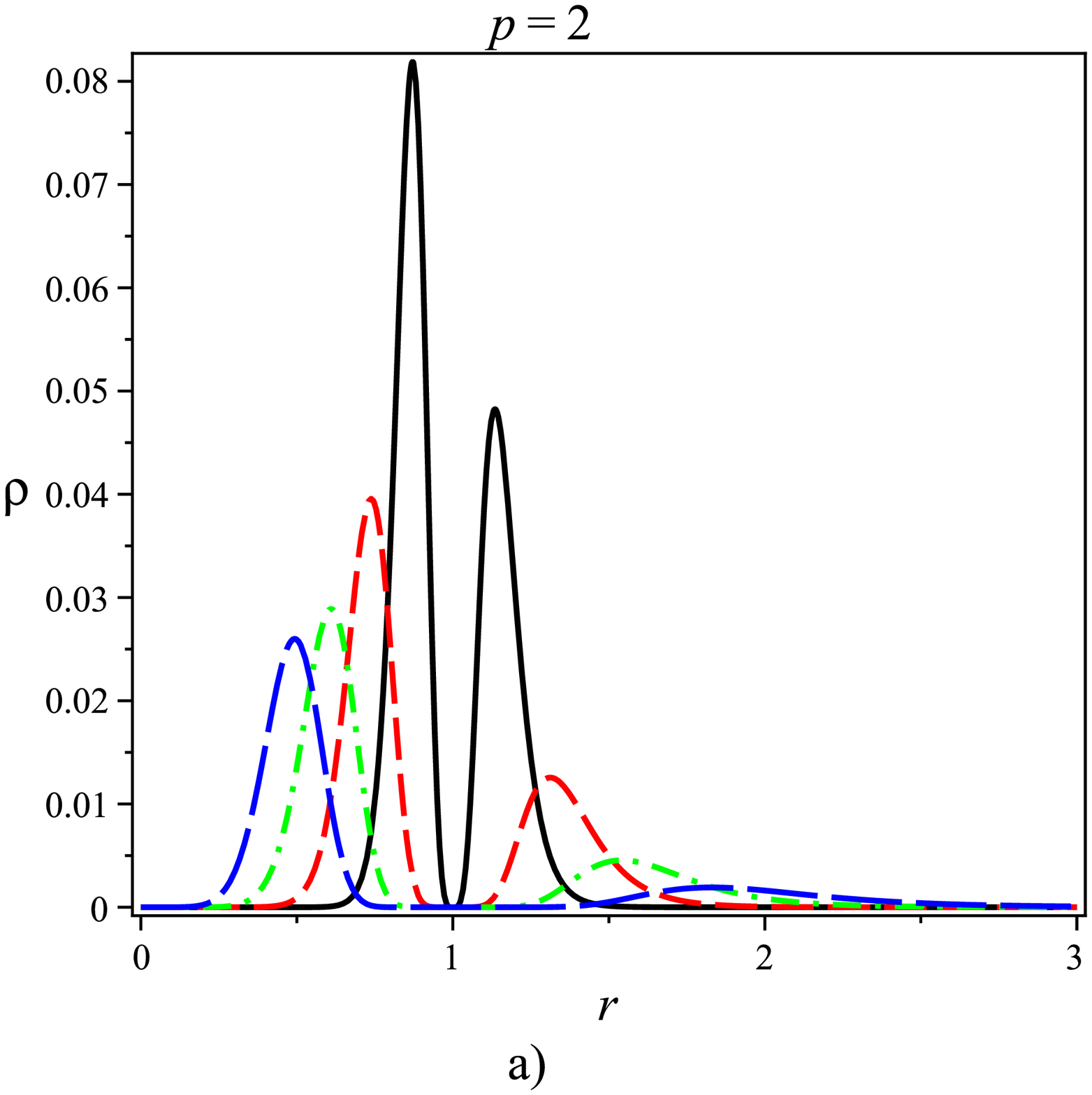}
\includegraphics[{angle=0,width=6.0cm}]{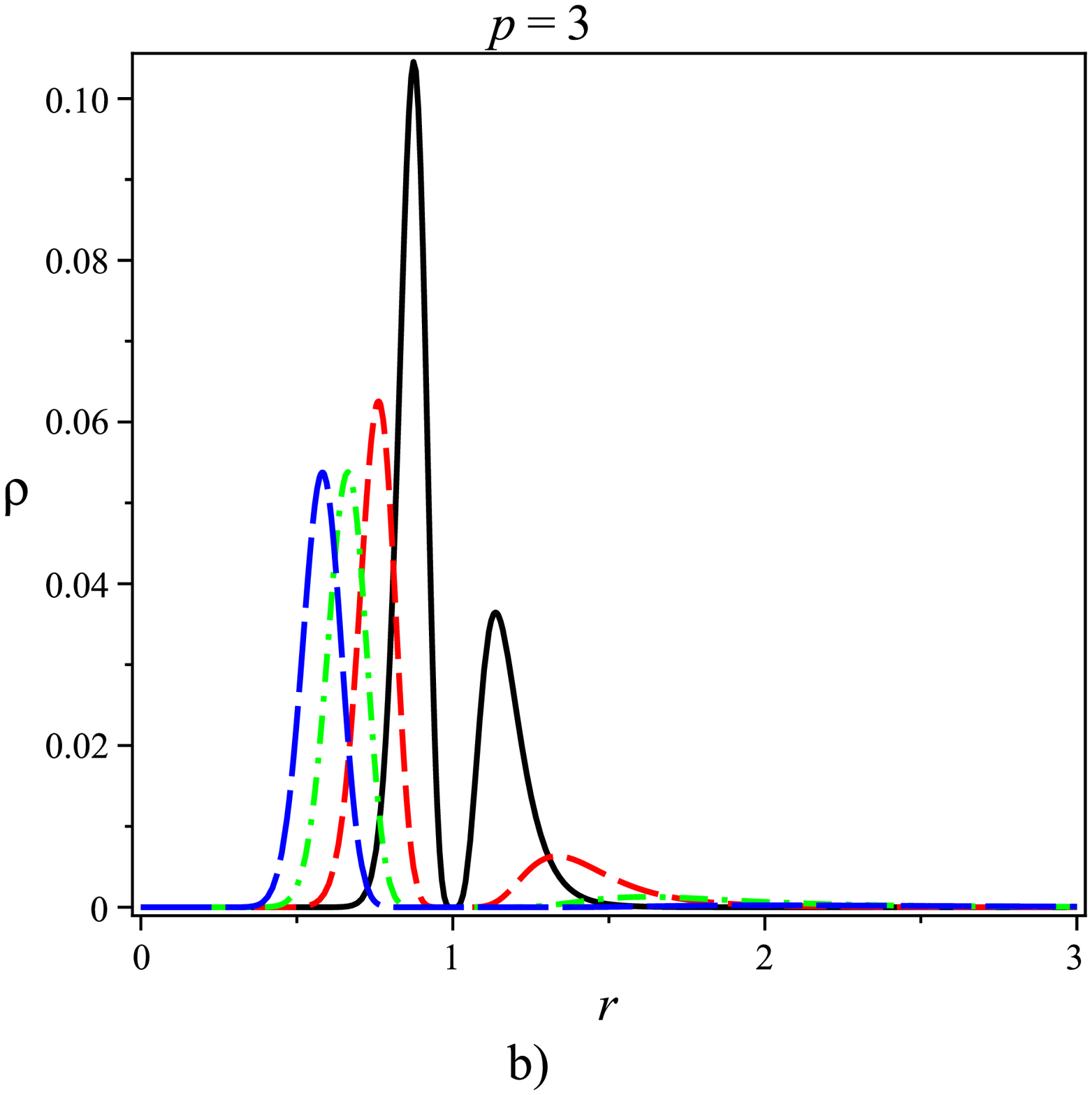}
\includegraphics[{angle=0,width=6.0cm}]{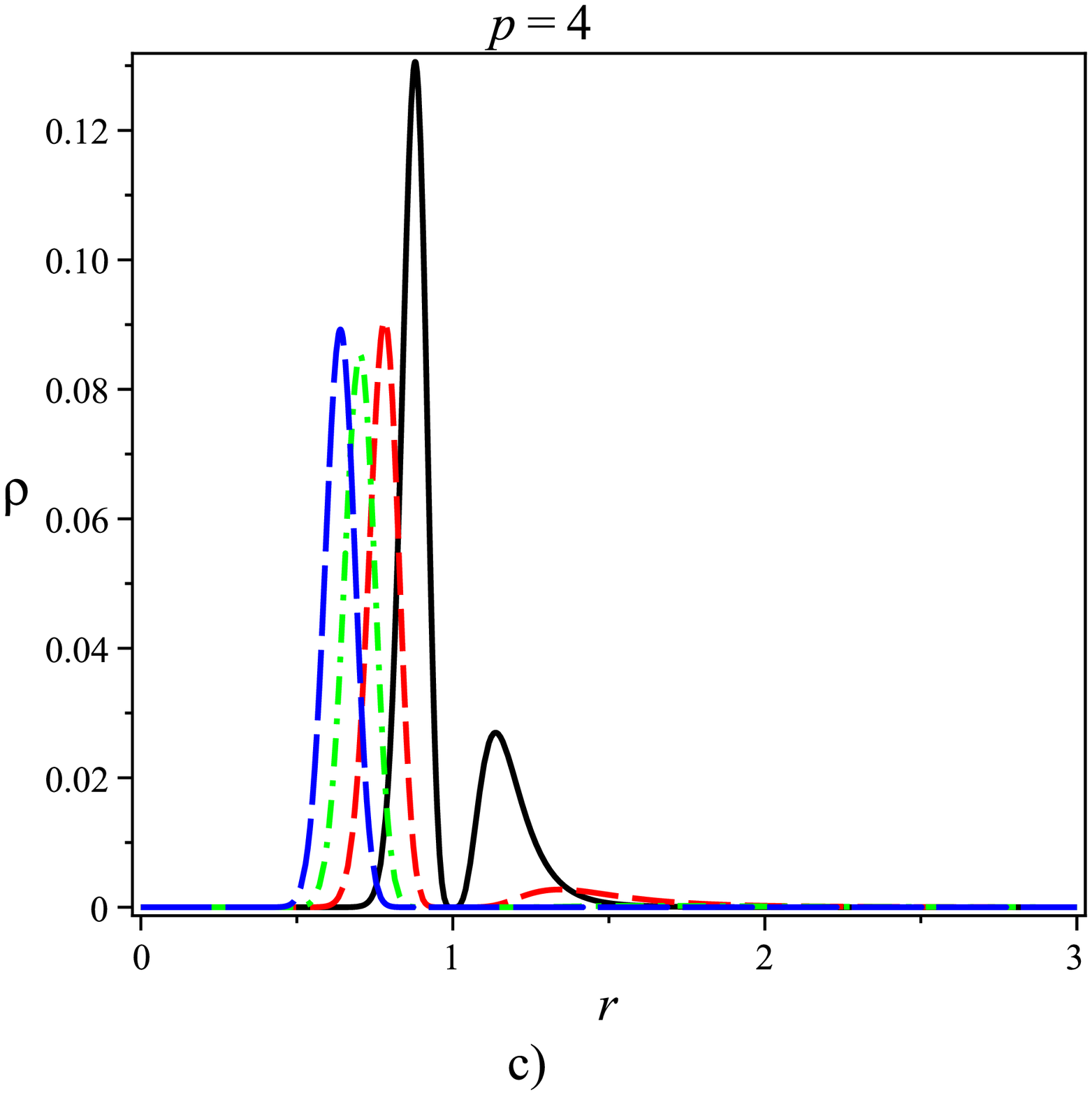}
}
\caption{A one-field model for $p$-balls in $(D,1)$-dimensions: function $\protect\rho(r)$. We fix $r_0=1$, $\lambda=30$. We have a) $p=2$, b) $p=3$ and c) $p=4$. Curves are for $q=3$ (black), 5 (red), 7 (green) and 9 (blue). }
\label{Dp_q_rho}
\end{figure}
Fig. \ref{Dp_q_rho} shows plots of the radial energy density $\rho(r)$ for fixed $\lambda, r_0$ and several values of $q\ge2$ and $p$. From the figures we see that the energy density is characterized by two peaks: a higher and thinner one, centered at $r<r_0$ and a lower and thicker one, centered at $r>r_0$. The distance between the peaks grows with $q$, enlarging the  region around $r=r_0$ where $\rho\sim 0$, at the price of reducing the height of the peaks. For fixed $q$, the effect of the increasing of $p$ is an increasing of height and thinness of the peak at $r<r_0$ and a corresponding decreasing of the peak at $r>r_0$.  The energy density for $q=1$ is characterized for a peak centered around $r=r_0$, and does not depend sensibly on $p$. Here we will consider the coupling $F(\phi)=\phi^2$, corresponding to the Schr\"odinger-like potential
\begin{eqnarray}
V & = & \frac{\jmath(\jmath + p-2)}{r^{2}} + \eta \tanh^{2q}(\eta_p), \,p=2,3,... .\\
\end{eqnarray}
\begin{figure}
\scalebox{0.9}{\includegraphics[{angle=0,width=6cm}]{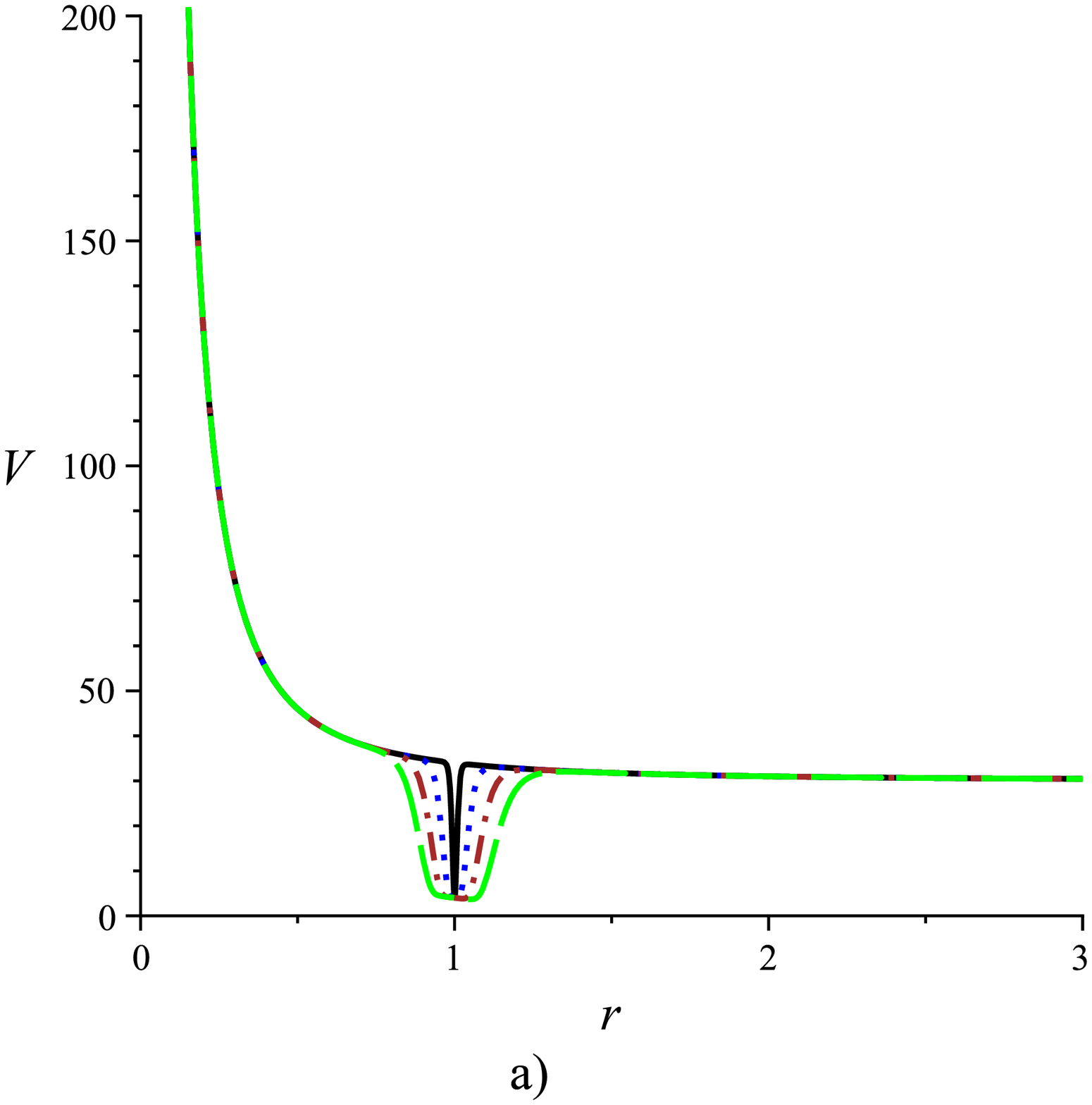}
\includegraphics[{angle=0,width=6cm}]{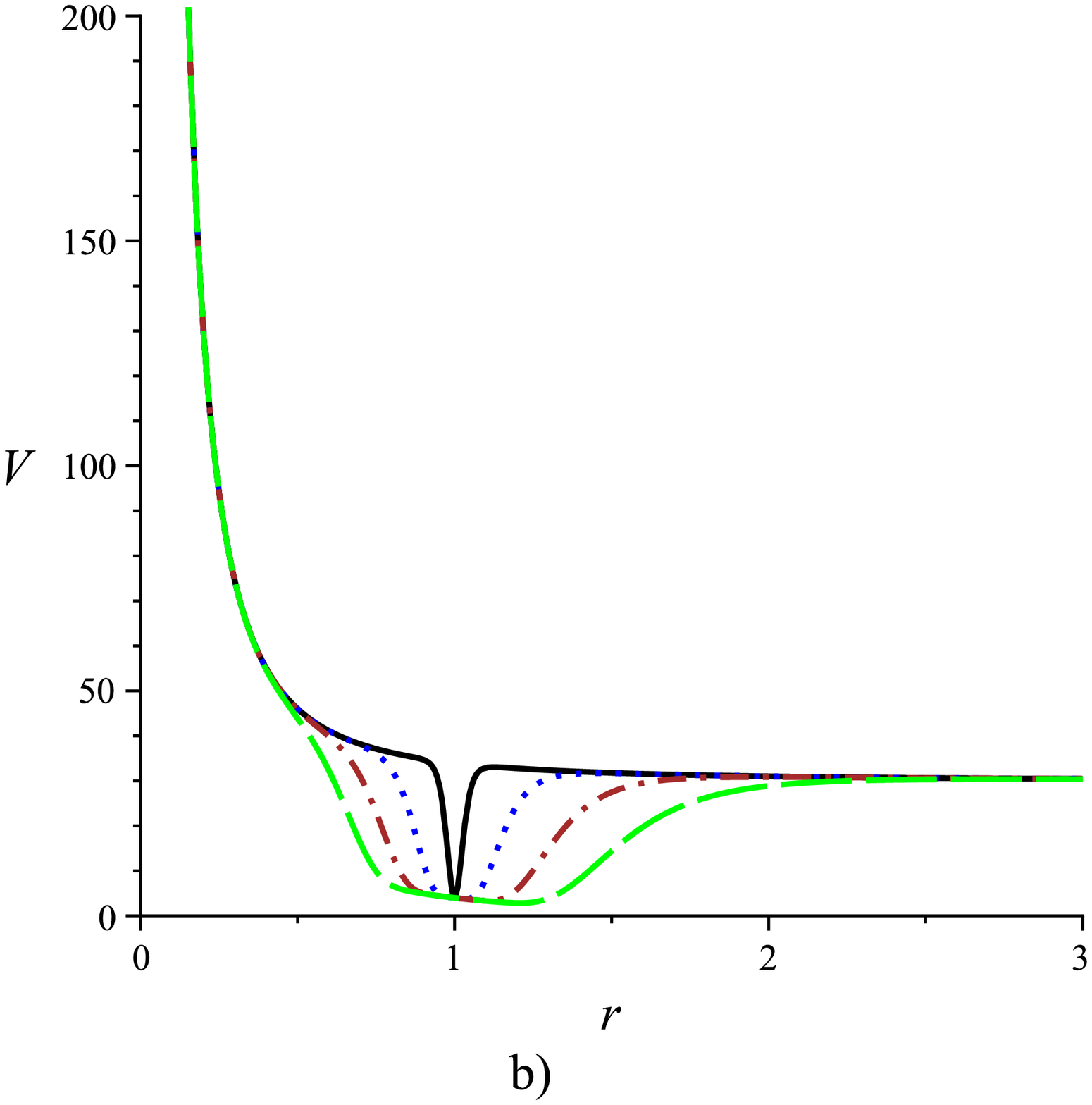}
\includegraphics[{angle=0,width=6cm}]{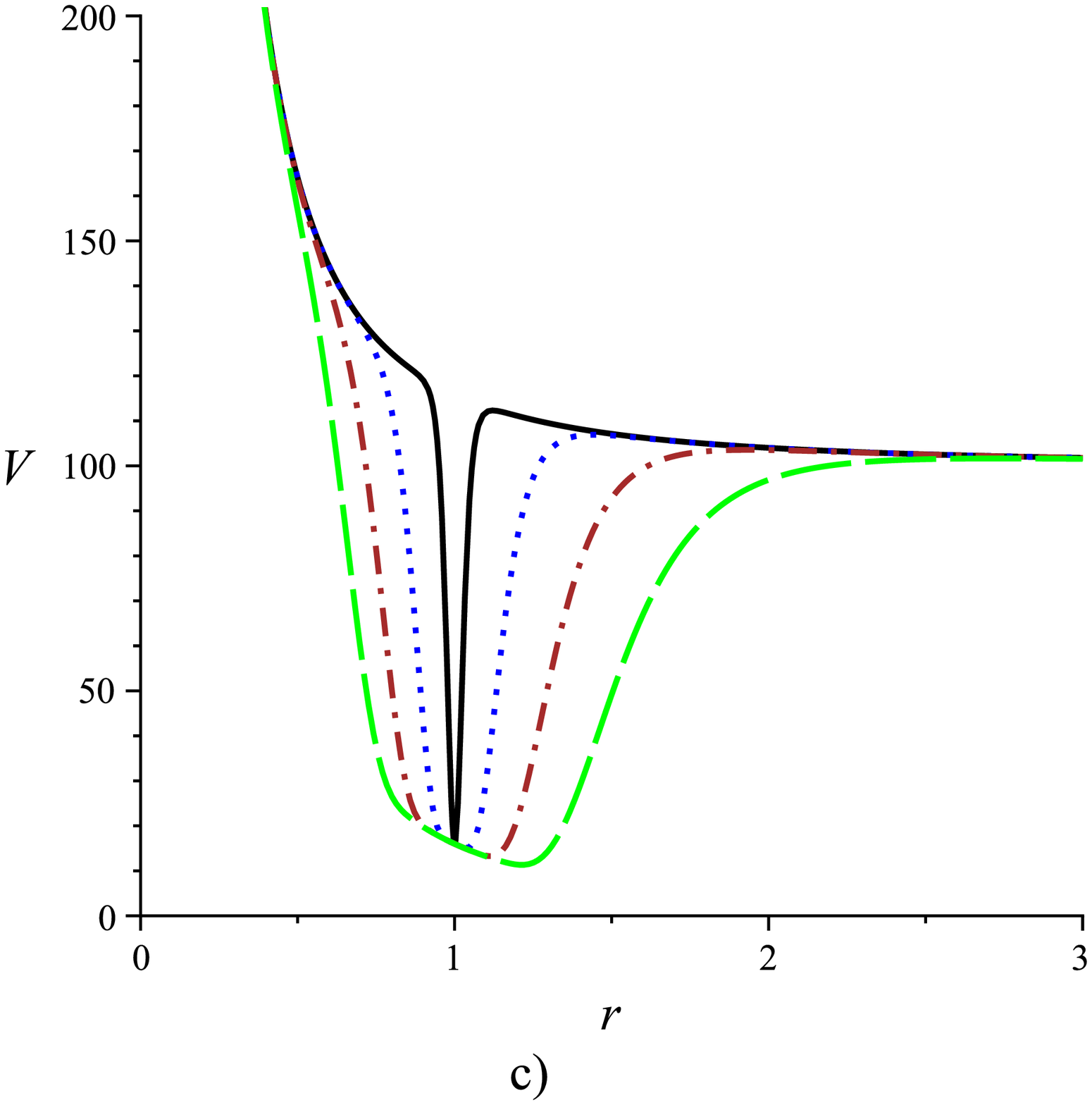}
}
\caption{A one-field model for $p-$balls in $(D,1)$-dimensions: Schr\"odinger potentials $V$ for $\jmath=2$, $r_0=1$, $p=2$ and $q=1$ (black line), $q=3$ (blue dotted line), $q=5$ (brown dashdotted line), $q=7$ (green longdashed line). Plots are for a) $\eta=30$, $\lambda=100$, b) $\eta=30$, $\lambda=30$ and c) $\eta=100$, $\lambda=30$.}
\label{Dp-q-potentials}
\end{figure}


\begin{figure}
\scalebox{0.9}{\includegraphics[{angle=0,width=6cm}]{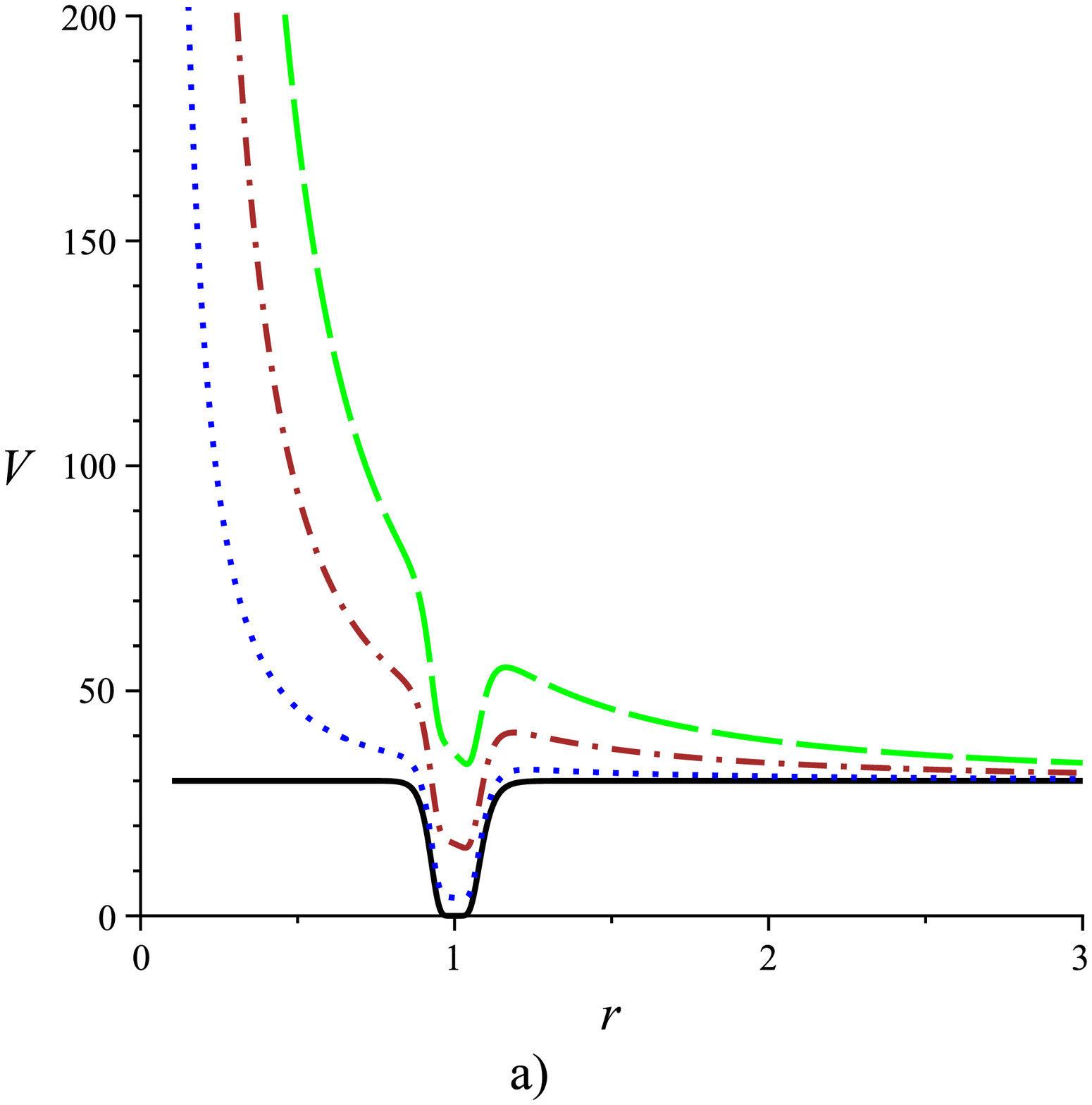}
\includegraphics[{angle=0,width=6cm}]{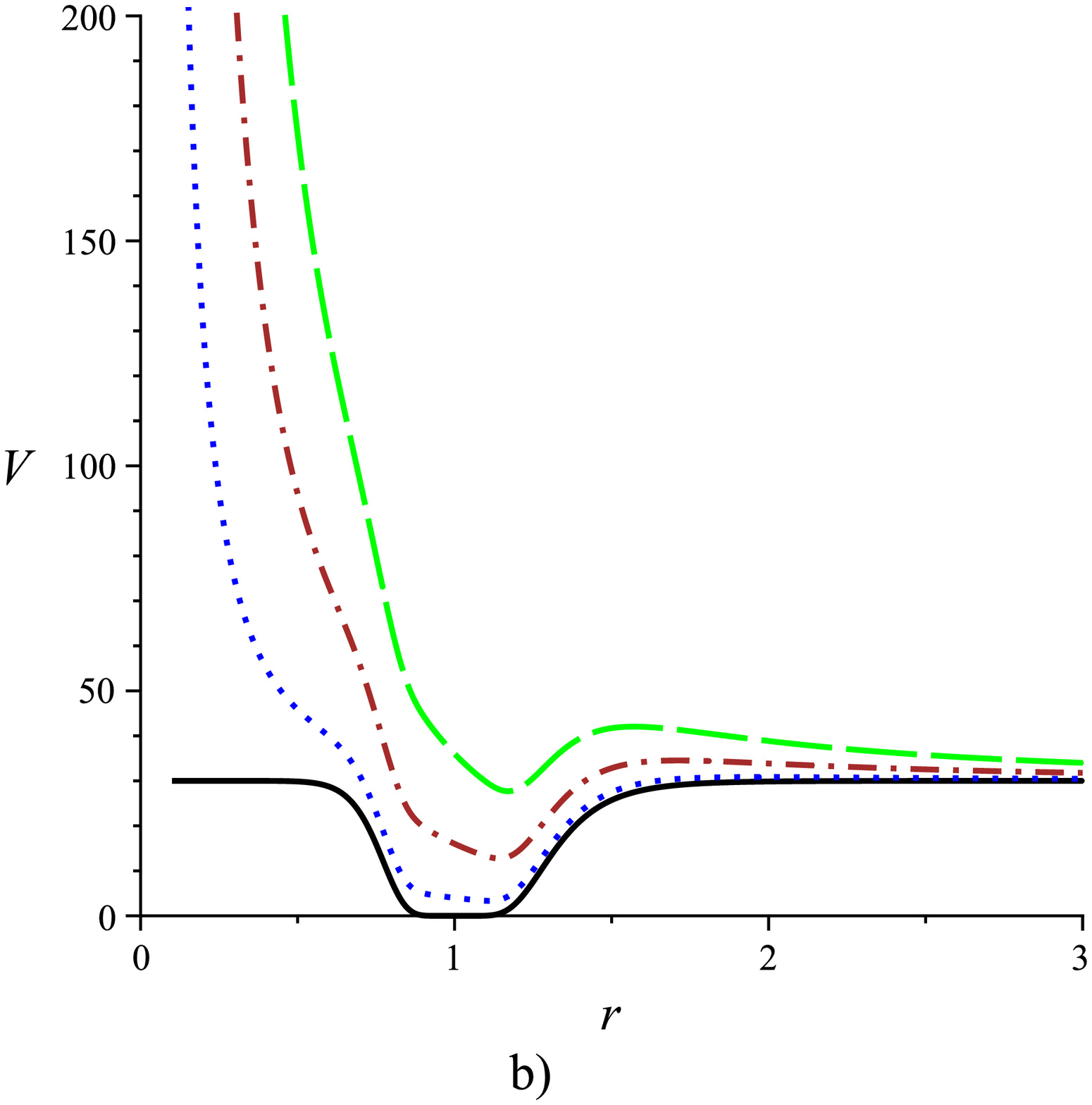}
\includegraphics[{angle=0,width=6cm}]{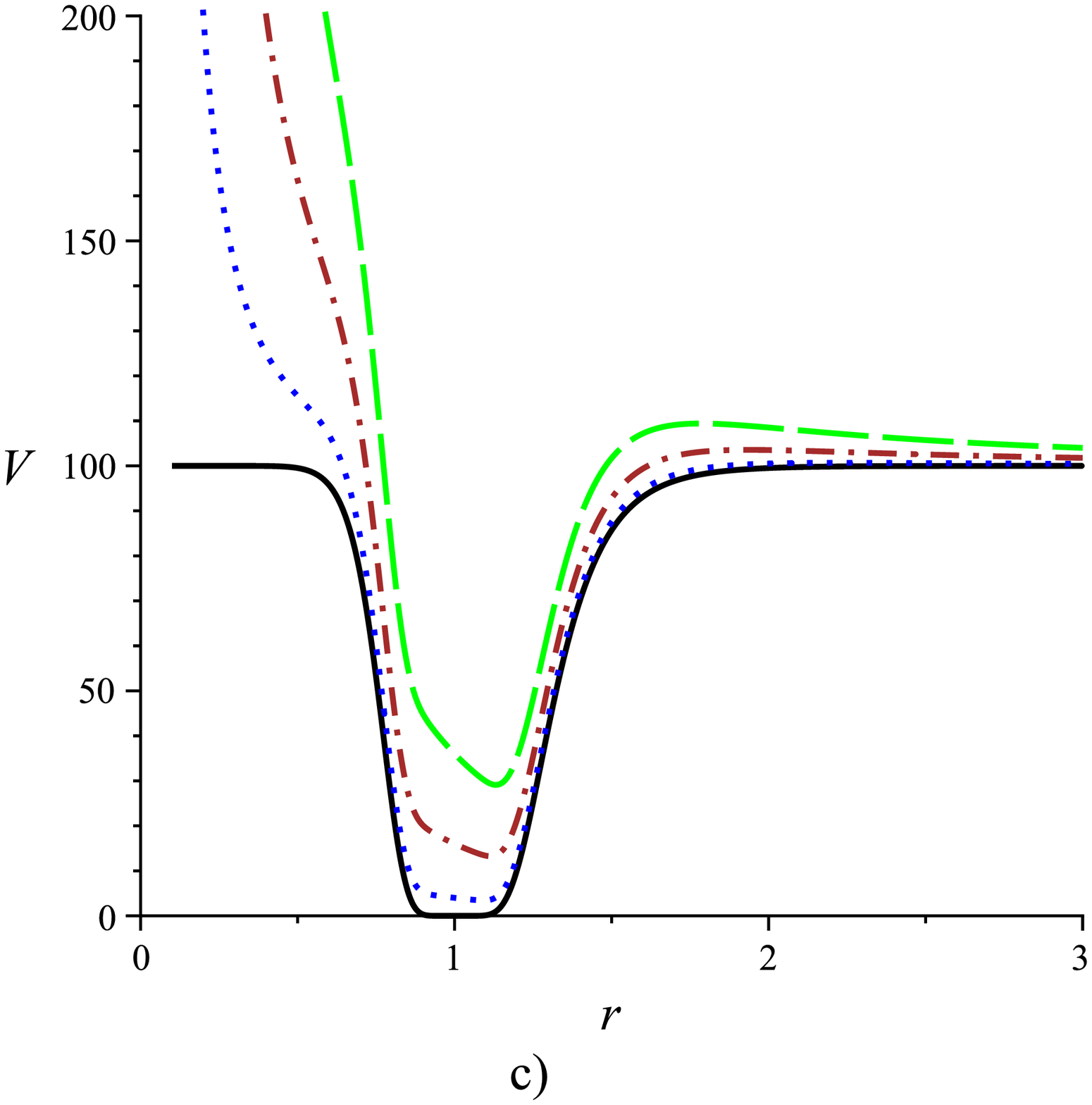}
}
\caption{A one-field model for $p-$balls in $(D,1)$-dimensions: Schr\"odinger potentials $V$ for $r_0=1$, $p=2$, $q=5$ and $\jmath=0$ (black line), $\jmath=2$ (blue dotted line), $\jmath=4$ (brown dashdotted line), $\jmath=6$ (green longdashed line). Plots are for a) $\eta=30$, $\lambda=100$, b) $\eta=30$, $\lambda=30$ and c) $\eta=100$, $\lambda=30$.}
\label{Dp-lambda-potentials}
\end{figure}


\begin{table}[]\scriptsize
\begin{tabular}{|l|c|c|c|c|c|c|c|c|c|}
\hline
                             & \multicolumn{2}{c|}{$q=1$} & \multicolumn{2}{c|}{$q=3$} & \multicolumn{2}{c|}{$q=5$} & \multicolumn{2}{c|}{$q=7$} & n \\ \hline
                             & $j=0$         & $j=2$        &     $j=0$         & $j=2$    &        $j=0$         & $j=2$     &  $j=0$         & $j=2$ & \\ \hline \hline
$\eta=30,\; \lambda=100$ & -- & -- & 28.6227 & -- & 24.8245 & 29.0266 & 20.5476 & 24.8500& 1   \\ \hline\hline
\multirow{2}{*}{$\eta=30,\; \lambda=30$} & 29.4788 & -- & 19.5224 & 23.8051 & 11.4668 & 15.5708 & 7.1019 & 10.9721 & 1   \\ \cline{2-10}
                             & -- & -- & -- & -- & -- & -- & 22.9866 & 27.0932& 2   \\ \hline\hline
\multirow{4}{*}{$\eta=100,\; \lambda=30$} & 90.8870 & 95.2469 & 41.9699 & 46.0534 & 20.1460 & 24.1420 & 11.2155 & 15.0803 & 1 \\ \cline{2-10}
  & -- & -- &  -- & -- & 66.4764 & 70.7870 & 40.1402 & 44.4125 & 2  \\ \cline{2-10}
  & -- & -- &  -- & -- & 99.3950 & -- & 73.4681 & 77.6431 & 3 \\ \cline{2-10}
  & -- & -- &  -- & -- & -- & -- & 95.9278 & 98.6800 & 4 \\ \hline
\end{tabular}
\caption{A one-field model for $p-$balls in $(D,1)$-dimensions: eigenvalues $M_{n\jmath}^2$, solutions of Eq. (\ref{eq_rho}) for $\jmath=0$ and $\jmath=2$. We fix $r_0=1$, $p=2$, with $q$, $\lambda$ and $\eta$ of Schr\"odinger potentials for $\jmath=2$ corresponding to Figs. \ref{Dp-q-potentials} and \ref{Dp-lambda-potentials}.}
\end{table}

Fig. \ref{Dp-q-potentials} shows some plots for $V(r)$ for fixed values of $p,r_0,\jmath$ and several values of $q$, $\eta$ and $\lambda$. We note that the potentials are strictly positive, with $V(r\to 0)=\infty$ and a minimum around $r=r_0$. The structure of the potential shows  that there are possibly bound states, to be investigated numerically. Comparing Figs. \ref{Dp-q-potentials}a-c we see that the increasing of $q$ or the decreasing of $\lambda$ enlarges the region around the local minimum, favoring the appearance of bound states. In addition, the increasing of $\eta$ turns the minimum deeper, also favoring bound states. This is confirmed with the eigenvalues obtained numerically, presented in Table I.

Fig. \ref{Dp-lambda-potentials} shows some plots for $V(r)$ for fixed values of $p,r_0,q$ and several values of $\lambda$, $\eta$ and $\lambda$. This figure shows that, for fixed parameters, an increasing in $\jmath$ decreases the possibility of occurrence of bound states. This is confirmed from the results of Tables I and II. The case $\jmath=0$ is special since there is no possibility of resonances. For larger values of $\jmath$, there is an increasing of a local maximum around $r=r_0$, increasing the possibility of occurrence of resonant states.

\begin{figure}
\scalebox{0.9}{\includegraphics[{angle=0,width=6cm}]{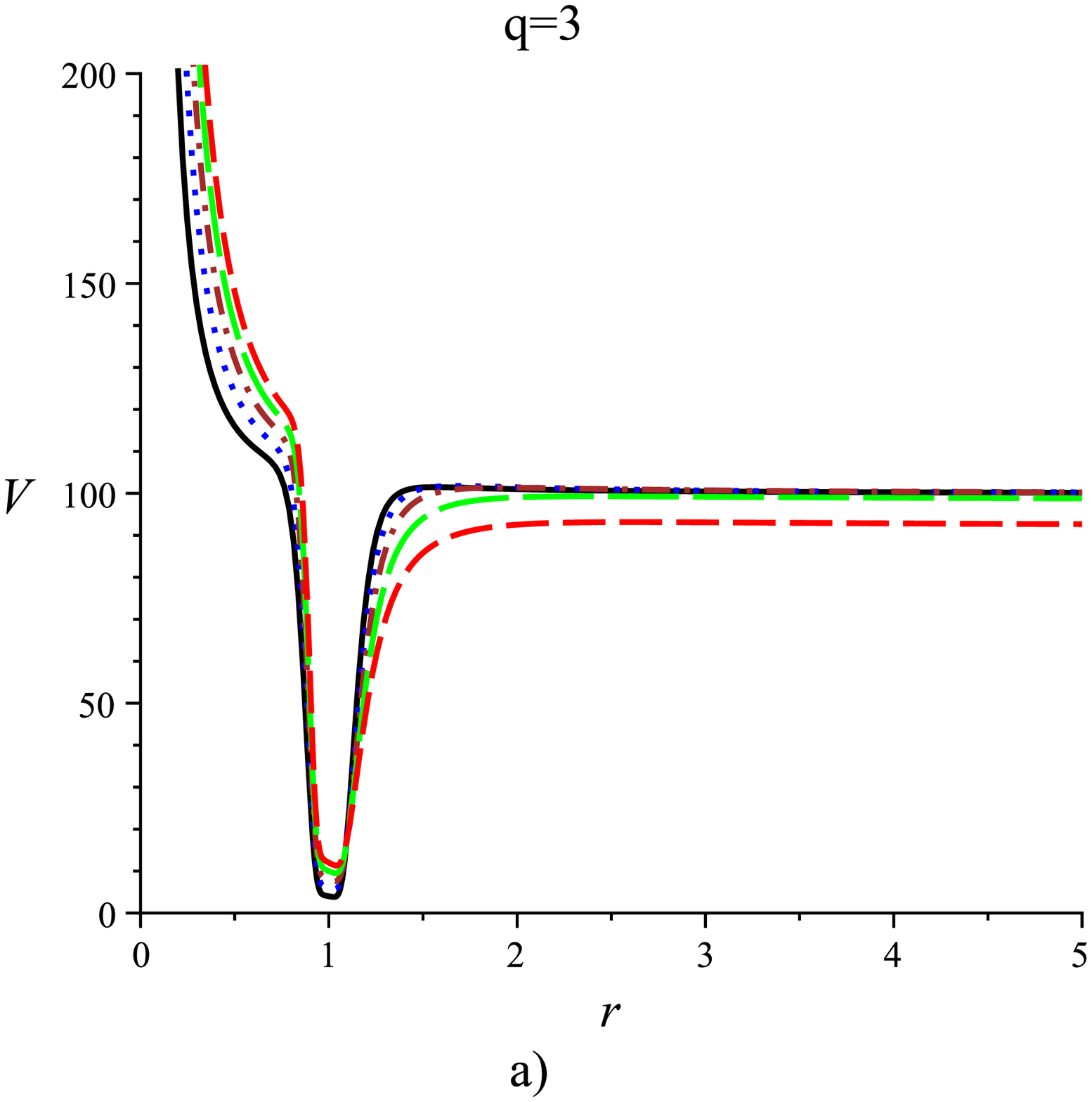}
\includegraphics[{angle=0,width=6cm}]{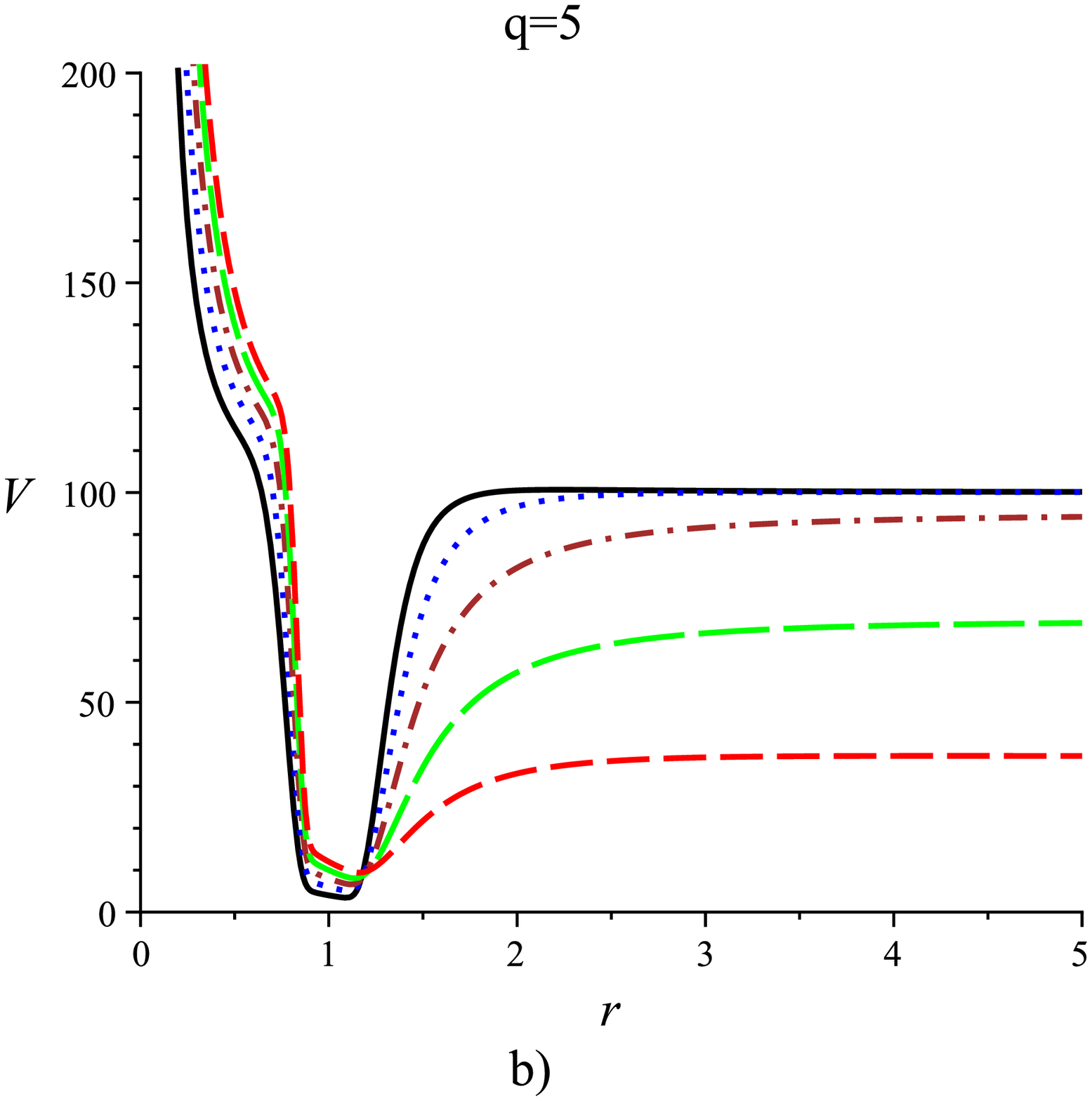}
\includegraphics[{angle=0,width=6cm}]{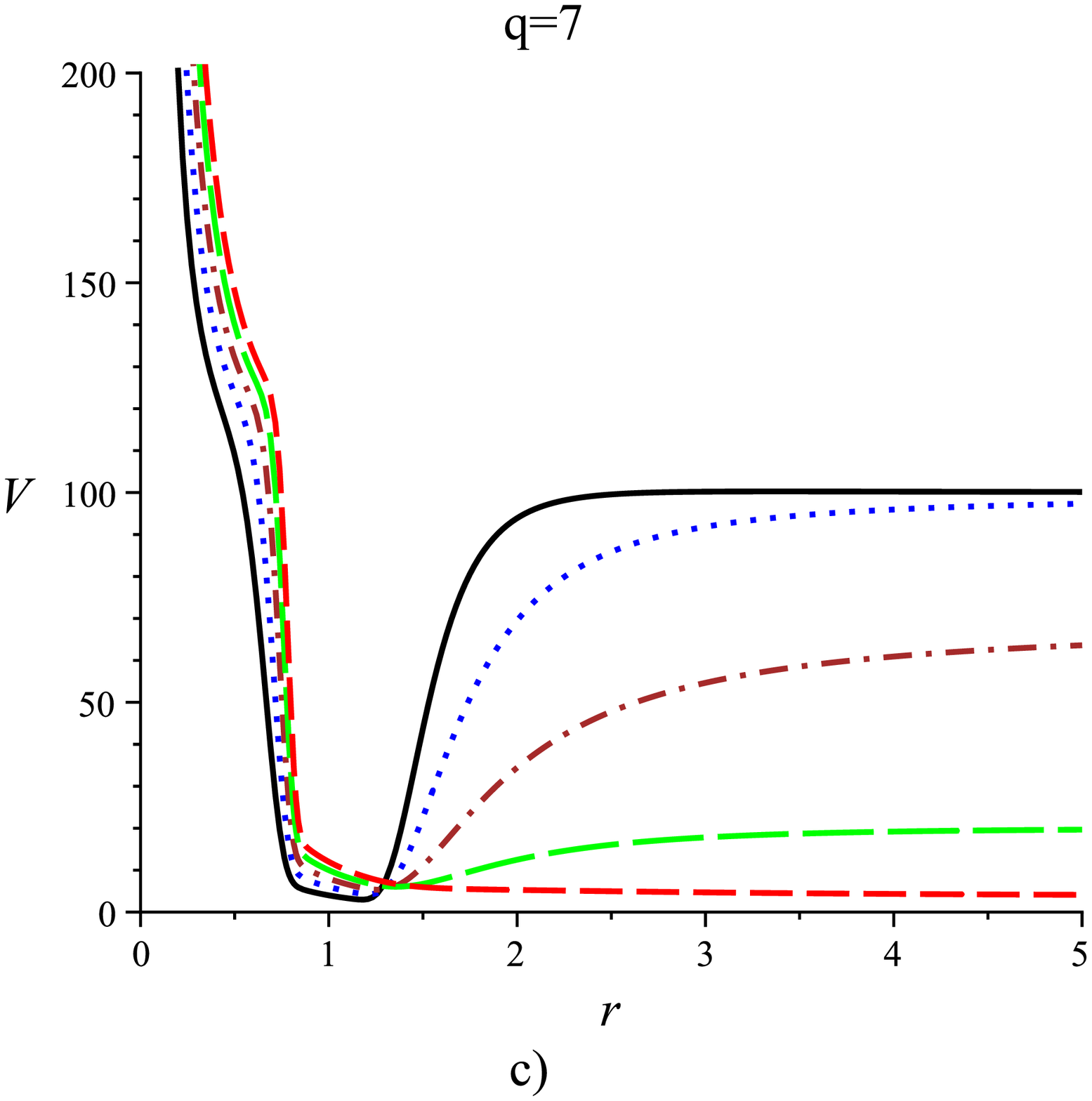}
}
\caption{A one-field model for $p-$balls in $(D,1)$-dimensions: Schr\"odinger potentials $V$ for $\jmath=2$, $r_0=1$, $\eta=100$, $\lambda=30$ and $p=2$ (black line), $p=3$ (blue dotted line), $p=4$ (brown dashdotted line), $p=5$ (green longdashed line) and $p=6$ (red dashed line). Plots are for a) $q=3$,  b) $q=5$ and c) $q=7$.}
\label{Dp-q-potentials2}
\end{figure}

Fig. \ref{Dp-q-potentials2} shows $V(r)$ for fixed values of $\eta,\lambda,r_0,\jmath$
and several values of $q$ and $p$. Comparing Figs. \ref{Dp-q-potentials2}a-c
we see that, for fixed parameters, the increasing of $p$ reduces the value of
$V(r\to\infty)$. On the other hand this occurs simultaneously with the
enlargement of the region around the local minimum  (more evident for
larger $q$). Concerning
to the influence for the occurrence of bound states, the former character
reduces the probability whereas the latter increases it. Then there is a
competition between both effects.
In particular Fig. \ref{Dp-q-potentials2}c shows that for $q=7$ and $p=6$ the
minimum of the potential disappears and there is no possibility of bound
states. This signals that for large  values of $q$ (i.e., $q\sim 7$),
intermediate values of $p$ are better for obtaining more bound states.
This analysis is confirmed from Table II, which shows that, for $q=5$ the
occurrence of bound states is more frequent for $p=4$ and $p=5$ whereas for
$q=7$ this occurs for $p=3$ and $p=4$.

\begin{table}[]\scriptsize
 \begin{tabular}{|c|c|c|c|c|c|c|c|c|c|c|c|c|c|}
\hline
& \multicolumn{2}{c|}{$p=2$} & \multicolumn{2}{c|}{$p=3$} &
\multicolumn{2}{c|}{$p=4$} & \multicolumn{2}{c|}{$p=5$} &
\multicolumn{2}{c|}{$p=6$} & \multicolumn{2}{c|}{$p=7$} & $n$ \\ \hline
& $j=0$ & $j=2$ & $j=0$ & $j=2$ & $j=0$ & $j=2$ & $j=0$ & $j=2$ & $j=0$ & $%
j=2$ & $j=0$ & $j=2$ &  \\ \hline\hline
\multirow{2}{*}{$q=3$} & 41.9699 & 46.0534 & 41.5186 & 47.4958 & 40.6852 &
48.4558 & 39.4634 & 48.9136 & 37.8409 & 48.8321 & 35.8017 & 48.1530 & 1 \\
\cline{2-14}
& -- & -- & 99.779 & -- & 97.8555 & -- & 93.7645 & -- & 87.0198 & -- &
77.3044 & -- & 2 \\ \hline\hline
\multirow{4}{*}{$q=5$} & 20.1460 & 24.1420 & 19.3599 & 24.9835 & 17.9279 &
24.9191 & 15.8604 & 23.8675 & 13.1655 & 21.6049 & 9.8750 & -- & 1 \\
\cline{2-14}
& 66.4764 & 70.7870 & 62.7023 & 68.2819 & 55.8200 & 62.1088 & 45.5658 &
51.7269 & 31.8084 & 36.0979 & -- & -- & 2 \\ \cline{2-14}
& 99.39.50 & -- & 93.7420 & 97.2674 & 81.3528 & 85.2906 & 61.7843 & 65.0786
& -- & -- & -- & -- & 3 \\ \cline{2-14}
& -- & -- & -- & -- & 91.2896 & 92.9968 & 67.7097 & 68.9111 & -- & -- & -- &
-- & 4 \\ \hline\hline
\multirow{6}{*}{$q=7$} & 11.2155 & 15.0803 & 10.2061 & 15.2934 & 8.4188 &
14.1784 & 5.9333 & 11.3420 & 2.9698 & -- & -- & -- & 1 \\ \cline{2-14}
& 40.1402 & 44.4125 & 35.3980 & 40.5926 & 26.9721 & 32.1424 & 15.0591 &
18.0844 & -- & -- & -- & -- & 2 \\ \cline{2-14}
& 73.4681 & 77.6431 & 61.9651 & 66.2717 & 42.8930 & 46.5882 & 19.0888 & -- &
-- & -- & -- & -- & 3 \\ \cline{2-14}
& 95.9278 & 98.6800 & 81.1822 & 84.3789 & 53.5926 & 56.0433 & -- & -- & -- &
-- & -- & -- & 4 \\ \cline{2-14}
& -- & -- & 91.9072 & 93.7907 & 60.0097 & 61.5195 & -- & -- & -- & -- & -- &
-- & 5 \\ \cline{2-14}
& -- & -- & 96.7928 & 97.4313 & -- & 64.5802 & -- & -- & -- & -- & -- & -- &
6 \\ \hline
\end{tabular}
\caption{A one-field model for $p-$balls in $(D,1)$-dimensions: eigenvalues $M_{n\jmath}^2$, solutions of Eq. (\ref{eq_rho}) for $\jmath=0$ and $\jmath=2$. We fix $r_0=1$, $\eta=100$, $\lambda=30$, with $p$ and $q$ of Schr\"odinger potentials for $\jmath=2$ corresponding to Fig. \ref{Dp-q-potentials2}.}
\end{table}

\section{A two-field model}
\label{sec_2field}
In this section we will consider the model
\be
W(\phi,\chi)=\lambda \bigg(\phi-\frac{1}{3}\phi^3-s\phi\chi^2 \bigg).
\label{superpotential}
\ee
With this choice of $W$, the potential $\tilde V(\phi,\chi)=(1/2)(W_\phi^2 + W_\chi^2)$ was introduced in Ref. \cite{BSR} to construct Bloch walls. The limit $s\to 0.5$ turns the two-field problem into a one-field one, recovering the model of an Ising wall. This can be better seen in the explicit solutions $\phi(\xi)$ and $\chi(\xi)$ bellow.
The equation of motion for the scalar fields, Eq. (\ref{phi_D}) is rewritten, after a change of variables $d\xi=1/r^{p-1} dr$, as
\begin{eqnarray}
\frac{d\phi}{d\xi}&=&W_\phi,\\
\frac{d\chi}{d\xi}&=&W_\chi,
\end{eqnarray}

\begin{figure}
\includegraphics[{angle=0,width=6.0cm}]{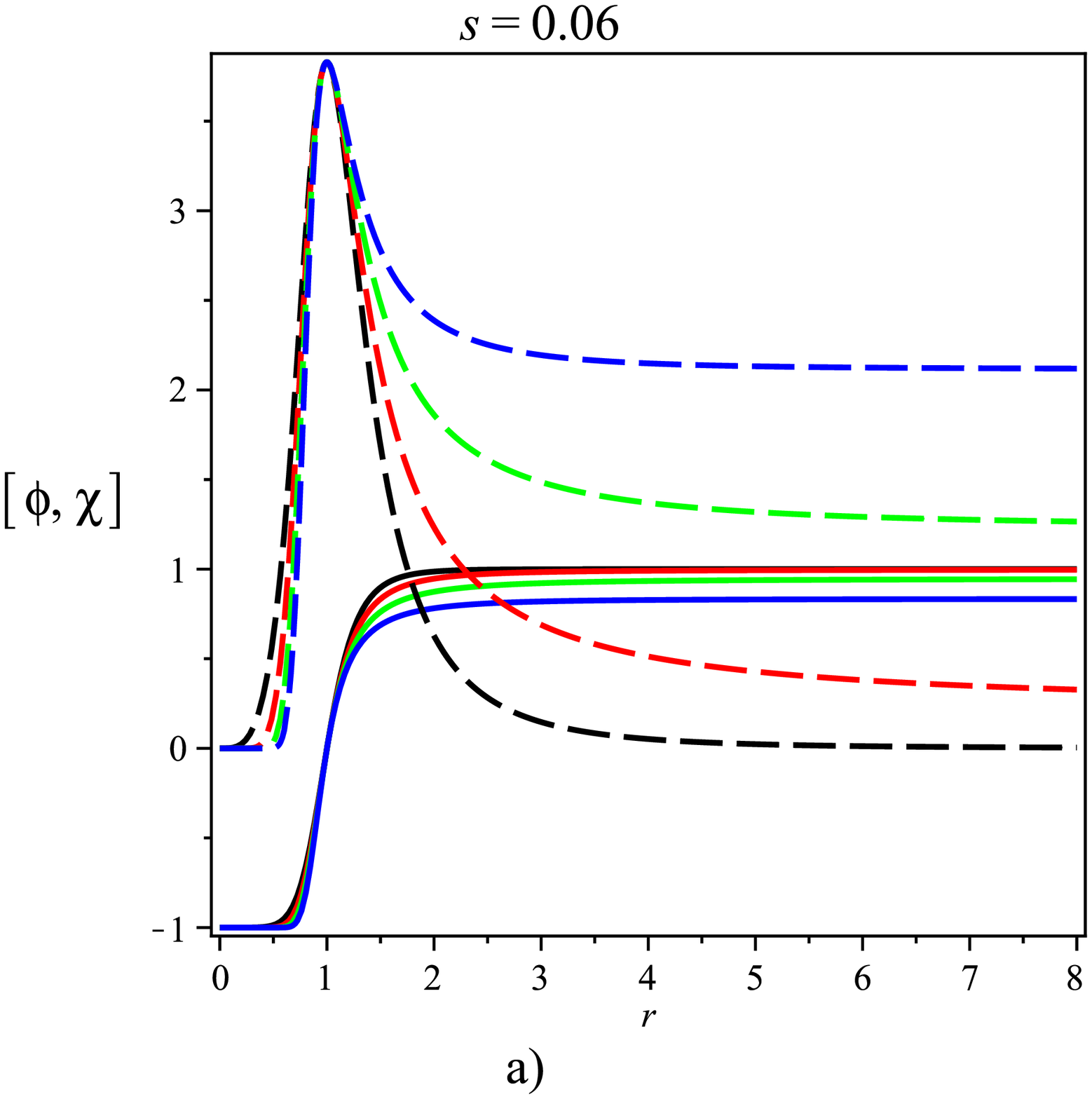}
\includegraphics[{angle=0,width=6.0cm}]{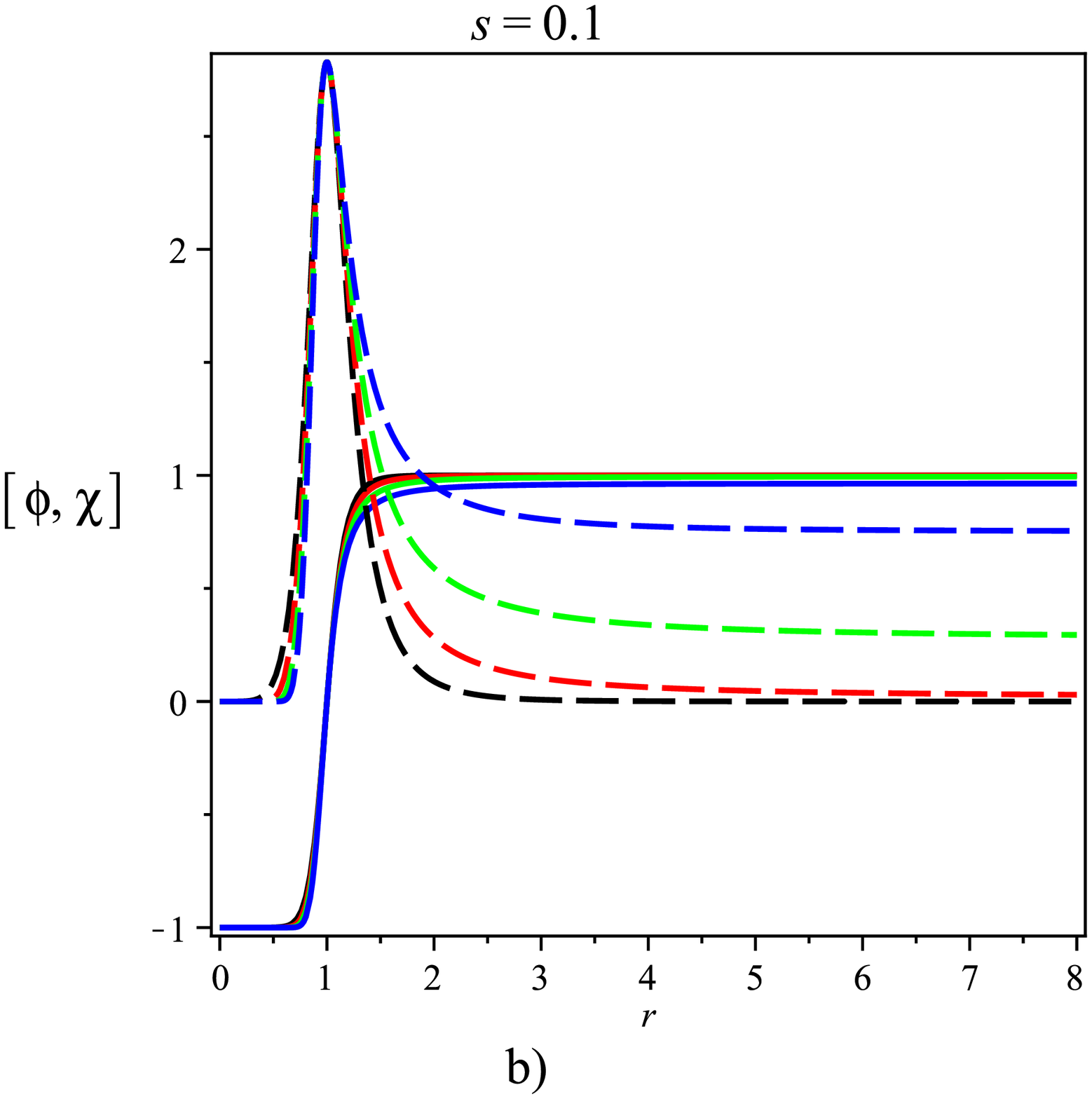}
\caption{A two-field model for $p-$balls in $(D,1)$-dimensions: functions $\protect\phi(r)$ (kink-like, solid lines) and $\protect\chi(r)$
(lump-like, traced lines). We fix $r_0=1$, $\lambda=30$. We have a) $s=0.06$ and b) $s=0.1$. Curves are for $p=2$ (black), 3 (red), 4 (green) and 5 (blue). }
\label{Dp_field}
\end{figure}
with solution
\begin{eqnarray}
\phi(\xi)&=&\tanh(2\lambda s\xi),\\
\chi(\xi)&=&\sqrt{\frac1s - 2}\sech(2\lambda s\xi).
\end{eqnarray}
Back to $r$ variable, explicit expressions for the scalar field profiles and consequently for the energy density can be easily attained. We have, for $p\ge2$,
\begin{eqnarray}
\phi(r)&=&\tanh(\tau_p), \\
\chi(r)&=&\sqrt{\frac1s - 2}\sech(\tau_p),\\
\rho(r) &=&\frac{\left( 2\lambda s\right) ^{2}}{r^{2p-2}}\mathrm{sech}^{4}(\tau_p)   \left\{ 1+\left( \frac{1}{s}-2\right) \mathrm{sinh}^{2}(\tau_p)\right\}.  \notag
\end{eqnarray}
where
\be
\label{taup}
\tau_p=2\lambda s \xi(r)
\ee
and $\xi(r)$ given by Eq. (\ref{xir_p2}), for $p=2$ and by Eq. (\ref{xir_p}), for p=3,4,....

Fig. \ref{Dp_field} shows plots of $\phi(r)$ and $\chi(r)$ for fixed $\lambda, r_0$ and several values of $s$ and $p$.  The scalar field $\phi(r)$ interpolates between $-1$ and $\phi_c$ , with $\phi_c=1$ for $p=2$ and $\phi_c=\tanh[2\lambda s/((p-2)r_0^{p-2})]$ for $p\neq 2$. The scalar field $\chi(r)$ interpolates between zero and $\chi_c$, with $\chi_c=0$ for $p=2$ and $\chi_c=\sqrt{(1/s)-2}\sech[2\lambda s/((p-2)r_0^{p-2})]$ for $p\neq 2$.
Starting from $p=2$, the larger is $p$, the lower is $\phi_c$ and the
larger is $\chi_c$. This means that with the increasing of $p$
the $\phi(r)$ and $\chi(r)$ configurations are, respectively, more departed
from a usual kink and lump configurations in $r$, centered at $r=r_0$.
Comparing Figs. \ref{Dp_field}a and \ref{Dp_field}b we see that, for all
other parameters fixed, larger values of $s$ make the profiles of $\phi(r)$
almost indistinguishable from a thin kink-like defect centered at $r=r_0$.
We have also verified that larger values of $\lambda$ turn the defect thinner
and turn $\phi_c$ and $\chi_c$ closer to 1 and 0, recovering the kink
and lump profiles for $\phi(r)$ and $\chi(r)$, respectively.
\begin{figure}
\includegraphics[{angle=0,width=6.0cm}]{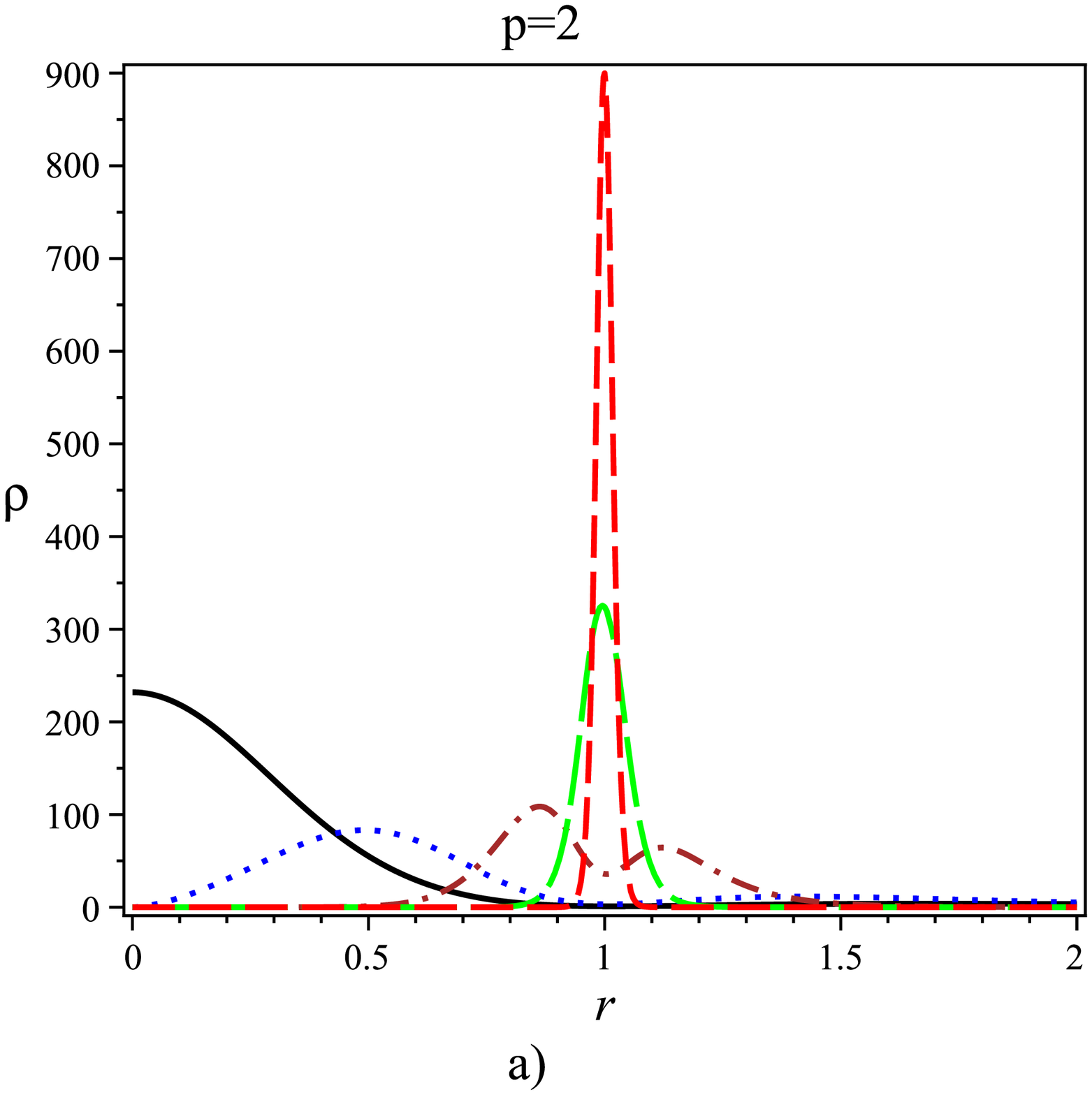}
\includegraphics[{angle=0,width=6.0cm}]{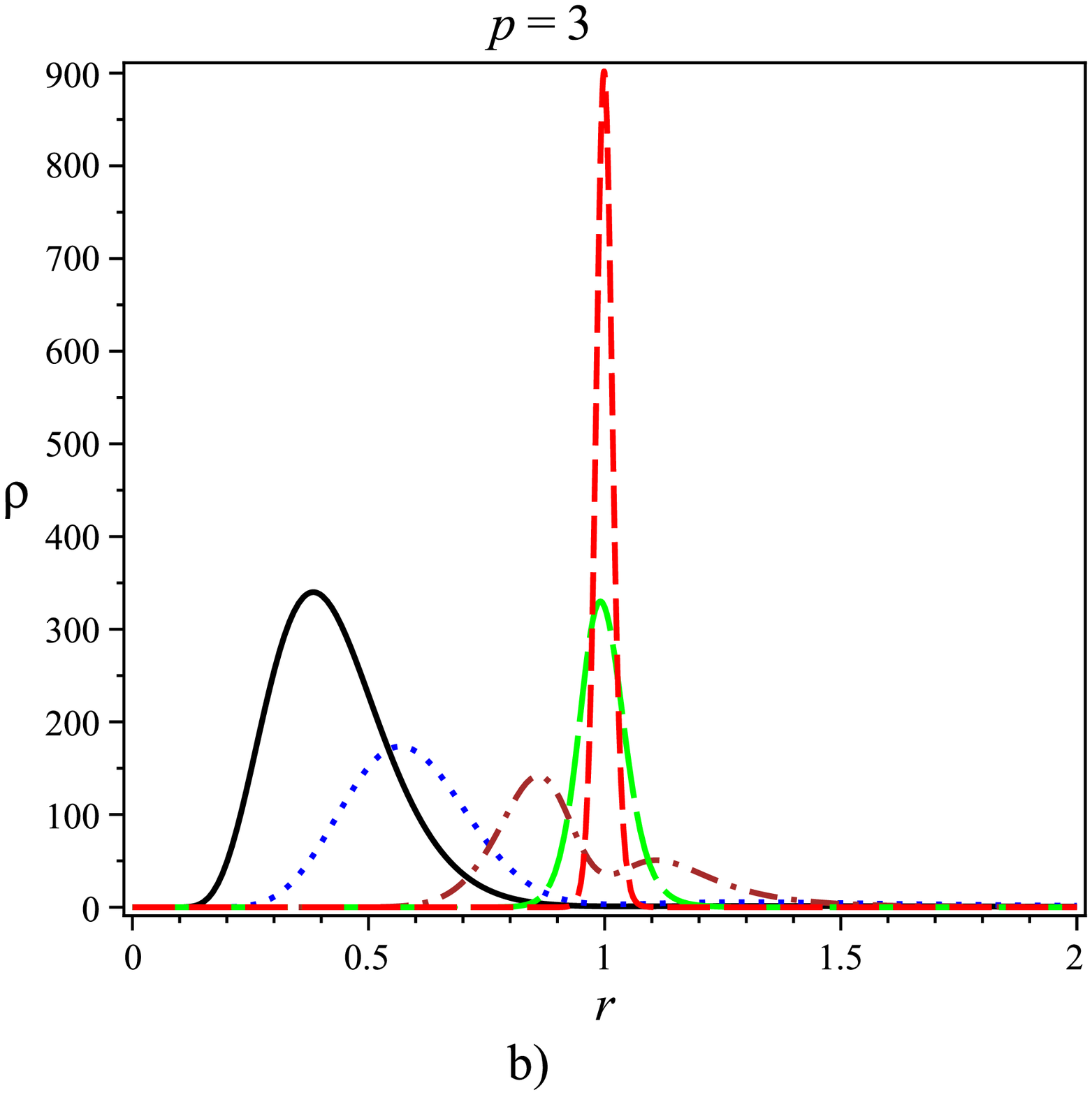}
\\
\includegraphics[{angle=0,width=6.0cm}]{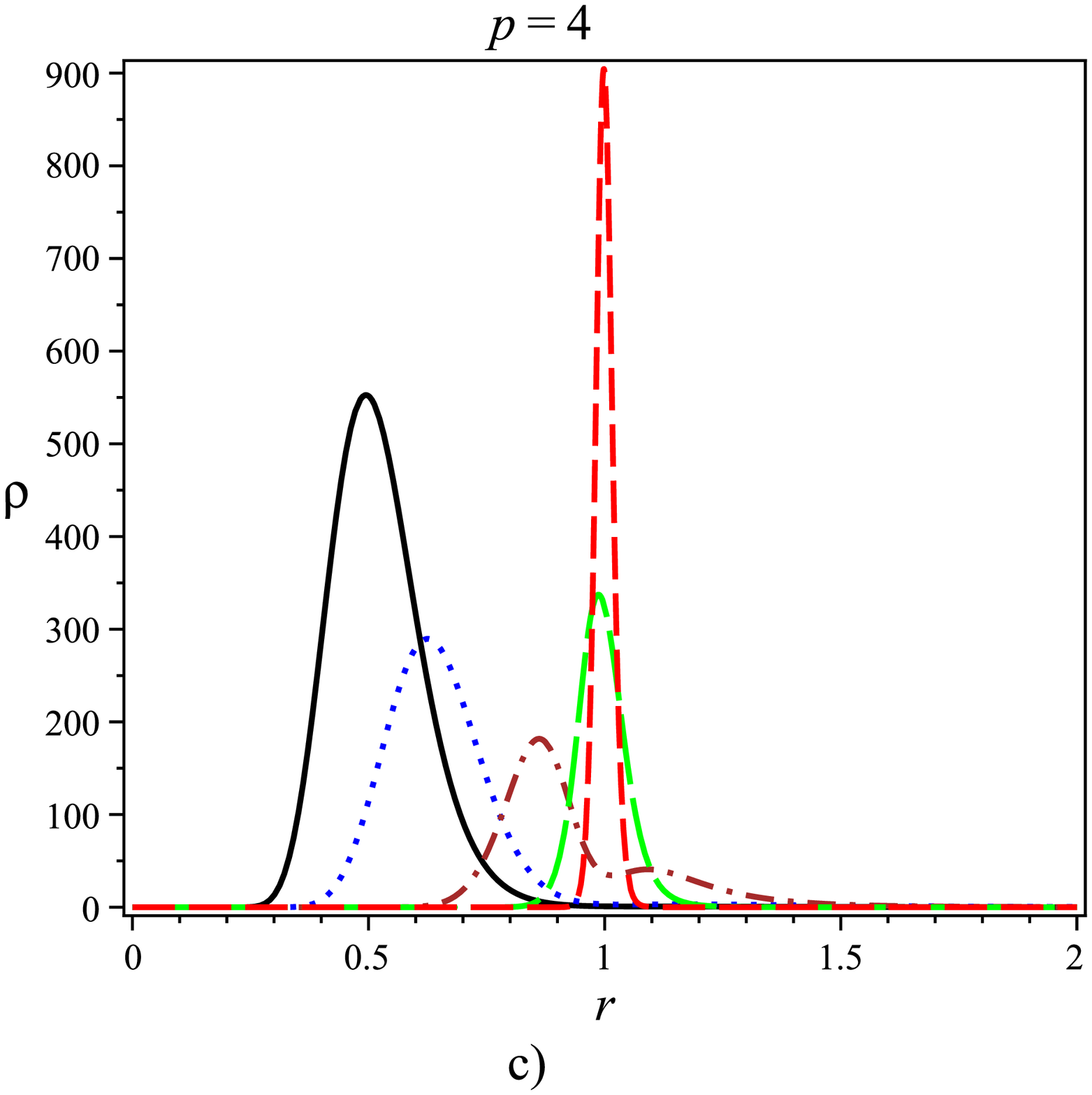}
\includegraphics[{angle=0,width=6.0cm}]{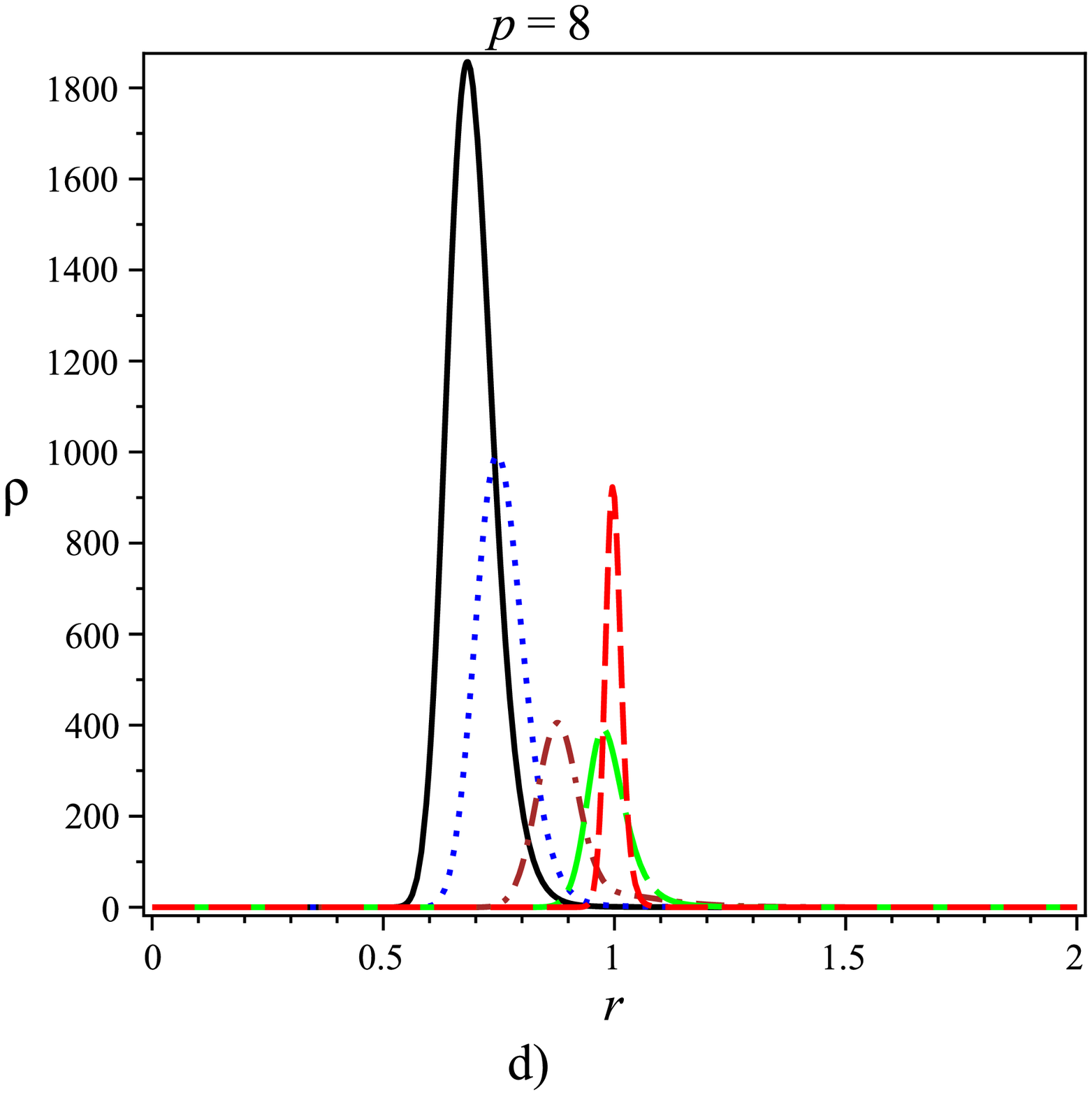}
\caption{A two-field model for $p-$balls in $(D,1)$-dimensions: energy density $\rho(r)$ for $r_{0}=1$ and $\protect\lambda =30$, $s=1/60$ (black line), $s=0.03$ (blue dotted line), $s=0.1$ (brown dash
dotted line), $s=0.3$ (green longdash line), $s=0.5$ (red dashed
line). Plots are for a) $p=2$, b) $p=3$, c) $p=4$ and d) $p=8$. }
\label{Dp_energy}
\end{figure}

\begin{figure}
\scalebox{0.9}{\includegraphics[{angle=0,width=6cm}]{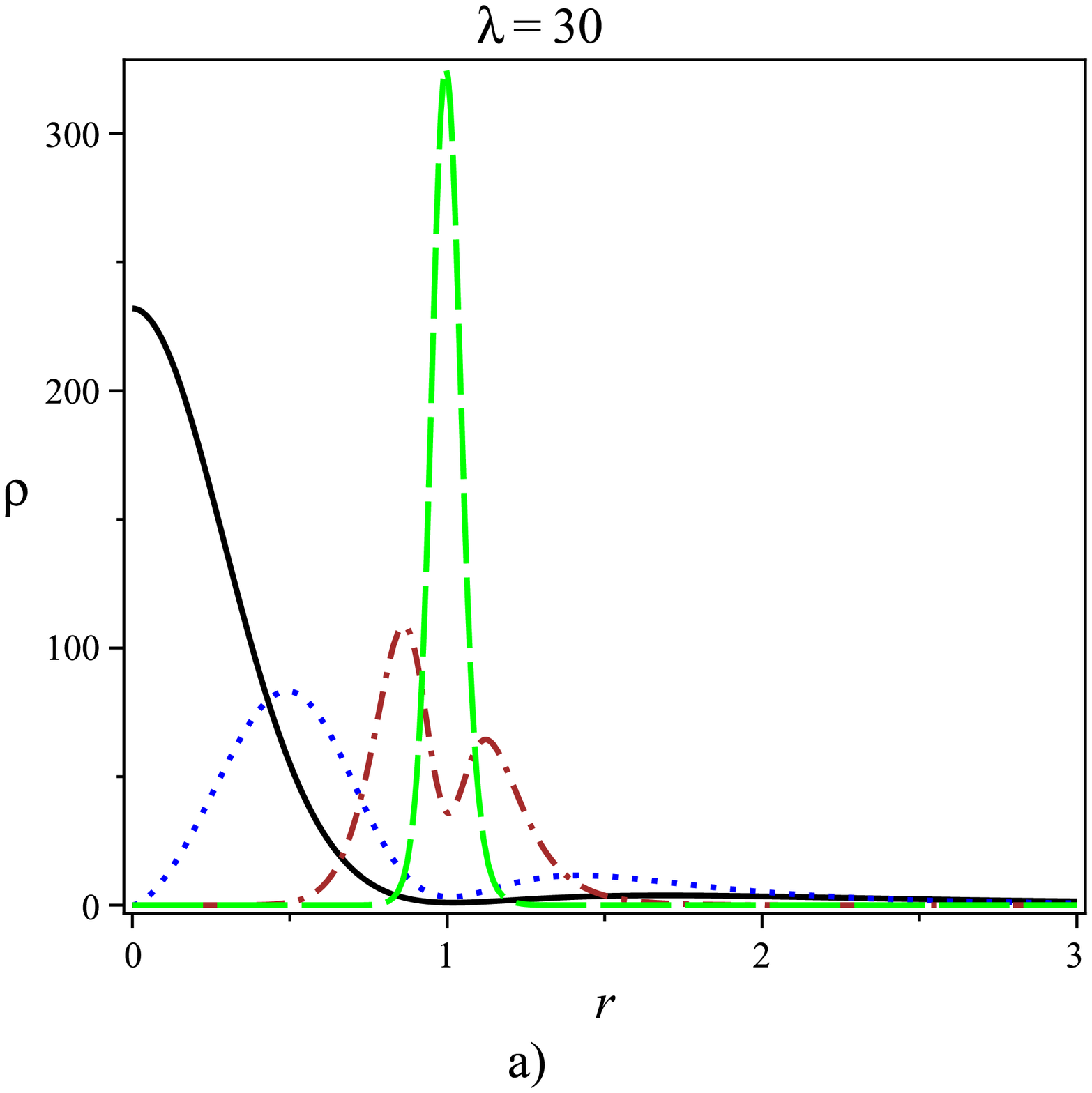}
\includegraphics[{angle=0,width=6cm}]{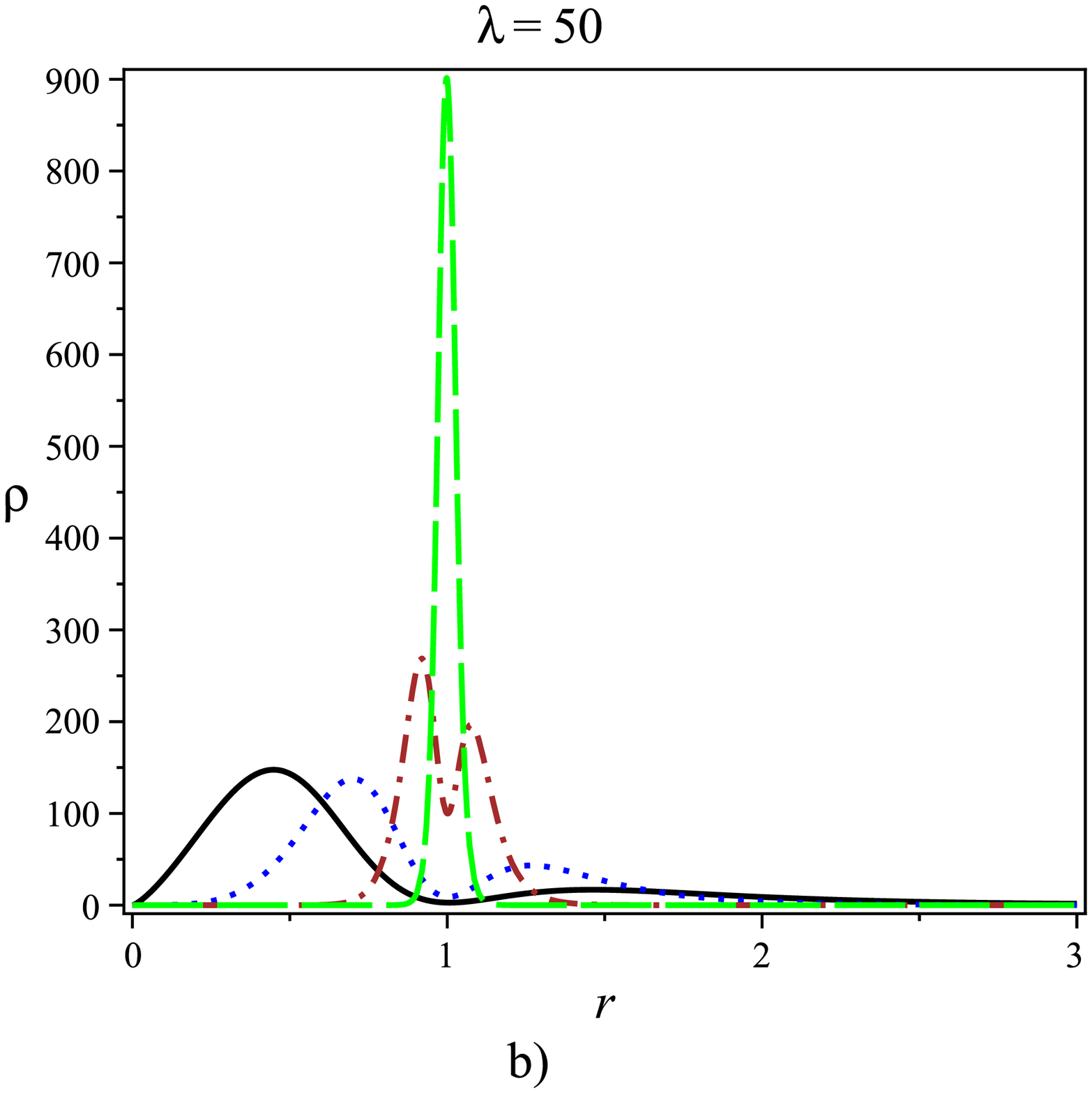}
\includegraphics[{angle=0,width=6cm}]{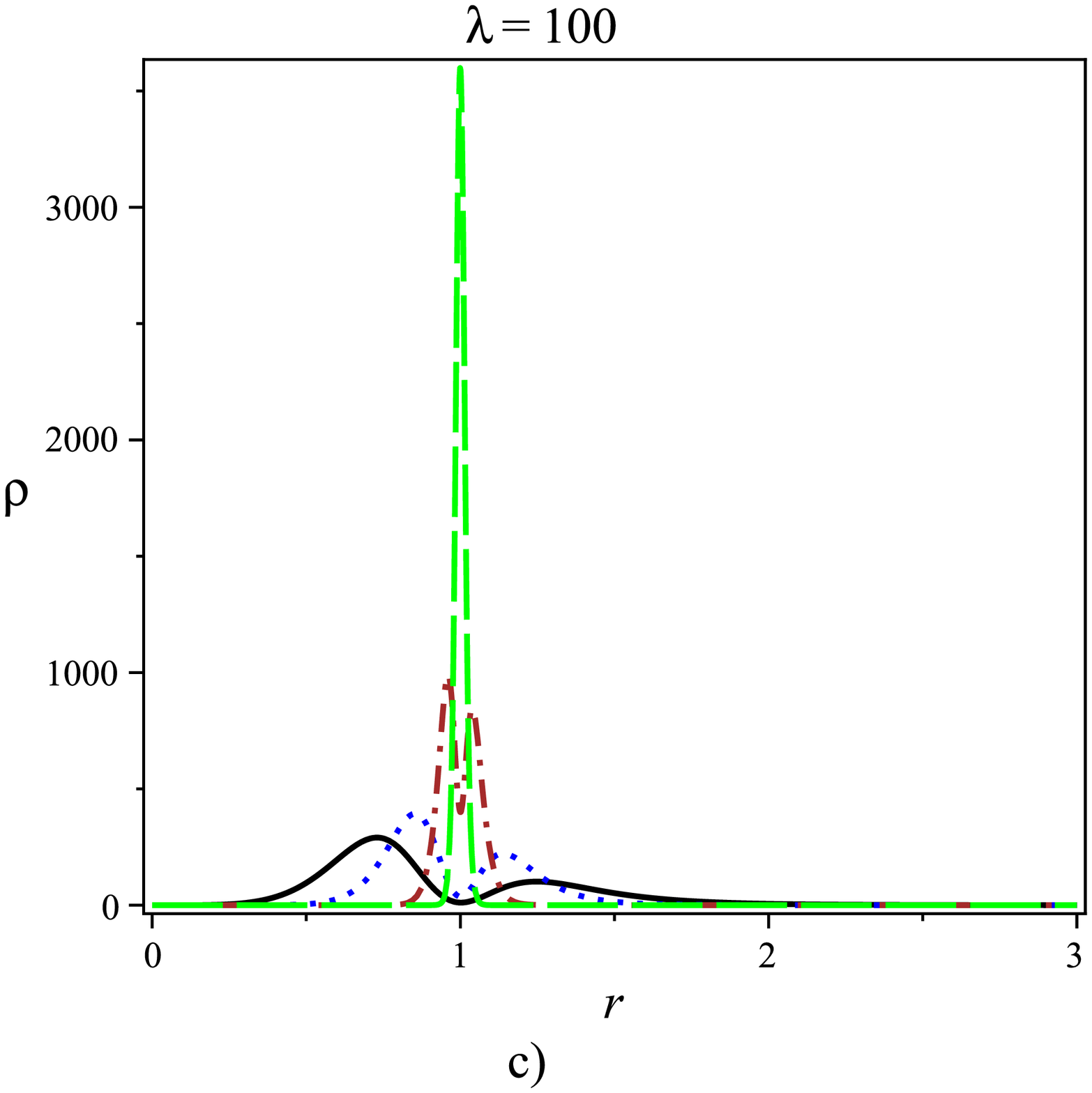}
}
\caption{A two-field model for $2-$balls in $(D,1)$-dimensions: Energy density $\rho(r)$ for $r_{0}=1$, $p=2$ and $s=1/60$ (black line), $s=0.03$ (blue dotted line), $s=0.1$ (brown dash
dotted line), $s=0.3$ (green longdash line).  Plots are for  a) $\lambda=30$, b) $\lambda=50$ and c) $\lambda=100$. }
\label{Dp_energy_lambda}
\end{figure}

Fig. \ref{Dp_energy} shows plots of the energy density $\rho(r)$ for fixed
$\lambda, r_0$ and several values of $s$ for $p=2,3,4$ and 8. From Fig.
\ref{Dp_energy}a we see that for $p=2$ the
behavior of the energy density changes from a lump centered in $r=0$
$\left(s=\frac{1}{2\lambda }\right) $ to a high peak centered around
$r_{0}$\ $\left( s=0.5\right) $. Comparing Figs. \ref{Dp_energy}a-d
we see that the peaks for $s\simeq 0.5$ do not depend on the number
$p$ of transverse dimensions. However, for lower values of $s$ the
behavior of $\rho(r)$ changes sensibly. Indeed, for small $s$ the
lump centered at $r=0$ occurs only for $p=2$. For $p\ge3$ there
appears a broad peak for centered at a $0<r<r_0$. The larger is
$p$, the higher and thinner is this peak. For $p\ge6$ the peak
for small $s$ turns to be the higher in comparison to those occurring
for larger values of $s$.  Fig. \ref{Dp_energy}d shows this effect for
$p=8$. The influence of $\lambda$ on the energy density can be seen in
Fig. \ref{Dp_energy_lambda} for $p=2$. Note that the increase of
$\lambda$ turns the energy density more centered around $r=r_0$.
At the same time this reduces the relative maximum of the energy
density for lower values of $s$ in comparison to the higher ones.
We noted a similar behavior with the variation of $\lambda$ for $p=3$.

\begin{figure}
\includegraphics[{angle=0,width=6cm}]{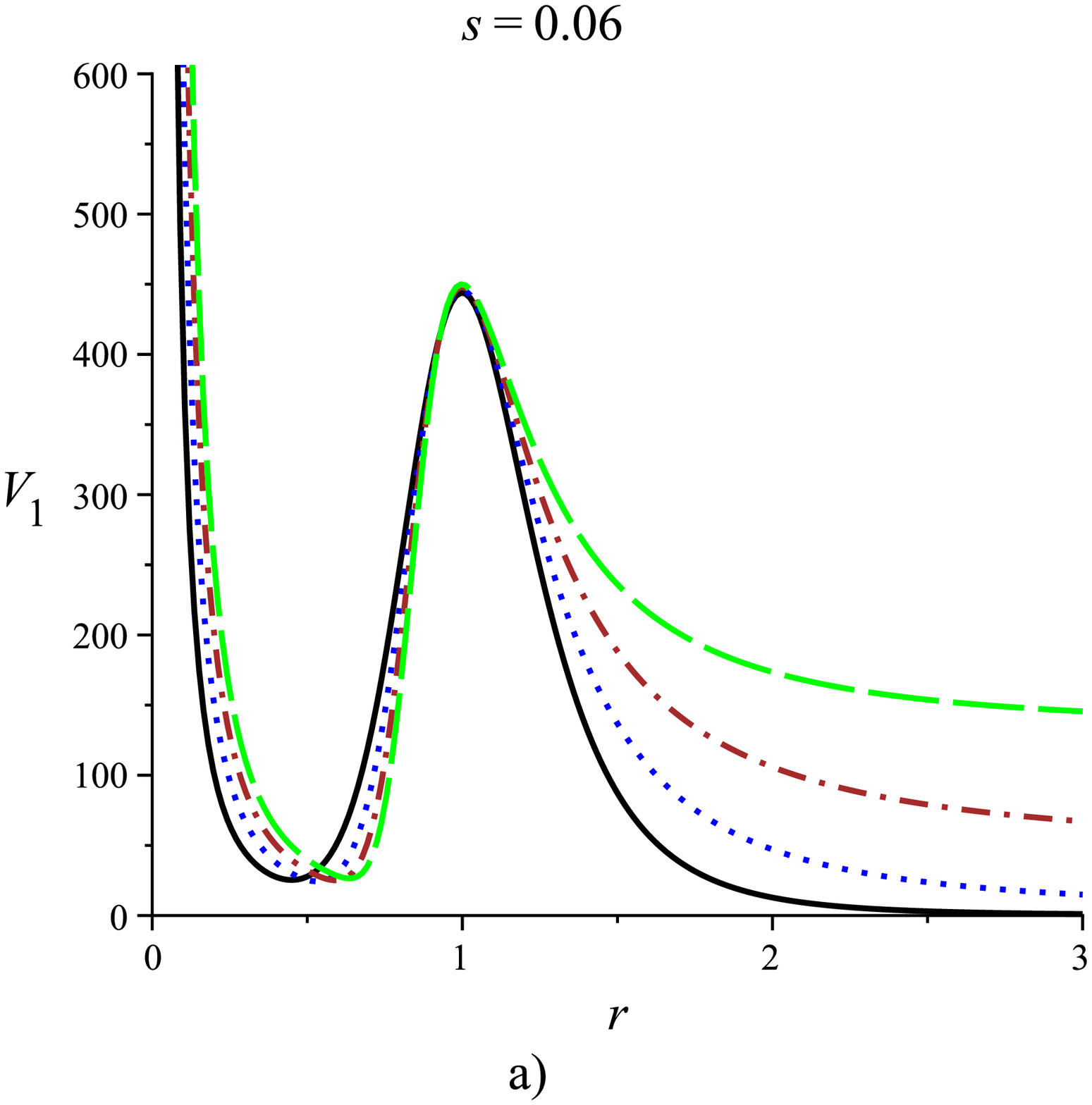}
\includegraphics[{angle=0,width=6cm}]{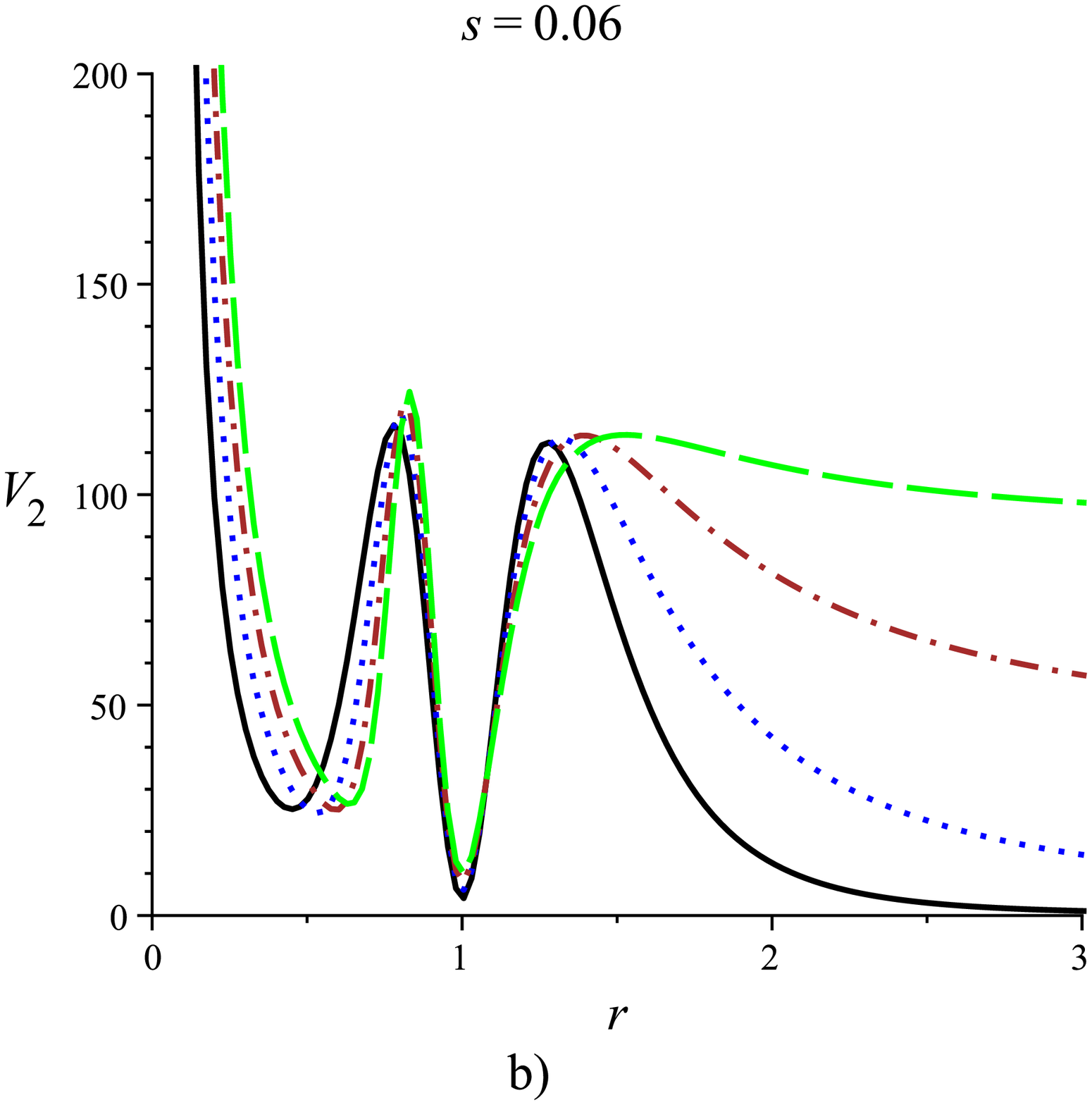}
\\
\includegraphics[{angle=0,width=6cm}]{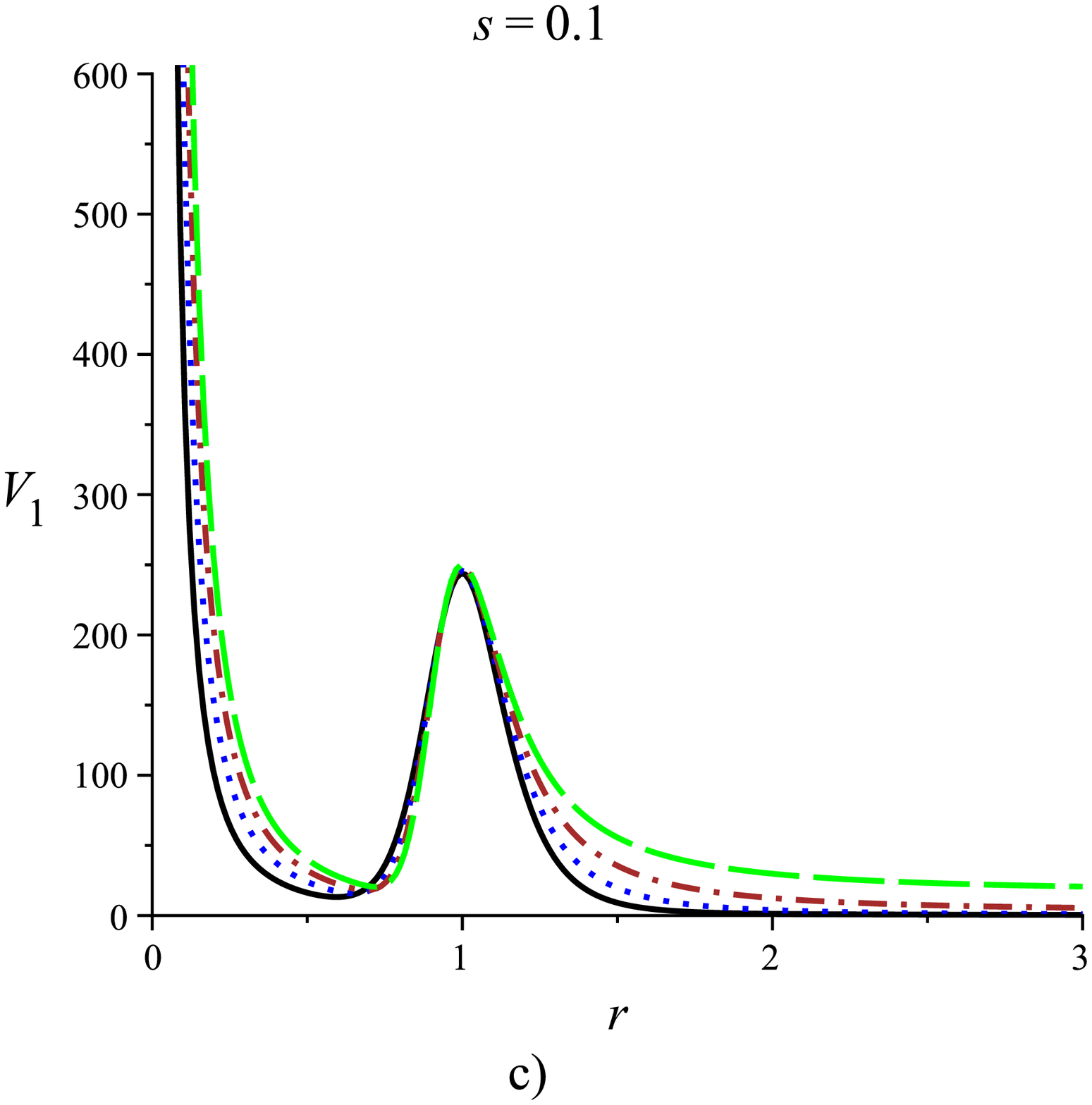}
\includegraphics[{angle=0,width=6cm}]{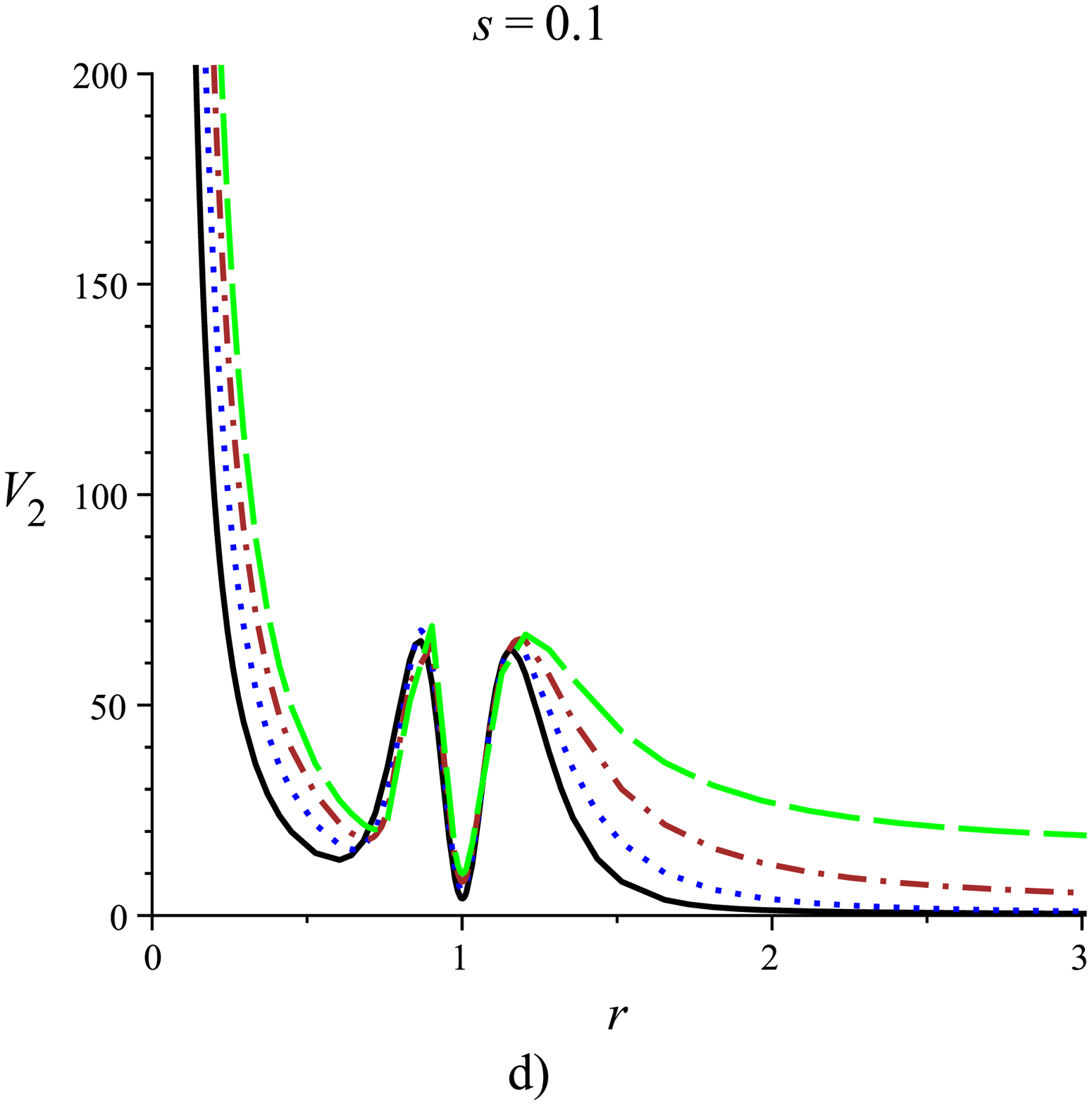}

\caption{A two-field model for $p-$balls in $(D,1)$-dimensions: Schr\"odinger potentials  $V_1(r)$ (left) and  $V_2(r)$ (right) for $\jmath=2$, $r_0=1$, $\eta=30$, $\lambda=30$
and $p=2$ (black line), $p=3$ (blue dotted line), $p=4$ (brown dash
dotted line), $p=5$ (green longdash line). Plots are for $s=0.06$ and $s=0.1$.}
\label{Dp-potentials}
\end{figure}

\begin{table}[tbp]\scriptsize
\begin{tabular}{|c|c|c|c|c|c|c|c|c|c|c|c|c|c|c|}
\hline
$V$ & $s$ & \multicolumn{2}{c|}{$p=2$} & \multicolumn{2}{c|}{$p=3$} &
\multicolumn{2}{c|}{$p=4$} & \multicolumn{2}{c|}{$p=5$} &
\multicolumn{2}{c|}{$p=6$} & \multicolumn{2}{c|}{$p=7$} & $n$ \\ \hline
&  & $j=0$ & $j=2$ & $j=0$ & $j=2$ & $j=0$ & $j=2$ & $j=0$ & $j=2$ & $j=0$ &
$j=2$ & $j=0$ & $j=2$ &  \\ \hline \hline
\multirow{4}{*}{$V_1$} & \multirow{4}{*}{$0.06$} & -- & -- & -- & -- &
18.4988 & -- & 16.9502 & 65.5898 & 15.9269 & 68.9682 & 15.1955 & 72.4943 & 1
\\ \cline{3-15}
&  & -- & -- & -- & -- & -- & -- & 66.3794 & -- & 62.7636 & 149.3757 &
60.0992 & 153.8414 & 2 \\ \cline{3-15}
&  & -- & -- & -- & -- & -- & -- & -- & -- & 137.6105 & -- & 132.5581 &
254.5088 & 3 \\ \cline{3-15}
&  & -- & -- & -- & -- & -- & -- & -- & -- & -- & -- & 228.5419 & -- & 4 \\
\hline \hline
\multirow{2}{*}{$V_1$} & \multirow{2}{*}{$0.10$} & -- & -- & -- & -- & -- &
-- & 13.3597 & -- & 12.9701 & -- & 12.6690 & 61.0235 & 1 \\ \cline{3-15}
&  & -- & -- & -- & -- & -- & -- & -- & -- & -- & -- & 50.0838 & -- & 2 \\
\hline\hline
\multirow{4}{*}{$V_2$} & \multirow{4}{*}{$0.06$} & -- & -- & -- & -- &
17.6796 & -- & 16.0810 & 57.2498 & 14.9927 & 58.2993 & 14.1899 & 59.1008 & 1
\\ \cline{3-15}
&  & -- & -- & -- & -- & -- & -- & 51.6688 & 71.1128 & 49.2798 & 73.4513 &
47.0188 & 75.5898 & 2 \\ \cline{3-15}
&  & -- & -- & -- & -- & -- & -- & 68.0193 & -- & 65.5464 & -- & 63.4274 & --
& 3 \\ \cline{3-15}
&  & -- & -- & -- & -- & -- & -- & -- & -- & 108.7936 & -- & 100.9167 & -- &
4 \\ \hline\hline
\multirow{2}{*}{$V_2$} & \multirow{2}{*}{$0.10$} & -- & -- & -- & -- & -- &
-- & 11.9547 & -- & 11.4910 & -- & 11.1174 & 48.1639 & 1 \\ \cline{3-15}
&  & -- & -- & -- & -- & -- & -- & -- & -- & -- & -- & 38.0263 & -- & 2 \\
\hline
\end{tabular}
\caption{A two-field model for $p-$balls in $(D,1)$-dimensions: eigenvalues $M_{n\jmath}^2$, solutions of Eq. (\ref{eq_rho}) for $\jmath=0$ and $\jmath=2$. We fix $r_0=1$, $\eta=30$, $\lambda=30$, with Schr\"odinger potentials  for $\jmath=2$ corresponding to Fig. \ref{Dp-potentials}.}
\end{table}

In the following we will consider separately the couplings $F_1(\chi)=\chi^2$ and $F_2(\phi,\chi)=(\phi\chi)^2$. The corresponding Schr\"odinger-like potentials are
\begin{eqnarray}
V_1 & = & \frac{\jmath(\jmath + p-2)}{r^{2}} + \eta \bigg(\frac{1}{s}-2 \bigg) \sech^2(\tau_p), \,p=2,3,...\\
V_2 & =  & \frac{\jmath(\jmath + p-2)}{r^{2}} + \eta \bigg(\frac{1}{s}-2 \bigg) \tanh^2(\tau_p) \sech^2 (\tau_p),\, p=2,3,... .
\end{eqnarray}
Fig. \ref{Dp-potentials} shows some plots for $V_1(r)$ and $V_2(r)$ for fixed values of $\eta,\lambda,r_0,\jmath$ and several values of $p$. For all values of $p$ the potentials are strictly positive. Also, for $p=2$ we have $V_1(r\to\infty)=0$ and $V_2(r\to\infty)=0$, showing that bound states are absent. For $p\ge3$ we have $V_1(r\to\infty)=\eta F_1(\phi_c,\chi_c)\neq0$ and $V_2(r\to\infty)=\eta F_2(\phi_c,\chi_c)\neq0$ and one can investigate the existence of bound states.

\begin{figure}
\includegraphics[{angle=0,width=6cm}]{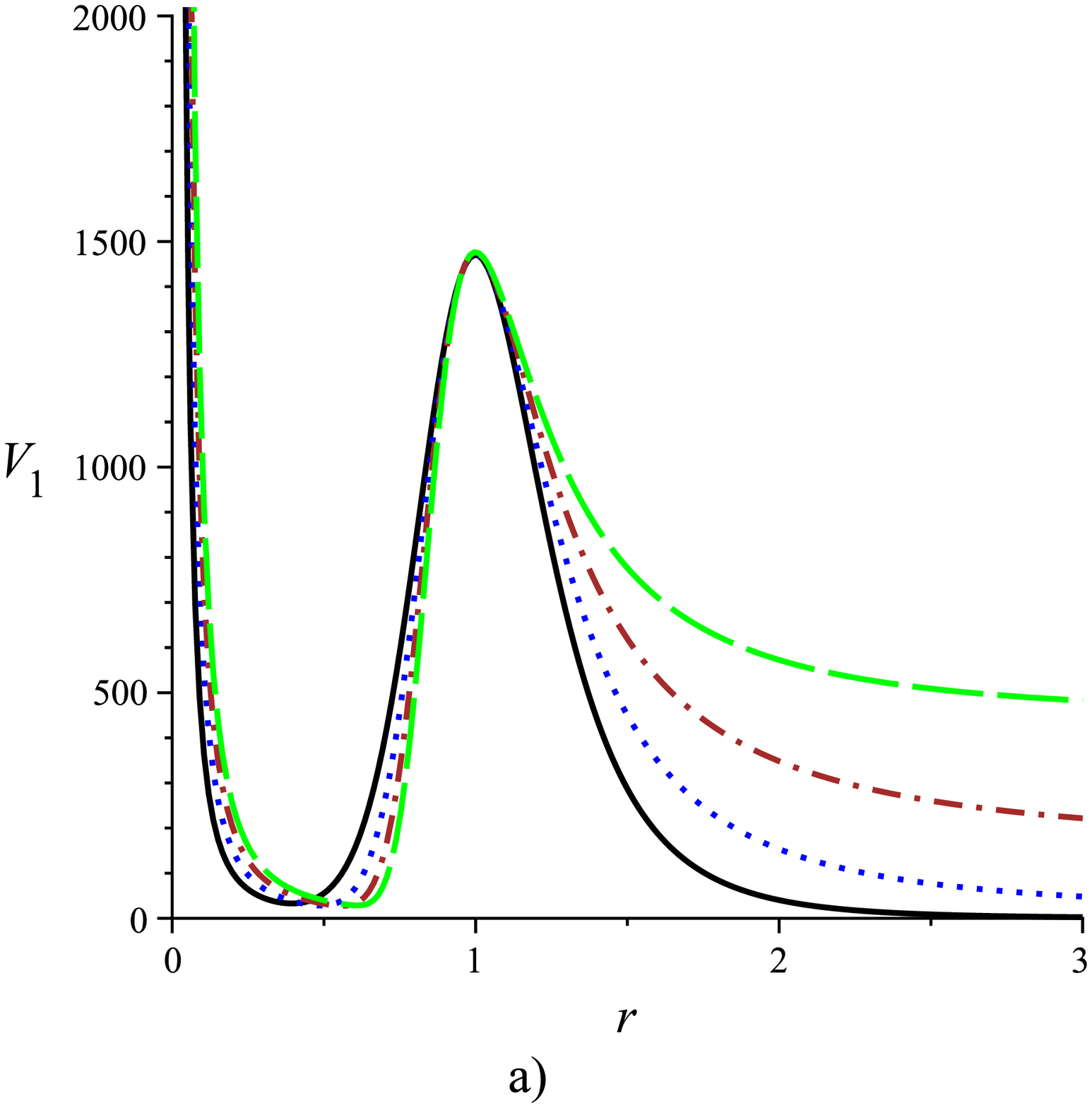}
\includegraphics[{angle=0,width=6cm}]{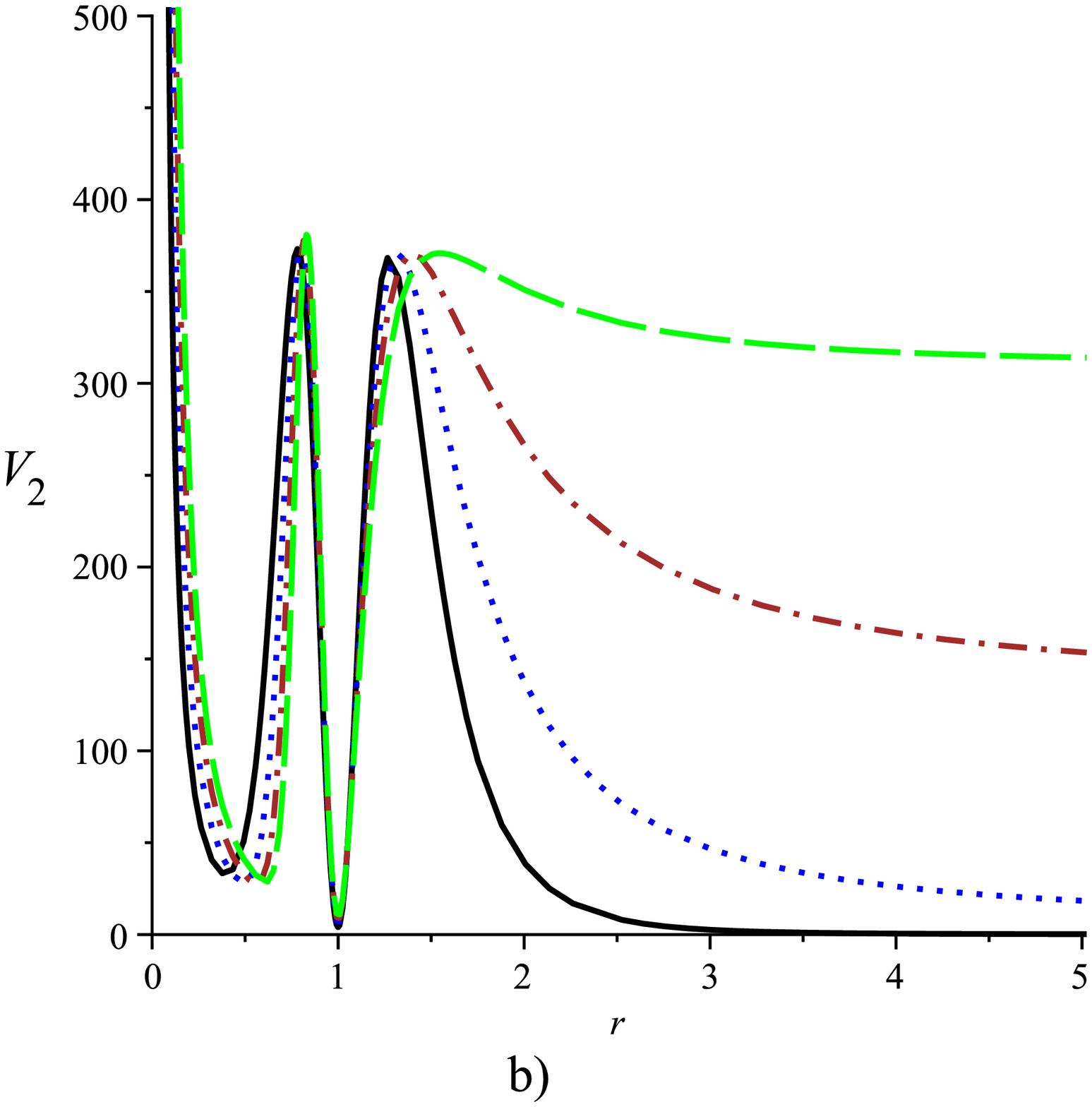}
\caption{A two-field model for $p-$balls in $(D,1)$-dimensions: Schr\"odinger potentials a) $V_1(r)$ (left) and b) $V_2(r)$ (right) for $\jmath=2$, $r_0=1$, $\eta=100$, $\lambda=30$, $s=0.06$
and $p=2$ (black line), $p=3$ (blue dotted line), $p=4$ (brown dash
dotted line), $p=5$ (green longdash line). }
\label{Dp-potentials-lambda30}
\end{figure}

First of all we consider the case $\jmath=2$. For the potentials shown in Fig. \ref{Dp-potentials}, and also for higher values of $p$, the existence of bound states was investigated and some results could be found according to Table III. { For $s=0.06$ and $s=0.1$ we could not find bound states for neither $V_1$ nor $V_2$ when $p=2,3,4$. Bound states start to appear for $p\ge5$ for $s=0.06$ and $p\ge7$ for $s=0.1$. The results show that lower values of $s$ are better for the occurrence of bound states. As explained above, $s\to 0.5$ recover a one-field $\phi$ model. Then, the presence of a second scalar field $\chi$ contributes to trapping spin-0 particles. In other words, $p-$balls with more internal structure are more able to trap scalar particles.
Also, for fixed $s$ and larger number $p$  of transverse dimensions, bound states occur with larger masses for potential $V_1$ than for potential $V_2$.  This shows that, for a trapping mechanism, the quadratic coupling $\chi^2$ is better that the quartic one $\phi^2\chi^2$.
Moreover, a multidimensional $p-$ball with  larger $p$ is a better trapping
mechanism, as $V_1(r\to\infty)$ and $V_2(r\to\infty)$ grows with $p$.
This is confirmed in Table III. Indeed, the larger is $p$, the greater
is the number of bound states. 
Also, the asymptotic values $V_1(r\to\infty)$ and $V_2(r\to\infty)$ grow
with $\lambda$ and $\eta$. }This signals that in order to grow the
probability for the occurrence of bound states one must have
$\eta,\lambda \gg1$ and decrease the ratio $\lambda/\eta$. This can be
seen in Fig. \ref{Dp-potentials-lambda30} where $\lambda/\eta=3/10$
(compare with Fig. \ref{Dp-potentials} where $\lambda/\eta=1$).
Corresponding eigenvalues are described in Table IV.  From the results
for $\lambda=30$ we see that for $\lambda/\eta=3/10$ and $s=0.06$ there
occur bound states for $p\geq 4$  (compare with the results of
Table III for $\lambda/\eta=1$ where bound states appear for $p\geq 4$).

\begin{table}[tbp]\scriptsize
\begin{tabular}{|c|c|c|c|c|c|c|c|c|c|c|c|c|c|}
\hline
$V$ & \multicolumn{2}{|c|}{$p=2$} & \multicolumn{2}{|c|}{$p=3$} &
\multicolumn{2}{|c|}{$p=4$} & \multicolumn{2}{|c|}{$p=5$} &
\multicolumn{2}{|c|}{$p=6$} & \multicolumn{2}{|c|}{$p=7$} & $n$ \\ \hline
& $j=0$ & $j=2$ & $j=0$ & $j=2$ & $j=0$ & $j=2$ & $j=0$ & $j=2$ & $j=0$ & $%
j=2$ & $j=0$ & $j=2$ &  \\ \hline\hline
\multirow{7}{*}{$V_1$} & -- & -- & -- & -- & 21.5827 & 72.5884 & 19.3059 &
74.4329 & 17.8608 & 77.1262 & 16.8559 & 80.2321 & 1 \\ \cline{2-14}
& -- & -- & -- & -- & 84.3439 & 166.9563 & 76.1917 & 166.9618 & 70.8185 &
168.9305 & 67.0053 & 171.9820 & 2 \\ \cline{2-14}
& -- & -- & -- & -- & -- & -- & 167.7651 & 285.8629 & 157.0522 & 286.6631 &
149.2038 & 289.1227 & 3 \\ \cline{2-14}
& -- & -- & -- & -- & -- & -- & 289.9211 & 427.9080 & 273.7365 & 427.8130 &
261.4261 & 429.7415 & 4 \\ \cline{2-14}
& -- & -- & -- & -- & -- & -- & 438.2871 & -- & 417.4132 & 589.5117 &
400.9835 & 591.3007 & 5 \\ \cline{2-14}
& -- & -- & -- & -- & -- & -- & -- & -- & 584.4205 & -- & 564.7141 & 770.9510
& 6 \\ \cline{2-14}
& -- & -- & -- & -- & -- & -- & -- & -- & -- & -- & 749.3110 & -- & 7 \\
\hline\hline
\multirow{7}{*}{$V_2$} & -- & -- & -- & -- & 21.3410 & 71.6610 & 19.0653 &
73.3757 & 17.6114 & 75.8862 & 16.5924 & 78.7597 & 1 \\ \cline{2-14}
& -- & -- & -- & -- & 82.9807 & 127.0889 & 74.9047 & 127.4501 & 69.5199 &
127.3683 & 65.6502 & 126.8865 & 2 \\ \cline{2-14}
& -- & -- & -- & -- & 119.3195 & -- & 117.8118 & 163.8790 & 115.8253 &
165.8453 & 113.3861 & 168.7750 & 3 \\ \cline{2-14}
& -- & -- & -- & -- & -- & -- & 164.1016 & 273.4499 & 153.9482 & 272.8946 &
146.5262 & 270.4678 & 4 \\ \cline{2-14}
& -- & -- & -- & -- & -- & -- & 273.8062 & 310.4326 & 259.9589 & 300.8025 &
248.3106 & 290.6058 & 5 \\ \cline{2-14}
& -- & -- & -- & -- & -- & -- & 304.4750 & -- & 290.8448 & -- & 275.0988 &
339.6651 & 6 \\ \cline{2-14}
& -- & -- & -- & -- & -- & -- & -- & -- & 359.1479 & -- & 332.6537 & -- & 7
\\ \hline
& -- & -- & -- & -- & -- & -- & -- & -- & -- & -- & 345.4081 & -- & 8 \\
\hline
\end{tabular}
\caption{A two-field model for $p-$balls in $(D,1)$-dimensions: eigenvalues $M_{n\jmath}^2$, solutions of Eq. (\ref{eq_rho})  for $\jmath=0$ and $\jmath=2$. We fix $r_0=1$, $\eta=100$, $\lambda=30$, $s=0.06$, with Schr\"odinger potentials for $\jmath=2$ corresponding to Fig. \ref{Dp-potentials-lambda30}.}
\end{table}

\begin{figure}
\includegraphics[{angle=0,width=6cm}]{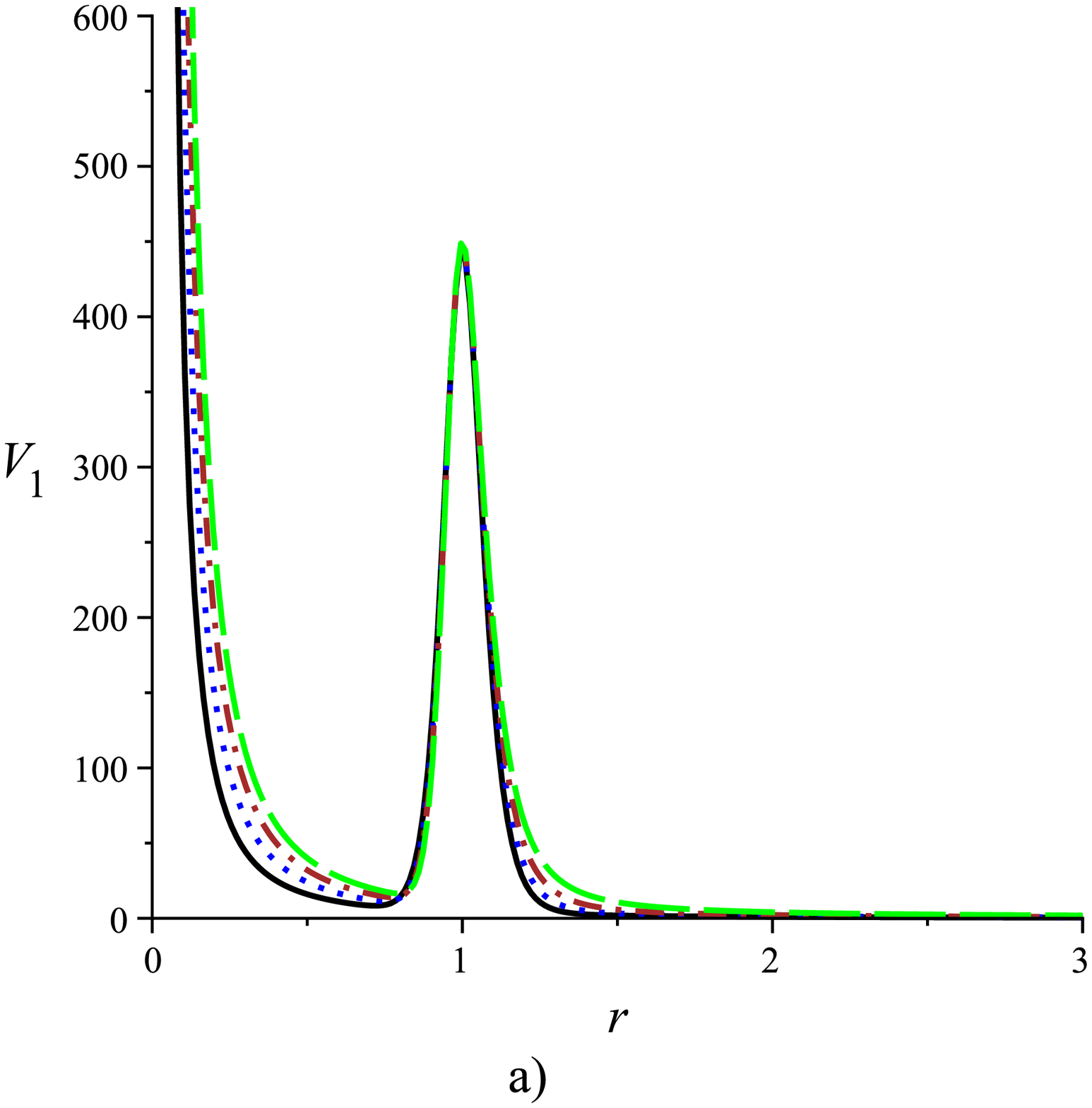}
\includegraphics[{angle=0,width=6cm}]{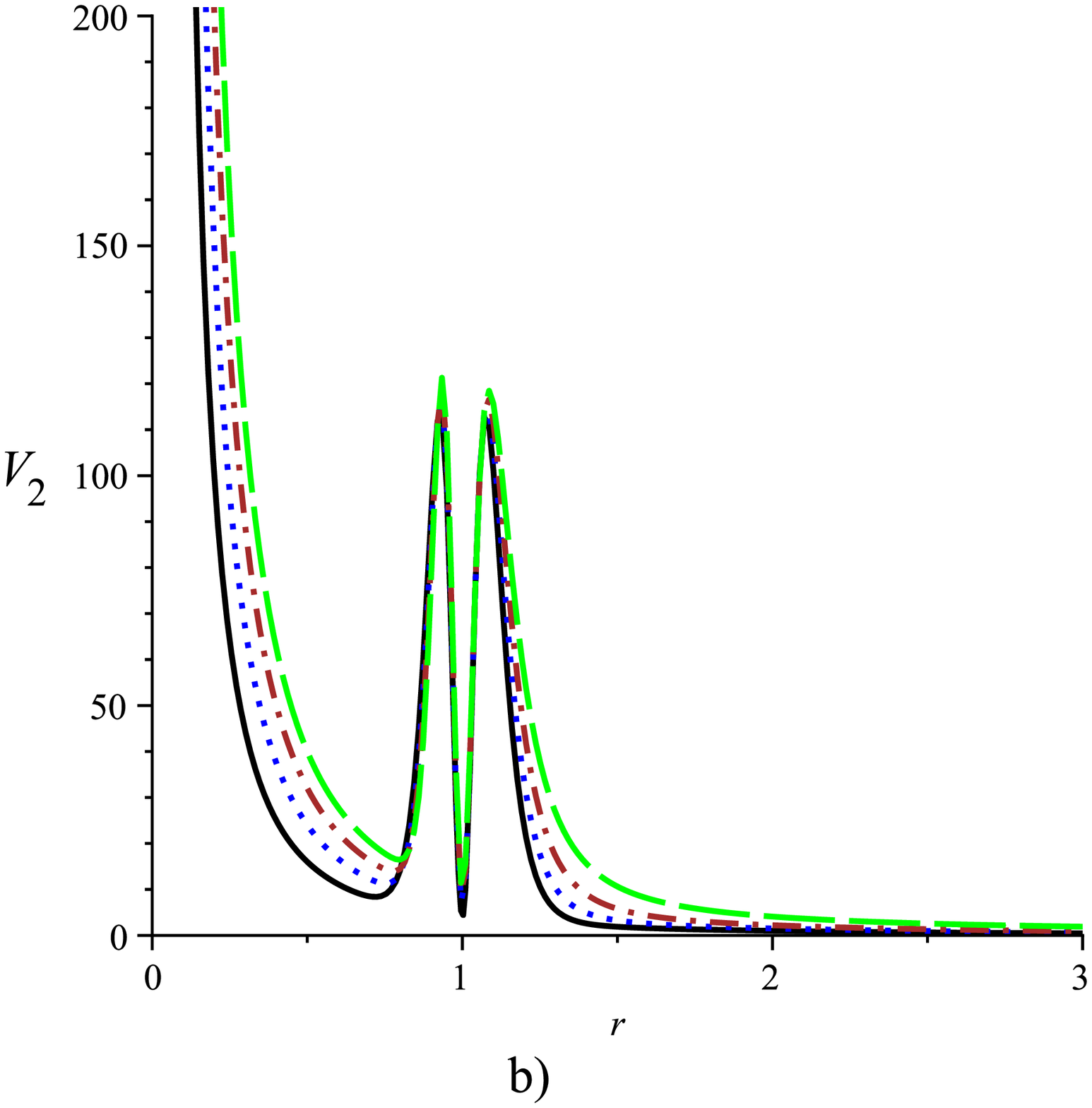}
\caption{A two-field model for $p-$balls in $(D,1)$-dimensions: Schr\"odinger potentials a) $V_1(r)$ (left) and b) $V_2(r)$ (right) for $\jmath=2$, $r_0=1$, $\eta=30$, $\lambda=100$, $s=0.06$
and $p=2$ (black line), $p=3$ (blue dotted line), $p=4$ (brown dash
dotted line), $p=5$ (green longdash line).}
\label{Dp-potentials-lambda}
\end{figure}

 Now note that the potentials for coupling $F_1$ are characterized by a local maximum at $r=r_0$ whereas for $F_2$ there is a local minimum at $r=r_0$ surrounded by two local maxima. The higher local peak for $V_1$ in comparison to the two local ones for $V_2$ suggests that, with the same set of parameters, resonant states are most probable with quadratic coupling $F_1$ than with quartic coupling $F_2$. This is in accord to the behavior of couplings concerning to the occurrence of bound states. We also see that the height of the local maxima grows with the decreasing of $s$, favoring the appearance of resonances. Then  we expect the presence of a second scalar field $\chi$ to be important for the increasing in the number and lifetime of resonances.  We also found that for all other parameters fixed, the best choice for  reducing the asymptotical value of $V_1$ and simultaneously increasing the difference between the local maxima and minima is to keep $\eta,\lambda \gg1$  and increase the ratio $\lambda/\eta$. This can be seen in Fig. \ref{Dp-potentials-lambda}, for $\lambda/\eta=10/3$ { (compare with Fig. \ref{Dp-potentials}, where $\lambda/\eta=1$ and with Fig. \ref{Dp-potentials-lambda30}, where $\lambda/\eta=3/10$)}. However, this occurs at the price of making the barrier thinner. Then we expect that an increasing of $\lambda/\eta$ increases the chances for getting a larger number of resonances, but with lower lifetimes.

\begin{figure}
\includegraphics[{angle=0,width=5cm}]{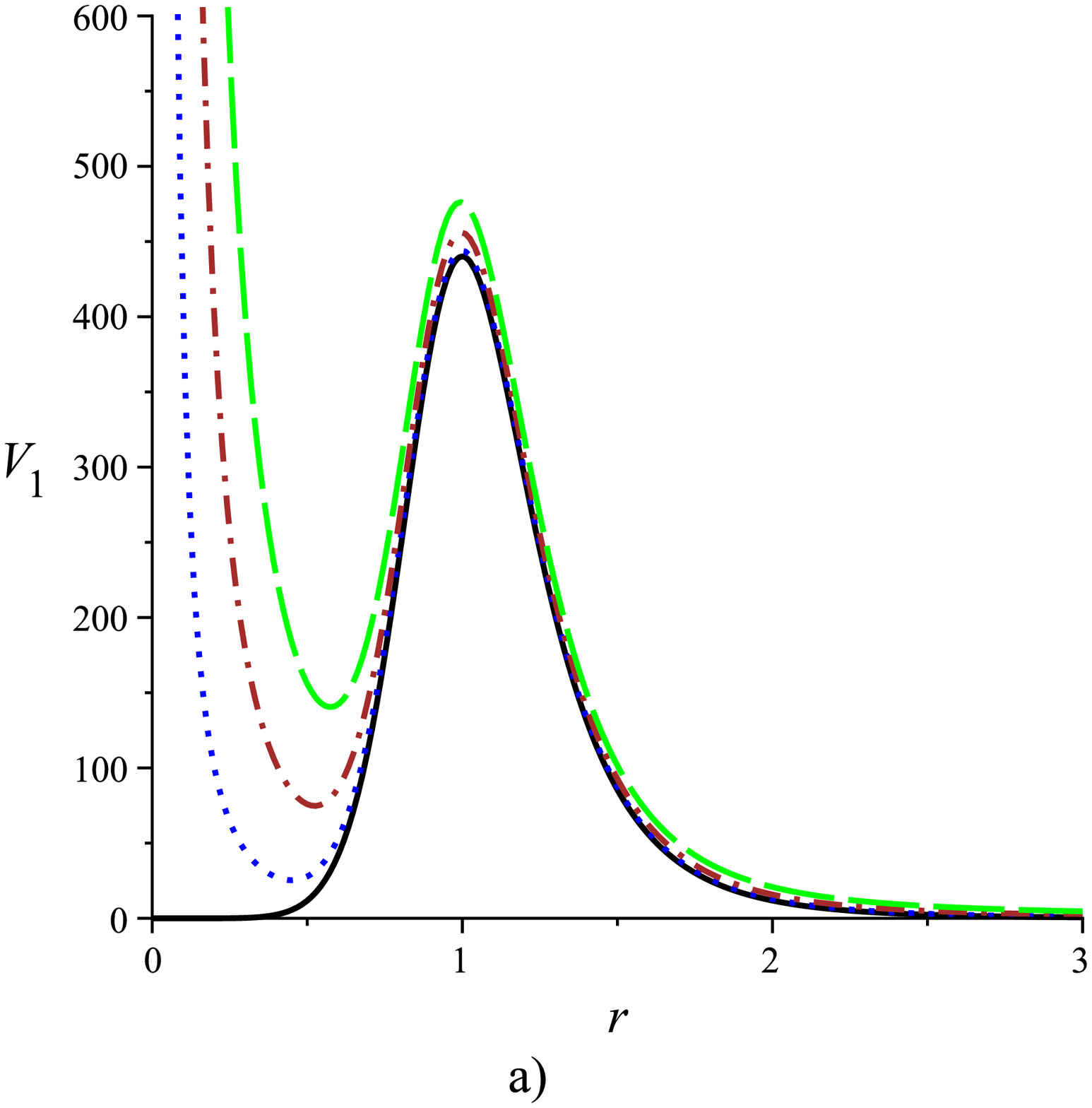}
\includegraphics[{angle=0,width=5cm}]{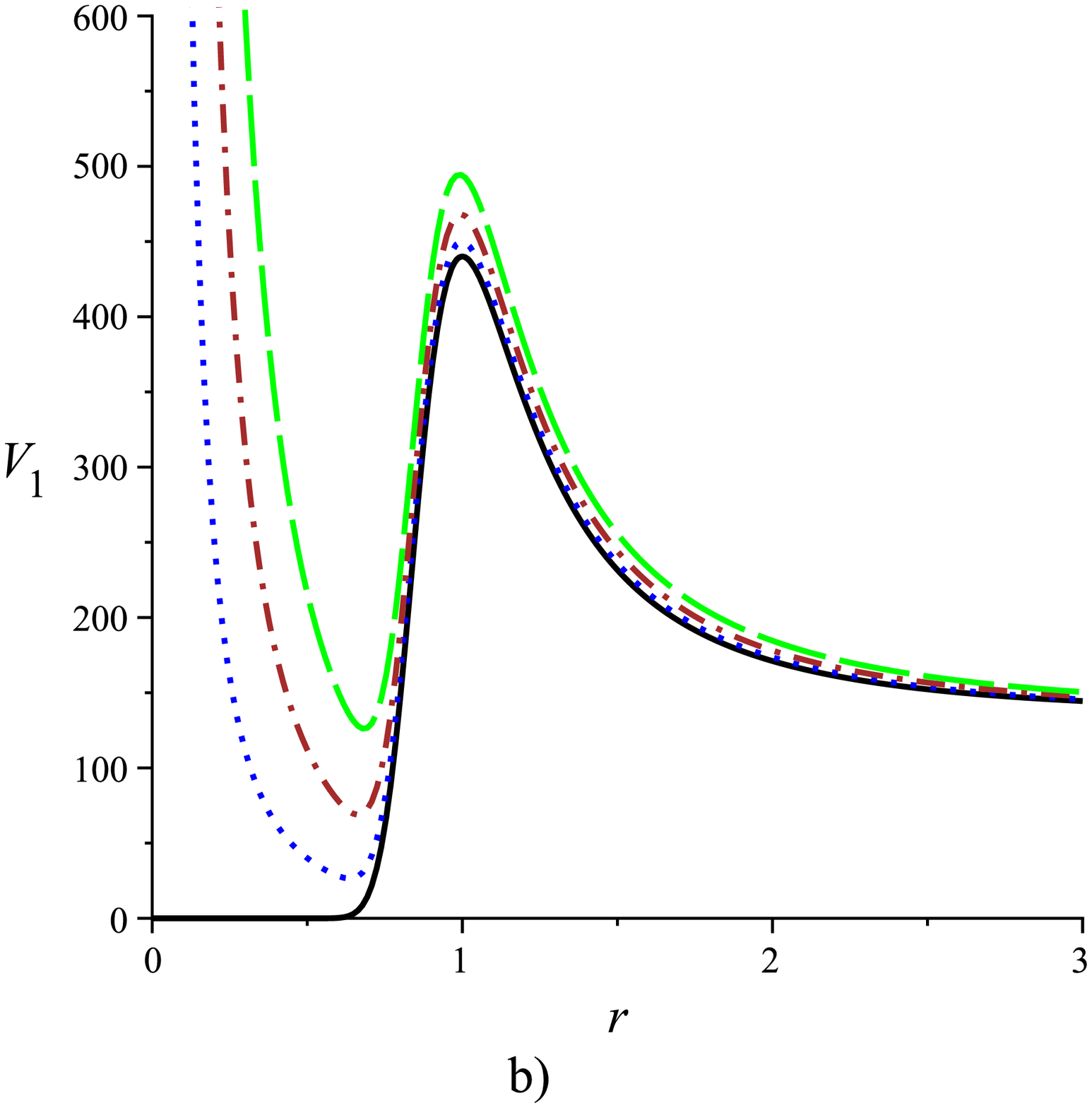}
\includegraphics[{angle=0,width=5cm}]{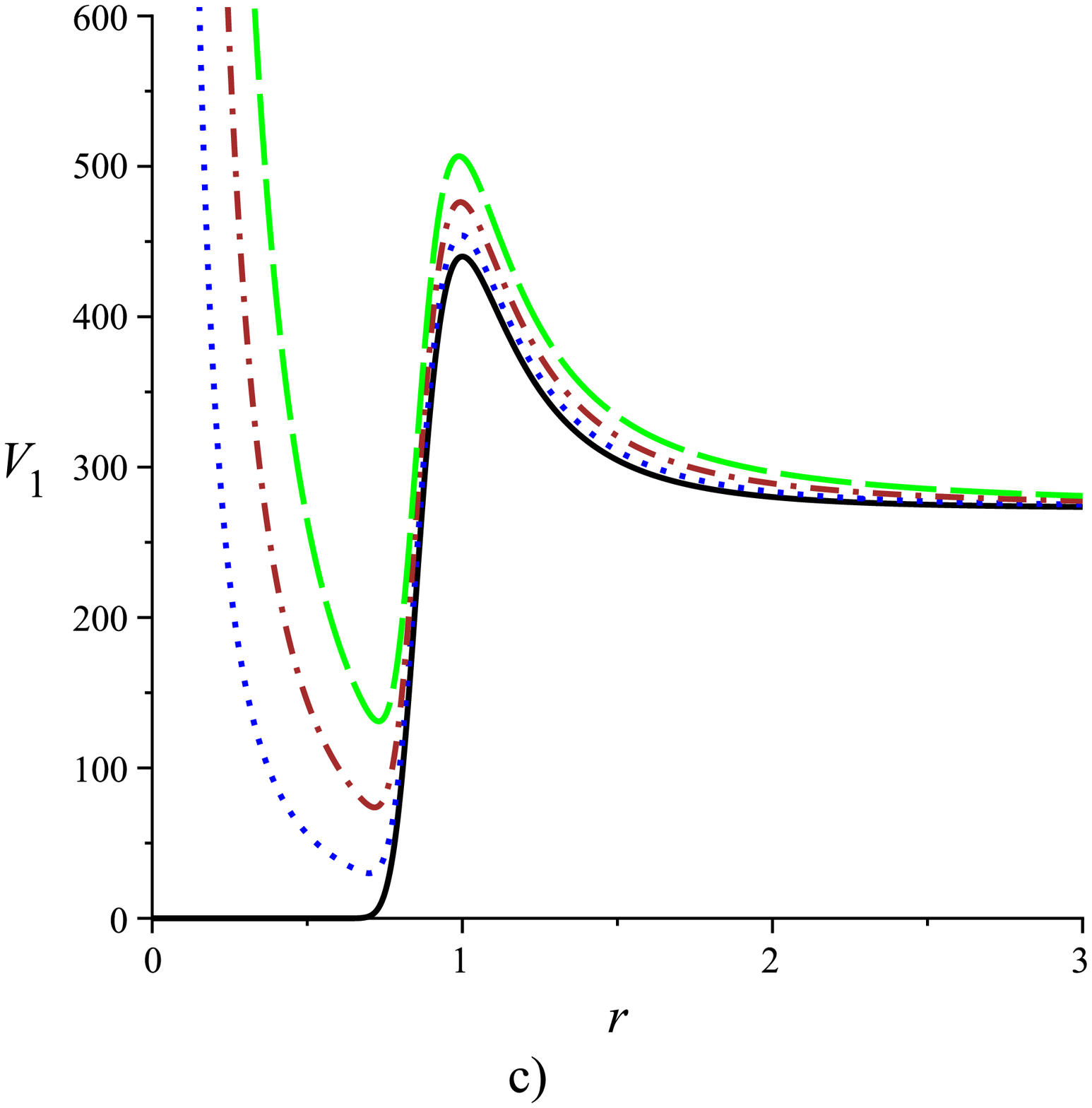}
\caption{A two-field model for $p-$balls in $(D,1)$-dimensions: Schr\"odinger potentials  $V_1(r)$ for a) $p=2$, b) $p=5$, c) $p=7$ and $r_0=1$, $\eta=30$, $\lambda=30$, $s=0.06$,
 $\jmath=0$ (black line), $\jmath=2$ (blue dotted line), $\jmath=4$ (brown dash
dotted line), $\jmath=6$ (green longdash line).}
\label{Dp-j-potentials}
\end{figure}

The influence of the variation of the angular momentum $\jmath$ can be seen in Fig. \ref{Dp-j-potentials}, where we present some plots for the potential $V_1$. The potentials for $\jmath\ge1$ are characterized by a local minimum and local maximum whose separation decreases with $\jmath$. This signals that the increasing of $\jmath$ reduces the possibility of occurrence of bound and resonant states. Case $\jmath=0$ is special since we have $V_1=0$ at $r=0$, being the case with highest possibility for occurrence of such states. Concerning to bound states this is confirmed from the results of Tables III and IV, where one can compare cases $\jmath=0$ and $\jmath=2$. Then former analysis and conclusions of the Schr\"odinger potential and bound states made for $\jmath=2$ also apply for general values of $\jmath$.
Similar analysis for potential $V_2$ leads to the same conclusion: lower values of $\jmath$ are favored for occurrence of bound and resonant states.

Now we will consider specifically the effect of the number of longitudinal and transverse dimensions on the resonance effect.
For the couplings $F_1$ and $F_2$, $\lambda s\geq 1/2$ and $\jmath\ge1$ the
potentials $V_{sch}$ for $r\ll r_{0}$ are dominated by the contributions of the angular momentum
proportional to $1/r^{2}$,
\begin{equation}
V(r) \approx \frac{\jmath(\jmath + p-2)}{r^{2}},
\end{equation}%
and the nonsingular solutions\ in $r=0$ are given by Eq. (\ref{Dp_rho_rsmall}), used for calculating the relative probability $P$.
\begin{figure}
\includegraphics[{angle=0,width=6cm}]{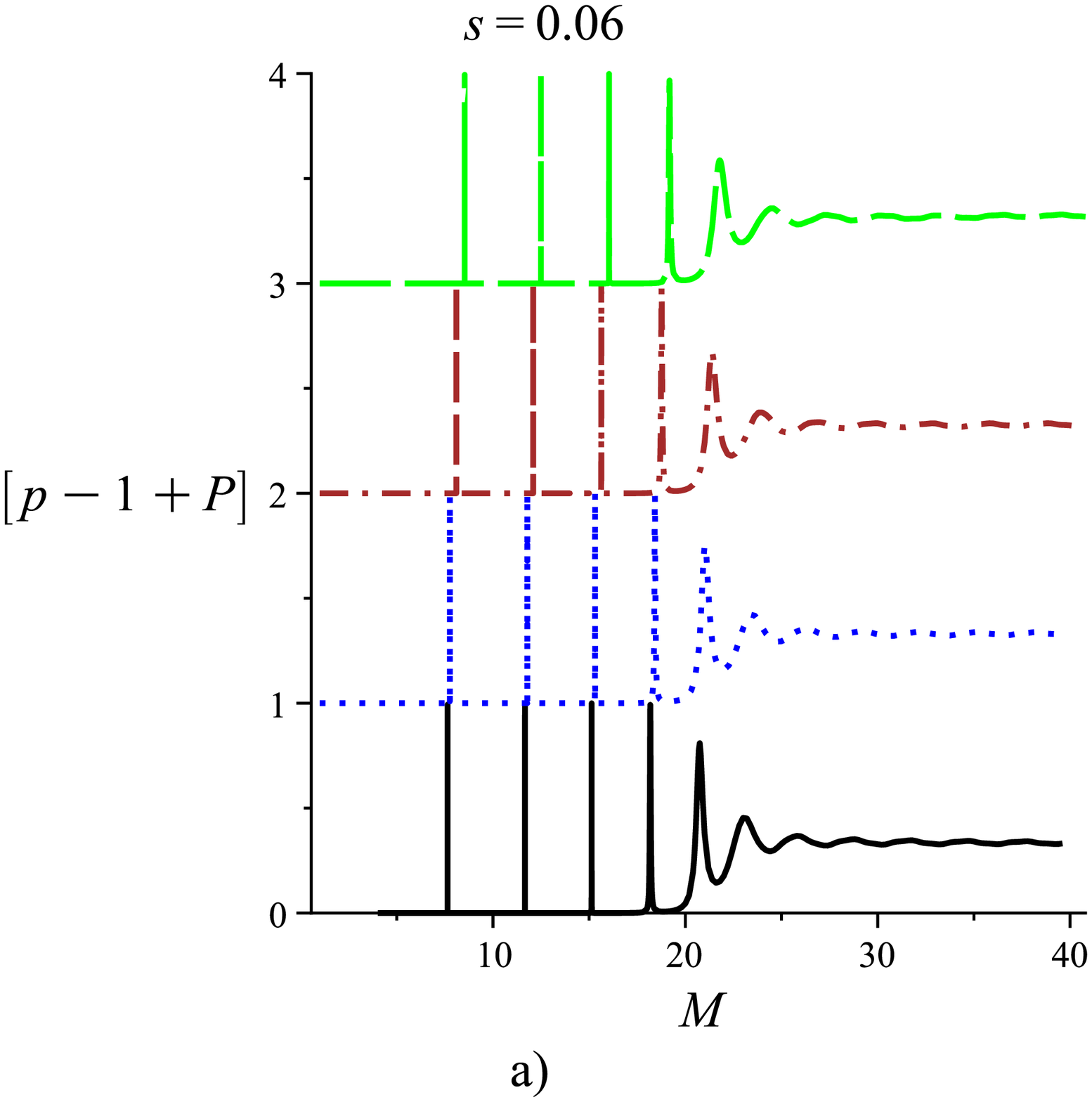}
\includegraphics[{angle=0,width=6cm}]{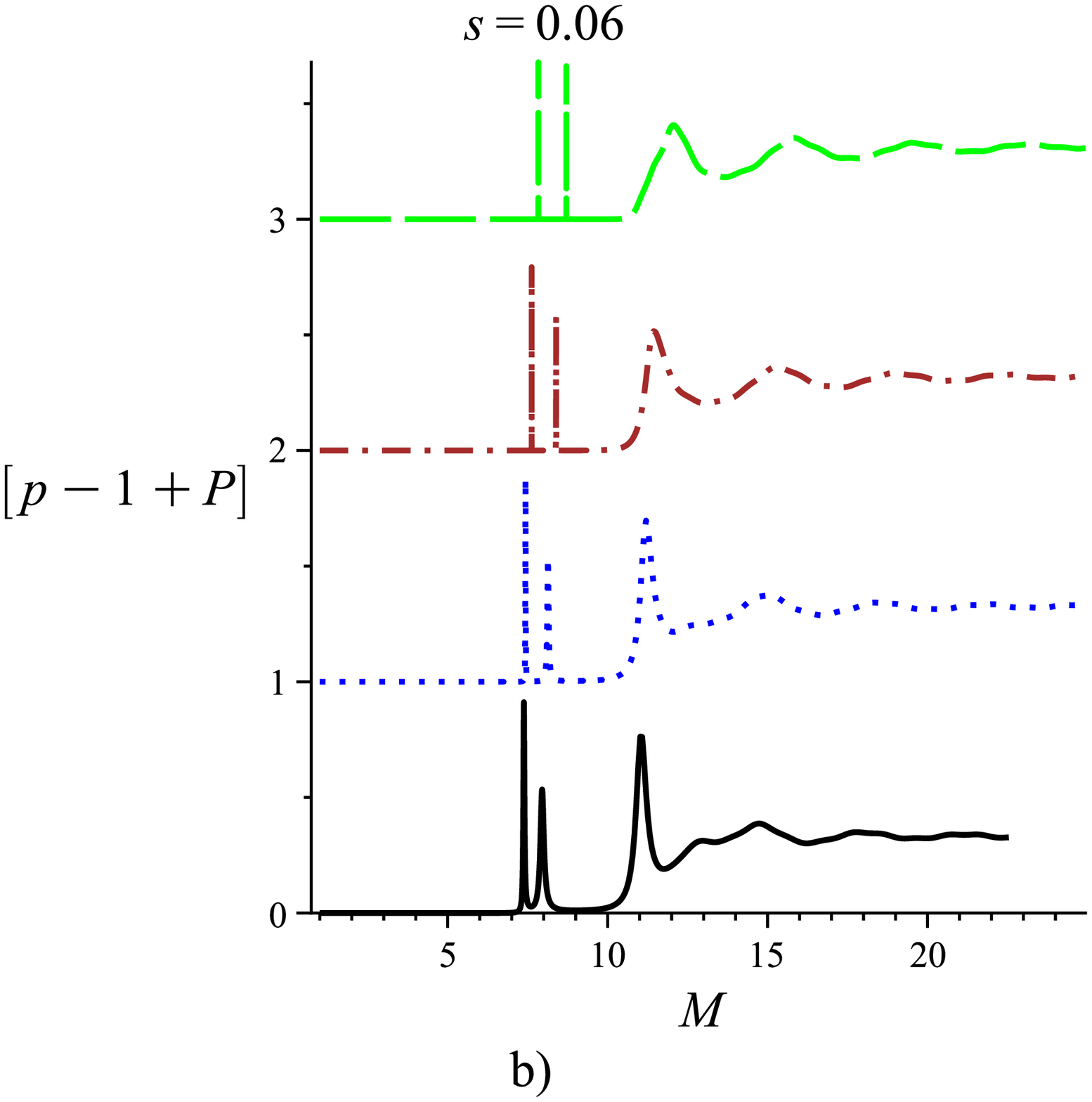}
\\
\includegraphics[{angle=0,width=6cm}]{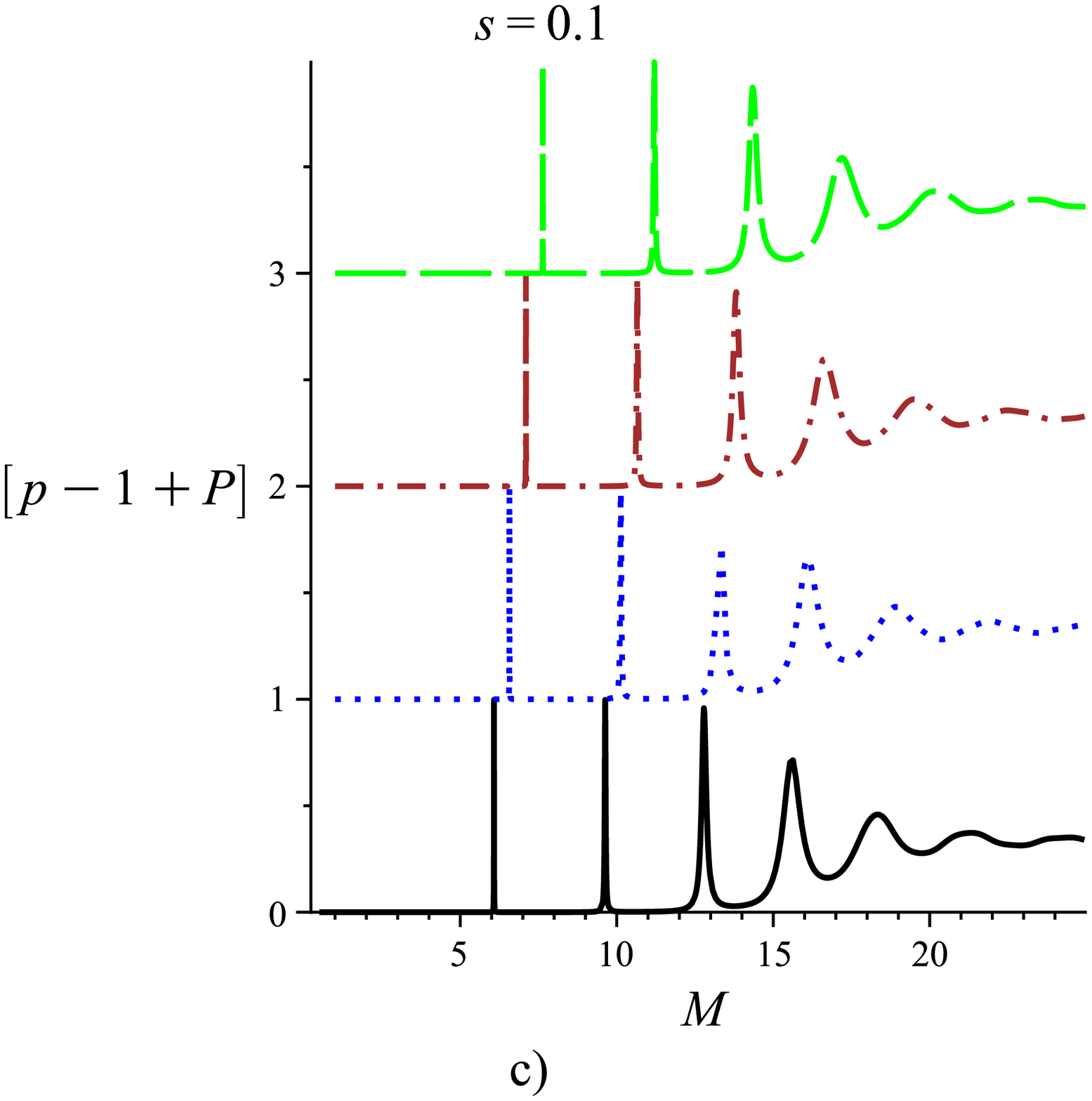}
\includegraphics[{angle=0,width=6cm}]{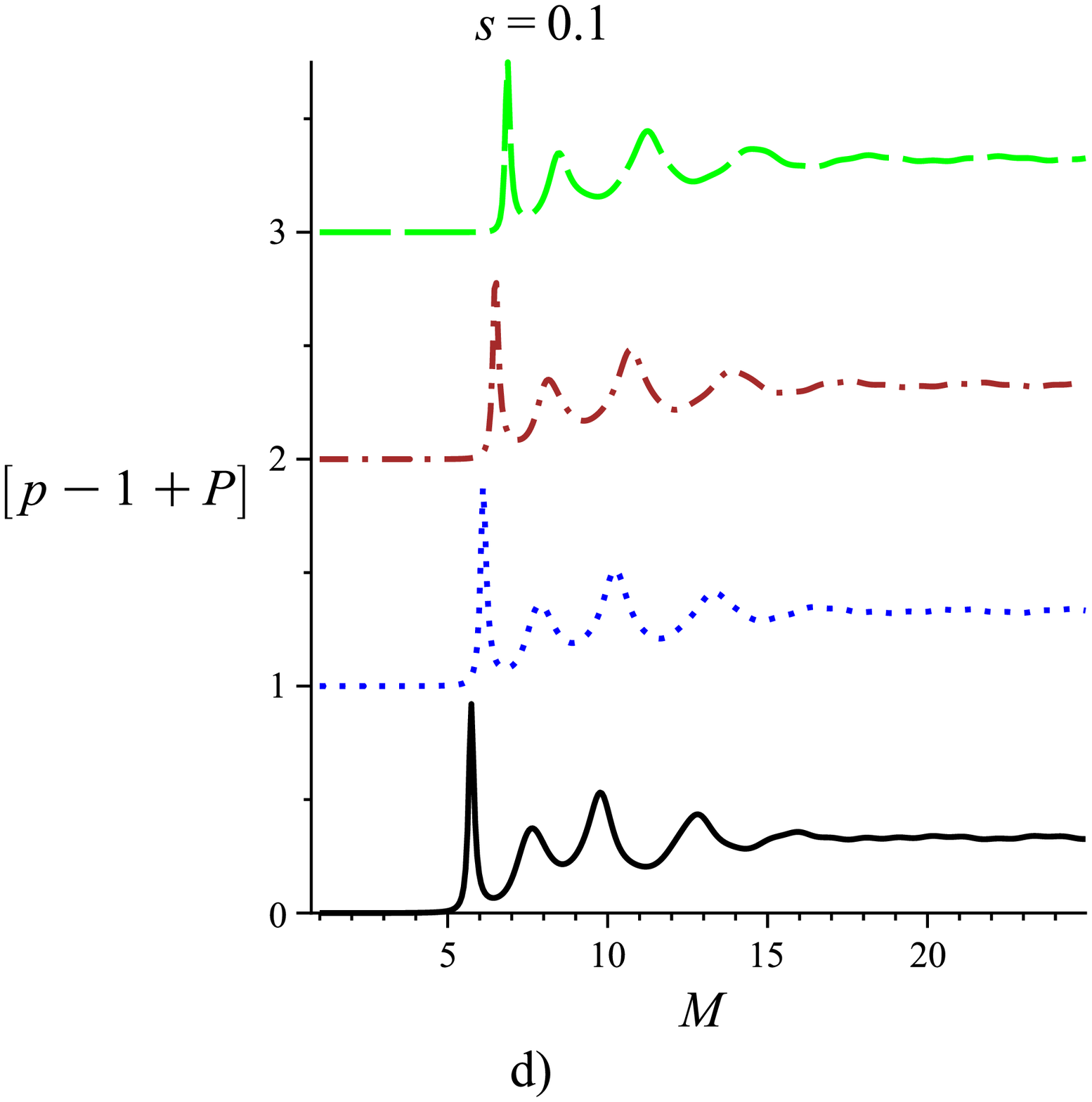}
\caption{A two-field model for $p-$balls in $(D,1)$-dimensions: $(p-1)+P$ as a function of $M$ for  coupling $F_1(\chi)$ (left) and $F_2(\phi,\chi)$ (right) for $\jmath=2$, $r_0=1$, $\eta=30$, $\lambda=30$,
and $p=2$ (black line), $p=3$ (blue dotted line), $p=4$ (brown dash
dotted line), $p=5$ (green longdash line). The plots are for $s=0.06$ and $s=0.1$, and correspond to potentials of Fig. \ref{Dp-potentials}.}
\label{probm}
\end{figure}

Fig. \ref{probm} depicts $P$ (rescaled for ease comparison) as a function of $M_{n\jmath}\equiv M$ for several values of $p$ and $s$, corresponding to the Schrodinger-like potentials of Fig. \ref{Dp-potentials}. The plots show several peaks of resonances, followed by a plateau for larger masses where $P=r_0/r_{max}$. From the figure we note that lower masses correspond to thinner peaks, or longer-lived resonances. The low-mass resonances are more difficult to be obtained numerically, due to the requirement of a larger number of digits of precision. Comparing Figs. \ref{probm}a and \ref{probm}c or Figs. \ref{probm}b and \ref{probm}d we see that lower values of $s$ correspond
to a larger number of resonance peaks. In addition, the peak separation is reduced for lower values of $s$. This shows that small values of $s$ are more effective for attaining resonances. The effect of the increasing in the number of extra dimensions $p$ is a displacement of the peak positions for larger masses, keeping the mass separation between the peaks almost unaltered. Figs. \ref{probm}b and \ref{probm}d shows the resonance peaks for coupling $F_2(\phi,\chi)$.
Comparing  this with Figs. \ref{probm}a and \ref{probm}c (related to $F_1(\chi)$), we see that for coupling $F_2(\phi,\chi)$ the number and masses of resonances is strongly reduced in comparison to the case of coupling $F_1(\chi)$.
Indeed, even for $s=0.06$, Fig. \ref{probm}b shows a pair of neighbor peaks, with only one with relative probability close to one. The sequence of almost equally spaced peaks present for coupling $F_1(\chi)$ (corresponding to Fig. \ref{probm}a) is now absent. This shows that coupling $F_2(\phi,\chi)$ is less effective for the occurrence of resonances.

\begin{figure}
\includegraphics[{angle=0,width=6cm}]{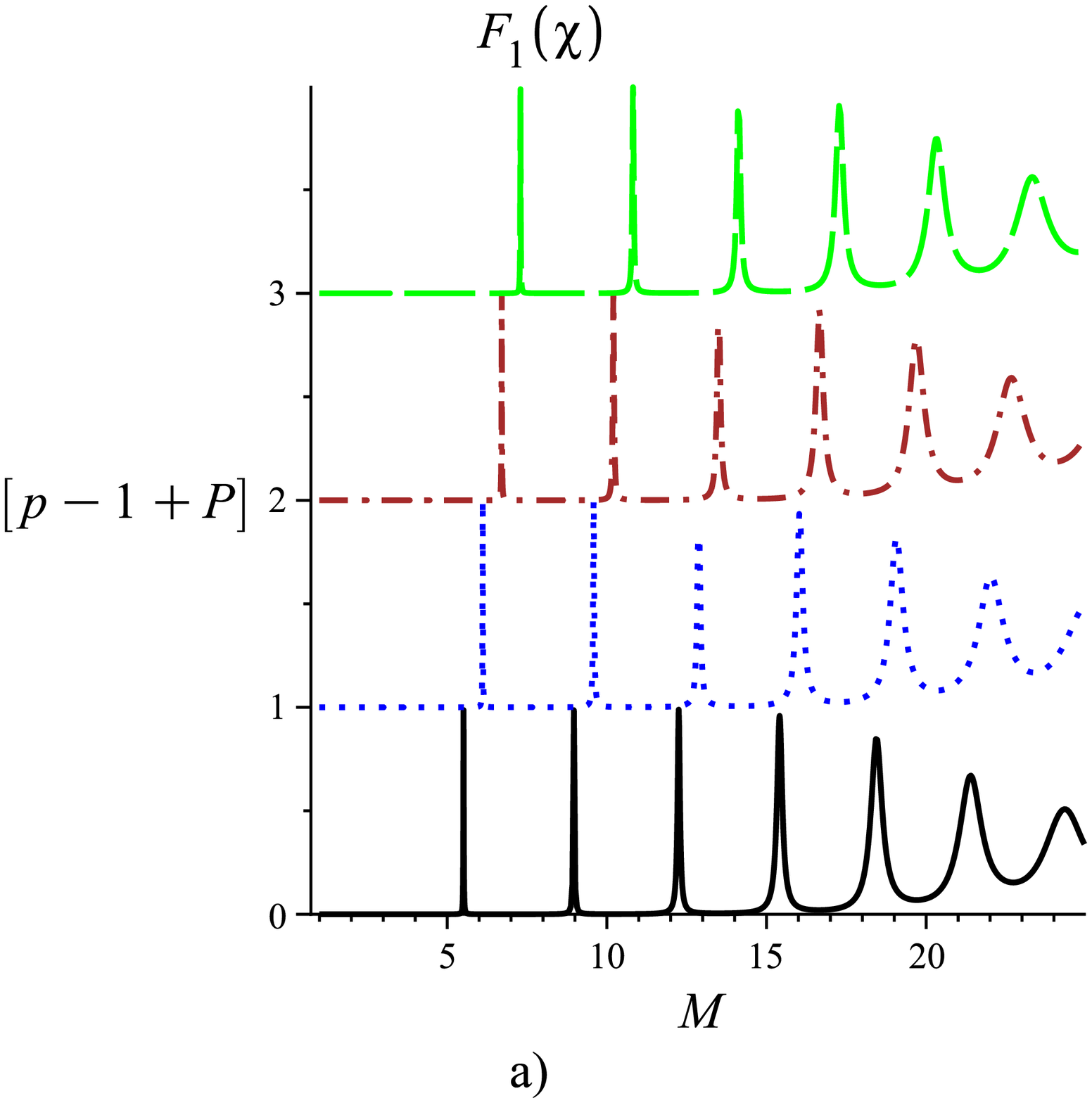}
\includegraphics[{angle=0,width=6cm}]{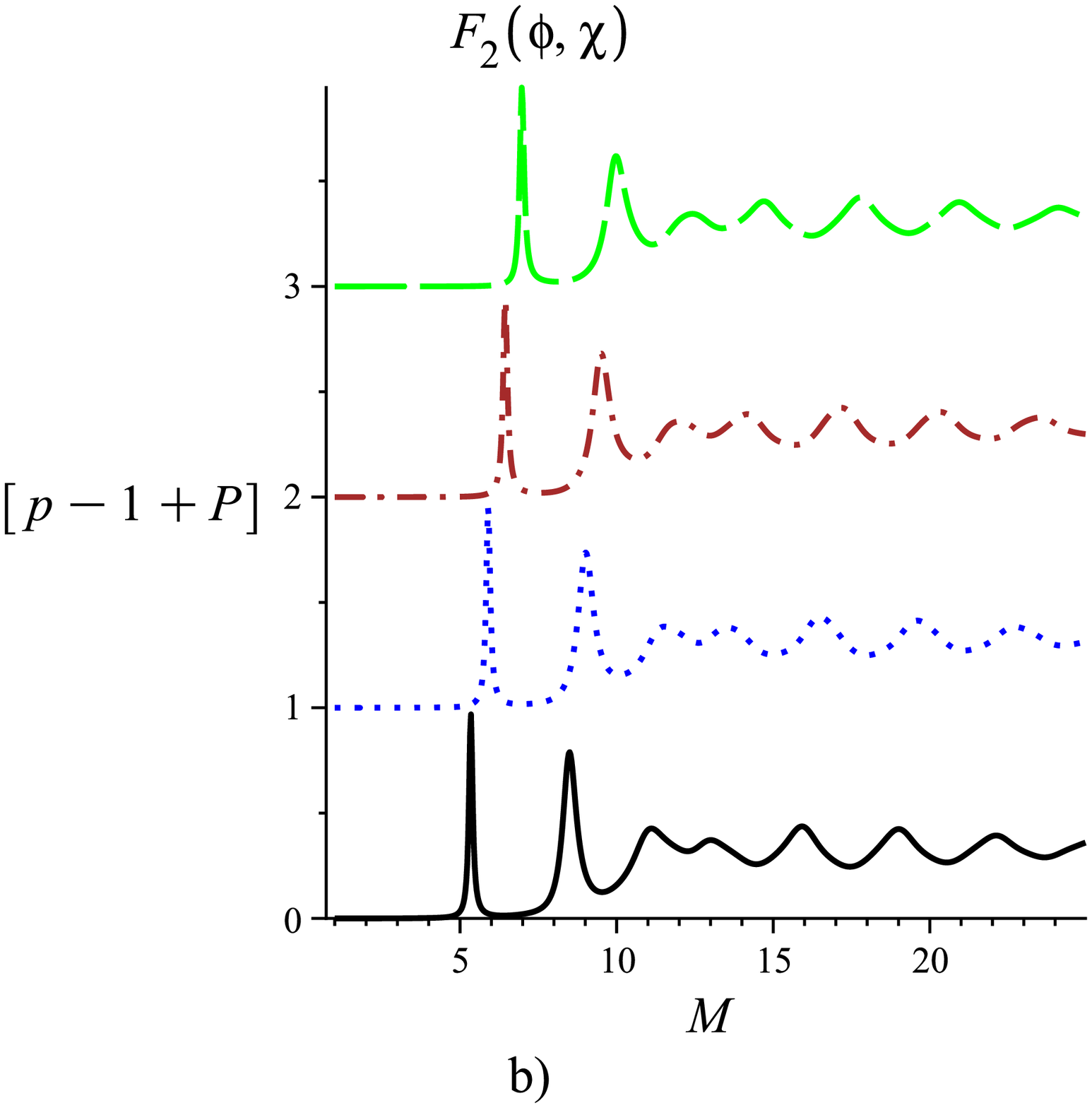}
\caption{A two-field model for $p-$balls in $(D,1)$-dimensions: $(p-1)+P$ as a function of $m$ for  coupling $F_1(\chi)$ (left) and $F_2(\phi,\chi)$ (right) for $\jmath=2$, $r_0=1$, $\eta=30$, $\lambda=100$, $s=0.06$
and $p=2$ (black line), $p=3$ (blue dotted line), $p=4$ (brown dash
dotted line), $p=5$ (green longdash line). The plots correspond to potentials of Fig. \ref{Dp-potentials-lambda}.}
\label{probm_lambda100}
\end{figure}

 Fig. \ref{probm_lambda100} shows some resonance peaks corresponding to $\eta=30$ and $\lambda=100$. Comparing with Figs \ref{probm}, we note that the resonances observed are more easy to be numerically obtained, but also thicker, meaning lower lifetimes. This is in accord with the analysis of the influence of the ratio $\lambda/\eta$ for the Schrodinger-like potentials.

Concerning to the angular sector of the decomposition of the weak scalar field, the number of extra dimensions is the key point. In the remaining of this section we will consider separately some specific choices of the number $D$ of spatial dimensions up to $D=5$. The procedure for larger values of $D$ is straightforward.

\subsection{$p-$balls in $(3,1)$-dimensions}
\label{31-p2}
The simplest choice is to consider $p-$balls in $(3,1)$-dimensions. In this case the only possibility is to construct a radial defect with $p=2$ spatial dimensions. This is a tube-like defect and has already been studied by us in Refs. \cite{cgms1,cgms2}.
Requiring $\rho(r)$ finite in $r= 0$ restricts the parameters to satisfy $\lambda s \ge 1/2$
when $\lambda > 1$.  For large values of $\lambda $, there exists a value $%
s_{0}$ so that for $\frac{1}{2\lambda }<s<s_{0}$, the effect of the field $%
\chi $ is stronger and the defect appears as a thick tube structure whose
center is localized between the origin and $r_{0}$.  The decomposition of the weak scalar field $\Phi(t,x^1,x^2)$ is
\begin{eqnarray}
\Phi(t,x^1,r,\varphi) = \sum_{n\ell}  \xi_{n\ell}(t,x^1) \varsigma_{n,\ell}(r) Y_\jmath(\varphi),
\end{eqnarray}
where the spherical harmonic is
 $Y_\ell(\varphi)=e^{i\ell\varphi}$
and  $\varsigma_{n,\ell}(r)$ satisfies a $(2,1)$-dimensional Klein-Gordon equation.

\subsection{$p-$balls in $(4,1)$-dimensions}

In this case we have two possibilities:  i) to construct a radial defect with $p=2$ spatial transverse dimensions and $(2,1)$ longitudinal dimensions, or ii) to construct a radial defect with $p=3$ spatial transverse dimensions and $(1,1)$ longitudinal dimensions. In the following we will consider these two possibilities separately.
\subsubsection{$p-$balls in $(4,1)$-dimensions with $p=2$ transverse dimensions}
The defect is characterized by a potential which generates, respectively, kink-like and lump-like solutions for the scalar fields $\phi(r)$ and $\chi(r)$, as well as energy density $T_{00}(r)$ with the same profile found for the case analyzed in Sect. \ref{31-p2} for $(3,1)-$dimensions.
This is expected since we have the same number (two) of transverse dimensions. All would follow the same as in Sec. \ref{31-p2}: the decomposition of the spherical harmonics $Y_\ell$, Schr\"odinger-like potentials $V_{sch}$ and relative probabilities $P$. The difference is that in the present case of $(4,1)-$dimensions we have $(2,1)$ longitudinal dimensions. This is reflected in the longitudinal part $\xi_{n\ell}$ of the decomposition of the weak scalar field $\Phi(t,x^1,x^2)$,
\begin{eqnarray}
\Phi(t,x^1,x^2,r,\varphi) = \sum_{n\ell}  \xi_{n\ell}(t,x^1,x^2) \varsigma_{n,\ell}(r) Y_\ell(\varphi),
\end{eqnarray}
which now satisfy a $(2,1)$-dimensional Klein-Gordon equation.

\subsubsection{$p-$balls in $(4,1)$-dimensions with $p=3$ transverse dimensions}
\label{41-p3}
The defect is a 3-dimensional sphere. For larger values of $\lambda$ and $s$, the defects looks like as a thin ball centered around $r_0$ and the field $\phi$ has stronger contribution to the energy density. On the other hand, when we have larger values of $\lambda$ and lower values of $s$ are formed peaks between origin and $r_0$, which results in higher contribution of the $\chi$ field and the defect has a thicker structure.  The spherical harmonic is $Y_\ell (\varphi,\theta)=\sum_{-\ell}^\ell Y_{\ell m}(\varphi,\theta)$.

\subsection{$p-$balls in $(5,1)$-dimensions}
\label{51-p4}
In this case we have three possibilities: i) to construct a radial defect with $p=2$ spatial transverse dimensions and $(3,1)$ longitudinal dimensions. The procedure for the transverse dimensions is analogous to Sec. \ref{31-p2}, with the exception that now the longitudinal part $\xi_{n\ell}$ of the decomposition of the weak scalar field $\Phi$ satisfy a $(3,1)-$dimensional Klein-Gordon equation. ii) to construct a radial defect with $p=3$ spatial transverse dimensions and $(2,1)$ longitudinal dimensions. The results for the transverse dimensions is analogous to Sec. \ref{41-p3}, with the exception that now the longitudinal part $\xi_{n\ell}$ of the decomposition of the weak scalar field $\Phi$ satisfy a $(2,1)-$dimensional Klein-Gordon equation. iii) to construct a radial defect with $p=4$ spatial transverse dimensions and $(1,1)$ longitudinal dimensions.

\section{A three-field model}
\label{sec_3field}
In this section we will consider some three-field models. The numerical analysis of bound and resonant states follows the same prescription done in the two previous sections and we will not pursue in this direction here, focusing mainly in the analysis of the Schr\"odinger-like potentials. We start with a simple extension of the previous model, given by \cite{blw}
\be
W(\phi,\chi,\sigma)=\lambda \bigg(\phi-\frac{1}{3}\phi^3-s\phi(\chi^2+\sigma^2) \bigg),
\ee
where $s$ is a real parameter.
The equation of motion for the scalar fields, Eq. (\ref{phi_D}) is rewritten, after a change of variables $d\xi=1/r^{p-1} dr$, as
\begin{eqnarray}
\frac{d\phi}{d\xi}&=&W_\phi,\\
\frac{d\chi}{d\xi}&=&W_\chi,\\
\frac{d\sigma}{d\xi}&=&W_\sigma.
\end{eqnarray}
One solution connecting the minima $(\pm1,0,0)$ of the potential is \cite{blw}
\begin{eqnarray}
\phi(\xi)&=&\tanh(2\lambda s\xi),\\
\chi(\xi)&=&\sqrt{\frac1s - 2}\cos(\vartheta)\sech(2\lambda s\xi),\\
\sigma(\xi)&=&\sqrt{\frac1s - 2}\sin(\vartheta)\sech(2\lambda s\xi).
\end{eqnarray}
with $0\le s \le 0.5$ and $0 \le \vartheta <2\pi$, where now $\vartheta$ is a new parameter of the model.
Back to $r$ variable, explicit expressions for the scalar field profiles and consequently for the energy density can be easily attained. We have,
for $p\ge2$,
\begin{eqnarray}
\phi(r)&=&\tanh(\tau_p), \\
\chi(r)&=&\sqrt{\frac1s - 2}\cos(\vartheta)\sech(\tau_p),\\
\sigma(r)&=&\sqrt{\frac1s - 2}\sin(\vartheta)\sech(\tau_p),\\
\rho(r) &=&\frac{\left( 2\lambda s\right) ^{2}}{r^{2p-2}}\mathrm{sech}^{4}(\tau_p)   \left\{ 1+\left( \frac{1}{s}-2\right) \mathrm{sinh}^{2}(\tau_p)\right\}.  \notag
\end{eqnarray}
with $\tau_p$ given by Eq. (\ref{taup}). One can interpret the $\phi$ field as forming a host hypersphere, with the fields $\chi$ and $\sigma$ giving its internal structure. The balancing of the internal fields is given by the parameter $s$. We can also consider the real scalar fields $\chi$ and $\sigma$ as the real and imaginary part of a complex scalar field $\zeta$, with the model given by
\be
W(\phi,\zeta)=\lambda \bigg(\phi-\frac{1}{3}\phi^3-s\phi|\zeta|^2 \bigg).
\ee
A simple coupling is $F_3=|\zeta|^2=(\chi^2+\sigma^2)$. The explicit solutions $\chi(r),\sigma(r)$ shows that this coupling recovers the results obtained for $F_1(\chi)$ from Sect. \ref{sec_2field}. Another coupling is $F_4=a\phi^2+b_1\chi^2+b_2\sigma^2$, with $a,b_1,b_2>0$. Cases $a=b_2=0, b_1=1$ or $a=0, b_1=b_2=1$ recover coupling $F_1$. For coupling $F_4$, the Schr\"odinger-like potential is
\be
V_4  =   \frac{\jmath(\jmath + p-2)}{r^{2}} + \eta a \tanh^2(\tau_p) +  \eta b \bigg(\frac{1}{s}-2 \bigg) \sech^2 (\tau_p),\, p=2,3,... ,
\ee
with $b = b_1 \cos^2(\vartheta) + b_2\sin^2(\vartheta)$.
\begin{figure}
\includegraphics[{angle=0,,width=6cm}]{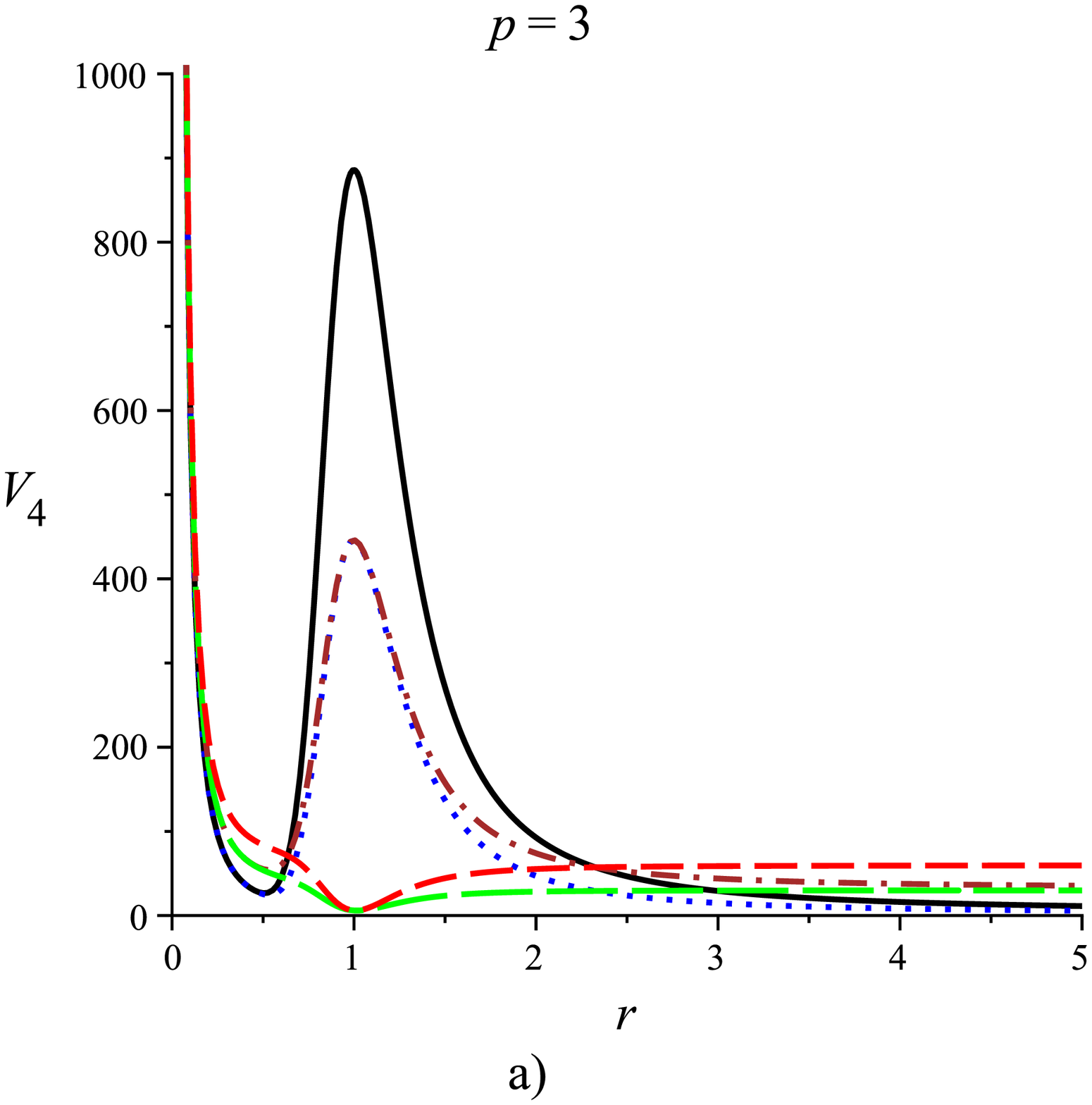}
\includegraphics[{angle=0,width=6cm}]{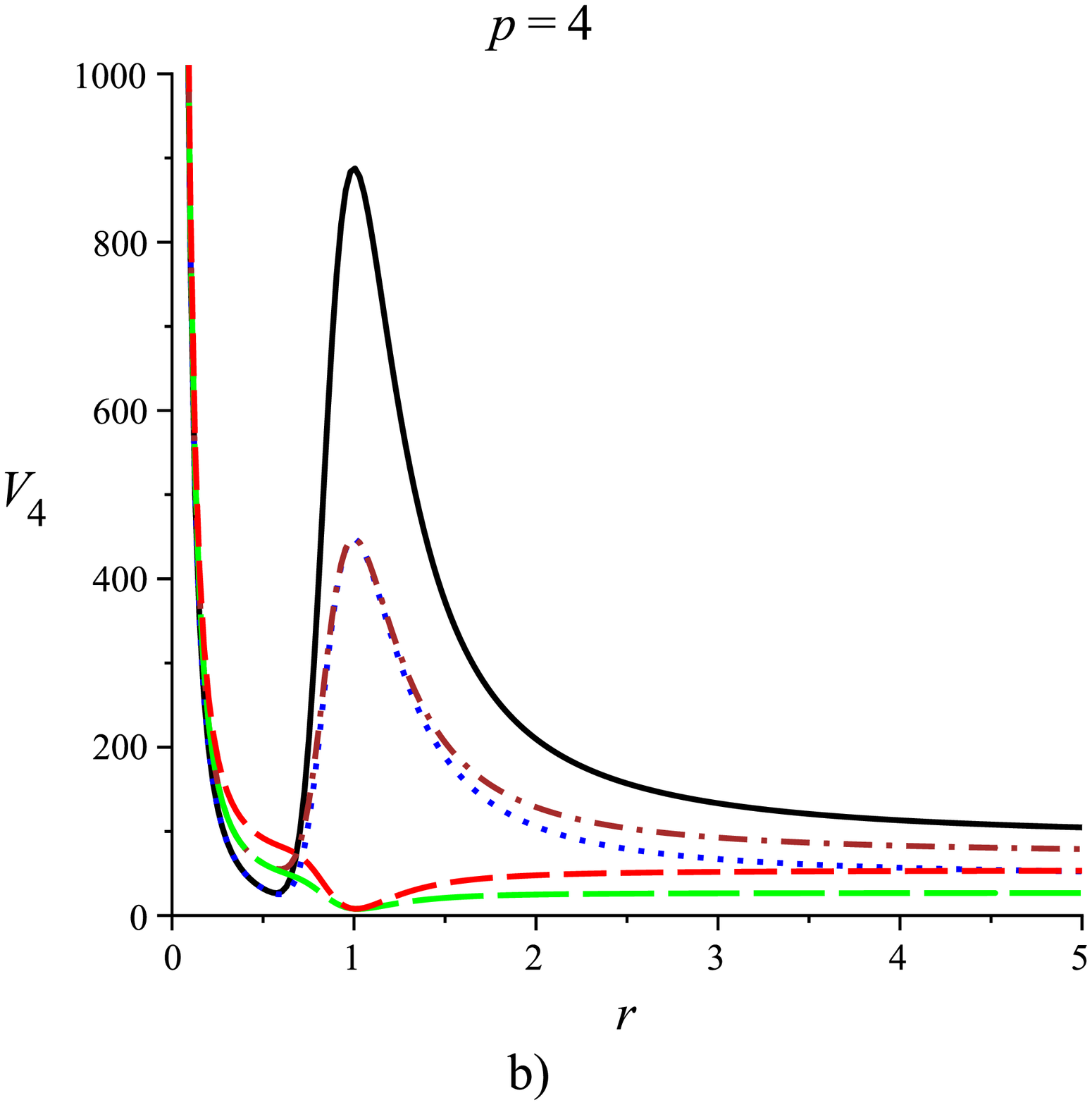}
\caption{Schr\"odinger potential $V_4$ for $\jmath=2$, $r_0=1$, $\eta=30$, $\lambda=30$, $s=0.06$
and $a=0,b=2$ (black line), $a=0,b=1$ (blue dotted line), $a=1,b=1$ (brown dash
dotted line), $a=1,b=0$ (green longdash line), $a=2,b=0$ (red dashed line). Plots are for $p=3$ (left) and $p=4$ (right).}
\label{figV4}
\end{figure}
Figure \ref{figV4} shows potential $V_4$ for $p=3$, $p=4$ and several values of parameters $a,b$. From the figure we see that a parameter $b\neq0, a=0$ results in a local maximum around $r=r_0$ that increases with $b$. With $b\neq0$, a value $a\neq0$ contributes to a small enlargement of the peak of $V_4$ around $r=r_0$. For  $b=0, a\neq0$ the local maximum disappears and only bound states are possible. The analysis shows that a parameter $b\neq0$ (meaning a quadratic coupling with fields $\chi$ or $\sigma$) is crucial for the occurrence of resonances. Also a quadratic coupling with field $\phi$ is of secondary importance, when compared with the effect of similar couplings with the other two fields that form the defect. A quadratic coupling with only the $\phi$ field has no effect concerning to resonances. This illustrates the importance of the secondary fields $\chi,\sigma$ that give to the defect an internal structure. Comparing Figs. \ref{figV4}a and \ref{figV4}b we see that for $p=3$ we have more possibility for resonant states in comparison to $p=4$. The increasing of $V_4(r\to\infty)$ with $p$ shows that an intermediate value of $p$ is better for attaining bound states, in a similar conclusion achieved in Sect. \ref{sec_1field} for one-field models. For $p=2$ the potential $V_4$ is a monotonically decreasing function, and there is neither bound nor resonant states.

Other couplings can be considered, but to further illustrate the generality of the construction of $p$-balls, here we choose to consider another model, restricted to a $Z_2\times Z_2$ symmetry in the $\chi$ and $\sigma$ axis \cite{blw}:
\be
W(\phi,\chi,\sigma)=\lambda \bigg(\phi-\frac{1}{3}\phi^3-s\phi(\chi^2+\sigma^2)  + rg\sigma^2\bigg).
\ee
For $s>0$ and $-1<g<1$ the corresponding potential $V(\phi,\chi,\sigma)$ has six minima given by (in units of $\xi$)
\begin{eqnarray}
v_{1,2}&=&(\pm1,0,0),\\
v_{3,4}&=&\biggl(0,\pm \sqrt{\frac1s},0\biggr),\\
v_{5,6}&=&\biggl(g,0,\mp \sqrt{\frac1s (1-g^2)}\biggr).
\end{eqnarray}
For $s\neq0$ the only solution connecting the minima $v_{1,2}$ is the one-field limit given by $\chi=\sigma=0$ and $\phi(\xi)=\tanh(\xi)$. Nontrivial solutions for the three scalar fields can be obtained connecting minima $v_{3,4}$ to $v_{5,6}$ and are given, with $sg^2=1$, by \cite{blw}
\begin{eqnarray}
\phi(\xi)&=&\frac g2(1+\tanh(\xi/g)),\\
\chi(\xi)&=&\pm \frac12 \sqrt{\frac1s}(1-\tanh(\xi/g)),\\
\sigma(\xi)&=&\pm \frac12 \sqrt{\frac1s(1-g^2)}(1+\tanh(\xi/g)).
\end{eqnarray}
Back to $r$ variable, and for $p\ge2$, we can obtain the following expressions for the scalar fields:
\begin{eqnarray}
\phi(r)&=&\frac g2(1+\tanh(\xi(r)/g)),\\
\chi(r)&=&\pm \frac12 \sqrt{\frac1s}(1-\tanh(\xi(r)/g)),\\
\sigma(r)&=&\pm \frac12 \sqrt{\frac1s(1-g^2)}(1+\tanh(\xi(r)/g)),
\end{eqnarray}
with $\xi(r)$ given by Eq. (\ref{xir_p2}), for $p=2$ or (\ref{xir_p}), for $p=2,3,....$

\begin{figure}
\includegraphics[{angle=0,width=6cm}]{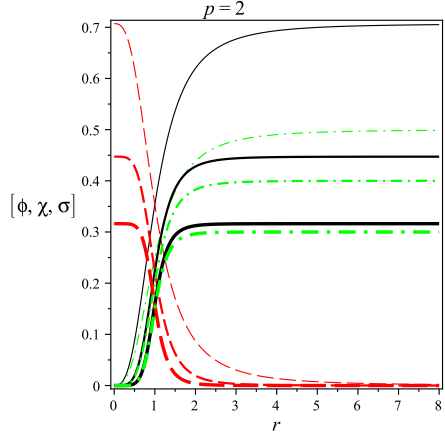}
\caption{$2-$balls in $(D,1)$-dimensions for three scalar fields: functions $\protect\phi(r)$ (black solid lines), $\protect\chi(r)$
(red traced lines) and $\sigma(r)$ (green dashdotted lines). We fix $r_0=1$ and $\lambda=30$ and couplings $sg^2=1$. We have $s=1.01$ (thinner lines), $s=5$ and $s=10$ (thicker lines). }
\label{fig_3field}
\end{figure}
\begin{figure}
\includegraphics[{angle=0,width=6cm}]{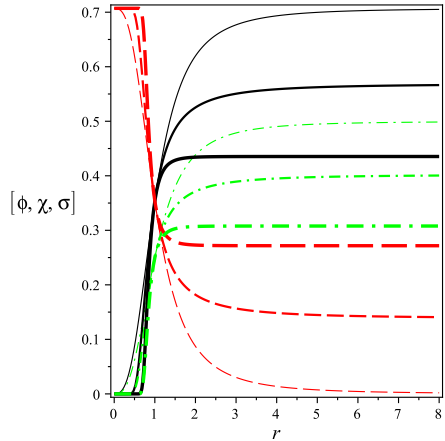}
\caption{$p-$balls in $(D,1)$-dimensions for three scalar fields: functions $\protect\phi(r)$ (black solid lines), $\protect\chi(r)$
(red traced lines) and $\sigma(r)$ (green dashdotted lines). We fix $r_0=1$ and $\lambda=30$, $s=2$ and $sg^2=1$. We have $p=2$ (thinner lines), $p=4$ and $p=8$ (thicker lines). }
\label{fig_3field_p}
\end{figure}

Figure \ref{fig_3field} shows plots of $\phi(r)$, $\chi(r)$ and $\sigma(r)$ for $p=2$ and fixed $\lambda, r_0$ and several values of $s$. From the figure we see that all the three fields have a kink-like structure around $r=r_0$. Note that an increasing in $s$ (and correspondingly a decreasing in $g$) results in a thinner defect. The same effect occurs with the increasing of $p$, as can be seen from Fig. \ref{fig_3field_p}.
For coupling $F_5=a\phi^2+b_1\chi^2+b_2\sigma^2$, with $a,b_1,b_2>0$, the Schr\"odinger-like potential is
\begin{eqnarray}
V_5  &=&   \frac{\jmath(\jmath + p-2)}{r^{2}} + \eta a \frac {g^2}4(1+\tanh(\xi(r)/g))^2 \nonumber \\
&&+  \eta b \frac14 {\frac1s}(1-\tanh(\xi(r)/g))^2 + \eta c \frac14 {\frac1s(1-g^2)}(1+\tanh(\xi(r)/g))^2 ,\, p=2,3,... ,
\end{eqnarray}
which can also be investigated for possible occurrence of bound and resonant states.


\section{Remarks and conclusions}
\label{sec_concl}

In this work we introduced p-balls as topological defects in $(D,1)$
dimensions constructed with $\mathcal{M}\ge 1$ scalar fields which depend radially on
only $2 \le p\le D-2$ spatial dimensions. Such defects are characterized by
an action that breaks translational invariance and are inspired on the
physics of a brane {with $D-p$ extra dimensions and $p$ transverse}
spatial dimensions. After presenting the general formalism, we have
found BPS {solutions living in the $p$ transversal dimensions} and
proved their stability.  {In order to analyze the localization
of a scalar field $\Phi$ (named a weak field because we
can neglect backreaction effects) in $D-p+1$-dimensions, we have considered a general coupling
between the weak field and the scalars fields generating the topological
defect. Our results have shown the existence of bound and/or resonant states
which were addressed after a convenient decomposition of the weak scalar
field in $D-p+1$ -dimensional spin-$0$ modes and its respective amplitudes in
the transverse $p$-dimensions. As usual the spin-$0$ modes satisfy a $D-p+1$
dimensional Klein-Gordon equation whereas the amplitudes are decomposed in an
angular part in terms of the generalized spherical harmonics and a radial part
satisfying a Schr\"odinger-like equation.}

We have particularized our analysis to the class of models where the
$p-$balls are formed with one, two and three scalar fields.  For the class
of one-field models we have considered a region of parameters where a larger
number of bound states are formed. It is characterized for $p-$balls with
larger internal structure (large $q$), intermediate number of extra dimensions
$3<p<5$, lower energy (lower coupling $\lambda$) and higher coupling parameter
$\eta$.  The two-field models resemble the Bloch brane model where we have
considered two type of couplings. A quadratic coupling $\chi^2$ is better
than the quartic one $\phi^2\chi^2$  concerning to the occurrence of either
bound or resonant states. We have verified the presence of the second scalar
field $\chi$ contributes to the increasing  of trapping spin-$0$ particles.
In these two-field models, the larger is $p$, the greater is the number of
bound states, but the number of resonances is roughly the same. We  have also  explored some three-field models, finding from
the analysis of the Schrodinger-like potential that in some cases an
intermediate value of $p$ is better for the occurrence of bound states.  

Concerning to the influence of $\jmath$, the probability of occurrence of bound states with $\jmath=0$ is higher in comparison to states with larger values of $\jmath$. This was verified numerically. On the other hand, for resonances the analysis of the Schr\"odinger potential shows that the behavior depends on the model. In our one-field model there is no such possibility for $\jmath=0$ whereas the possibility of occurrence of such states is low but grows with the increasing of $\jmath$. On the other hand for our two-field model the number of resonant states seems to accompany the tendency of bound states, with a decreasing in number with $\jmath$.

It is worthwhile to observe that we where able to establish a connection between the number of extra dimensions and the capability of trapping massive states. Such a study was attained by analyzing qualitatively the Schrodinger-like potentials and by performing numerical analysis. For all class of models considered, we could identify that initially there is an increasing of the number of bound states with $p$.  For the one-field model this occurs up to a certain number of transverse dimensions; in this way there is an optimal number of transverse dimensions for trapping states. On the other hand, for the two-field model considered, we have investigated up to $p=7$ and the number of bound states always grow with $p$, whereas the number of resonances seems to be independent of $p$. Also we have confirmed  that, for fixed $p$, the growing of the internal structure of the defect (connected with larger values of $q$ for the one-field model and smaller values of $s$ in the two-field and three-field models) lead to an increasing of the number of bound and/or resonance states.

Whether our results represent a general characteristic of the $p$-balls or the main conclusions are results of the particularities of the models here described deserves to be better understood. Advances in this direction will be reported elsewhere.


\begin{acknowledgments}
{The authors thank to the Brazilian agencies FAPEMA, CAPES and CNPq for financial support. We also to thank M. Hott for stimulating fruitful discussions concerning to stability analysis.}
\end{acknowledgments}


\end{document}